\documentclass[fleqn,11pt]{article}
\usepackage{graphicx}

\newcommand{\na}{\nabla}
\newcommand{\C}{\cdot}
\newcommand{\ep}{\epsilon}

\newcommand{\vphi}{\varphi}

\newcommand{\lra}{\longrightarrow}

\newcommand{\Gafunc}[1]{\Gamma(#1-\frac{D}{2})}
\newcommand{\quddiv}[1]{\frac{#1 N_c}{(4\pi)^{D/2}}
       (\frac{\mu^2}{m^2})^{\epsilon/2}\Gafunc{1}}
\newcommand{\logdiv}[1]{\frac{#1 N_c}{(4\pi)^{D/2}}
       (\frac{\mu^2}{m^2})^{\epsilon/2}\Gafunc{2}}

\newcommand{\pa}{\partial}

\newcommand{\td}{\tilde}

\begin{document}

\begin{center}
{\Large\bf Large $N_c$ Expansion in Chiral Quark Model of Mesons}\\[5mm]
Xiao-Jun Wang\footnote{E-mail address: wangxj@mail.ustc.edu.cn} \\
{\small
Center for Fundamental Physics,
University of Science and Technology of China\\
Hefei, Anhui 230026, P.R. China} \\
Mu-Lin Yan\\
{\small CCST(World Lad), P.O. Box 8730, Beijing, 100080, P.R. China}\\
{\small and}\\
{\small  Center for Fundamental Physics,
University of Science and Technology of China\\
Hefei, Anhui 230026, P.R. China}\footnote{mail
address}
\end{center}
\begin{abstract}
\noindent
\begin{center}
\begin{minipage}{5in}
{\small We study SU(3)$_L\times$SU(3)$_R$ chiral quark model of mesons
up to the next to leading order of $1/N_c$ expansion. Composite vector and
axial-vector mesons resonances are introduced via non-linear realization
of chiral SU(3) and vector meson dominant. Effects of one-loop graphs of
pseudoscalar, vector and axial-vector mesons is calculated systematically
and the significant results are obtained. We also investigate correction
of quark-gluon coupling and relationship between chiral quark model and
QCD sum rules. Up to powers four of derivatives, chiral effective
lagrangian of mesons is derived and evaluated to the next to leading order
of $1/N_c$. Low energy limit of the model is examined. Ten low energy
coupling constants $L_i(i=1,2,...,10)$ in ChPT are obtained and agree with
ChPT well.}
\end{minipage}
\end{center}
\end{abstract}

\section{Introduction}

The Chiral Quark Model (ChQM) and its extensions\cite{Wein79}-\cite{WY98}
have attracted much interest continually during the last two decades.
The reasons why the ChQM is so attractive are both because of its
elegant descriptions of the features of low-energy QCD and because of its
great successes in the different aspects of phenomenological predictions
on hadron physics (see the brief review on ChQM in section 2 of
this paper). However, so far, the studies on ChQM face two challenges
as soon as we start going beyond lowest order: 1) How to go beyond
the chiral perturbative theory(ChPT)? In order to capture physics
between perturbative QCD and ChPT\cite{GL84}-\cite{GL85b}, freedoms of
meson resonances have to be included into dynamics. It was well known
that, in this energy region, there are no well-defined method to yield a
convergence expansion. Therefore, we need some phenomenological models
including spin-1 meson resonances as well pseudoscalar meson to describe
the physics in this energy region. Although there are some
discussion\cite{Bejinen96a,Brise97}, role of mesons resonances in ChQM are
still uncertain. 2) How to go beyond large $N_c$ limit? The ChQM-studies
are limited to be on the level to catch merely the leading order effects
of large $N_c$ expansion. In other words, like other QCD-inspired models
(e.g., bag model, Skyrme model, pole model and so on),
only the leading order evaluations in ChQM are legitimate and practicable,
and there is no way to calculate the next to leading order
contributions in the model so far, even though there are some arguments on
suppression of $1/N_c$ in ChQM. This is a bad shortage. It makes the
calculations of ChQM uncontrolled approximations, in that there is no
well-defined way to put error bars on the predictions.
In this sense, any conclusions coming from ChQM would become uncertain.
In order to improve this unhappy situation of ChQM,
it is necessary and urgent to study contributions beyond the leading order
in ChQM. With these motivation, in this paper, we first extend ChQM to
include spin-1 meson resonances. Then we like to provide a systematical
study to illustrate the ChQM-calculations to be legitimate up to the next
to leading order of $1/N_c$ expansion.

The simplest version of ChQM which was originated by Weinberg\cite{Wein79},
and developed by Manohar and Georgi\cite{GM84} provides a QCD-inspired
description on the simple constituent quark model(M-G model). In view of
this model, in the energy region between the chiral symmetry-breaking
scale ($\Lambda_{CSSB}\simeq 2\pi f_\pi\simeq 1.2GeV$) and the confinement
scale ($\Lambda_{QCD}\sim 0.1-0.3 GeV$), the dynamical field degrees of
freedom are constituent quarks(quasi-particle of quarks), gluons and
Goldstone bosons associated with Chiral Symmetry Spontaneous Breaking(CSSB).
In this quasiparticle description, the effective coupling between gluon
and quarks is small and the important interaction is the coupling between
quarks and Goldstone bosons. Fields of pseudoscalar mesons in M-G
model are treated as composite fields of quarks and anti-quarks instead of
independent dynamical freedoms. Dynamics of pseudoscalar mesons is
described by an effective Lagrangian, which is derived via integrating out
the quark fields("freeze" the quark-freedom).

There are no meson resonances in orginal M-G model. From viewpoint
of chiral symmetry, role of the lowest meson resonances in low
energy effective theories has been analyzed systematically in
Ref.\cite{Ecker89}. The authors illustrate that all meson resonance
fields can be treated on same level: they carry non-linear
realizations of chiral $SU(3)$ which are uniquely determined by the
known transformation properties under the vectorial subgroup
$SU(3)_V$(octets and singlet). This is an attractive symmetry
property on meson resonances and shall be adopted in this present paper.
Of course, it is not necessary to describe the degrees of freedom of
vector and axial-vector mesons by antisymmetric tensor
fields\cite{Ecker89}, and there are other phenomenological
successful attempts to introduce spin-1 meson resonances as massive
Yang-Mills fields\cite{Li95a,Brise97,Bando}. In addition, it was
well known that vector meson dominant(VMD) is a very important
phenomenological feature in electro-weak interaction of hadrons. In
ChQM, it is nature to introduce spin-1 meson resonances via VMD and
is especially convenient to describe relevant degrees of freedom in
terms of familiar vector representation(see section 2). Similar to
M-G model, all freedoms of meson resonances in ChQM are treated as
composite fields of quark pairs, effective lagrangian describing
dynamics of mesons is derived via effects of quark loops.

According to large $N_c$ arguement\cite{tH74}, the color coupling between
quarks and gluon fields also yields the leading order contribution at
large $N_c$ limit. However, this contribution was omitted in orginal M-G
model and its many extensions. This problem has been first noticed in
ref.\cite{Esp90}, in that reference the authors investigated the effects
of leading gluon coupling($O(\alpha_s)$) in M-G model. Since it is
unknown how to perform analytically the
remaining integration over the gluon fields and this coupling become
rather unclear in presence of meson composite fields, the authors
suggested a possible way, which is to parameterize phenomenologically this
effective coupling and make a weak gluon-field expansion around the
resulting physical vacuum. In this present paper, in order to carry out a
complete study on $1/N_c$ expansion at leading order, the investigation in
ref.\cite{Esp90} will be extended to include pseudoscalar, vector and
axial-vector mesons. In ref.\cite{Esp90}, the authors showed that the
gluon coupling contributes to low energy coupling constants $L_3$, $L_5$,
$L_8$ and $L_{10}$ but does not contribute to other coupling constants of
ChPT at $O(p^4)$. In particuar, the correction due to gluon coupling
contributes to $L_3$ and $L_{10}$ is about $50\%$. It makes $L_3$ agree
with experiment data. This is an
important improvment so that the quark-gluon coupling in ChQM can not be
omitted simply. From these results, it seems to be little fortunate
ingredients in phenomenological success of those chiral quark model
without the gluon coupling. However, in this paper we will show that, in
case of presence of spin-1 meson resonances, the contribution of the gluon
coupling is suppressed due to mixing between axial-vector and pseudoscalar
mesons. The results are that gluon coupling corrects to $L_{10}$ about
$10\%$ and to $L_3$ about $3\%$ only. Simultaneously, $L_3$ and $L_{10}$
are compensated by effect of exchange of spin-1 meson resonances and,
hence, they agree with data well. Therefore, the gluon correction 
in presence of meson resonances is much smaller than one obtained in
ref.\cite{Esp90}. Thus, it becomes legitimate to treat quark-gluon
coupling as perturbation in the present model.

ChQM is a non-renormalizable effective field theory, since there are
too few free parameters to introduce enough counterterms to absorb
divergences from quark-loop and meson-loop integral. Therefore, we have to
parameterize the divergences from quark loops as well as meson loops,
i.e., we need to introduce cut-offs for regularizing these divergences.
We first discuss the divergences from quark loops. From viewpoint of QCD,
these divergences are caused by ``point particle'' approximation of
mesons. Hence the cut-offs to regularize these divergences can be treated
as description on scale of interaction of quark pairs in mesons. In this
paper, corresponding to spin zero and one, two cut-offs are needed
to describe internal interaction scale of $0^-$ and $1^\pm$ mesons. It is
well known that, at very low energy, there should be a cut-off
which correspond to energy scale of chiral symmetry spontaneous broken,
$\Lambda_{\rm CSSB}$. In presence of spin-1 mesons, another cut-off should
be introduced with energy increasing. Unfortunately, there are not any
practical methods helping us to understand properties of this cut-off from
QCD. It is only expected that, due to requirement of consistence, this
cut-off should be larger than $\Lambda_{\rm CSSB}$ and all spin-1
meson masses. In sect. 3, we will fit values of these two cut-offs
phenomenologically and show that they are consistent indeed.

As soon as we start studying the next to leading order of $1/N_c$
expansion, there are two contributions contained the same as at leading
order. One contribution is from meson loops and another contribution is
from quark-gluon coupling. It has to be recognized that it is too hard
to systematically study quark-gluon coupling at the next to leading order
of $1/N_c$ expansion. The reason is that we will face many uncertain
ingredients and very tedious calculation. In fact, even though at leaing
order of $1/N_c$ expansion, it is impossible to completely investigate the 
contribution from quark-gluon coupling, since theoretically, the numbers
of these kinds of feynman diagrams are infinite(see detail in sect. 4).
Fortunately, the dominant contribution can be captured here. The
phenomenological results show that the quark-gluon coupling is small even
at leading order. We can expect that it is legitimate to ignore the next
to leading order contribution of quark-gluon coupling.\footnote{In this
paper, the contribution of quark-gluon coupling denotes the exchange
effects of gluon in internal of quark loop. In this case gluon is {\sl
soft}(see discussion in sect. 8). Here we ignore another kind of gluon
effect which is exchange of gluon between two quark loops, since study on
this effect is beyond the method provided by ref.\cite{Esp90}. In this
case gluon can be {\sl hard}. It is well known that this effect is
suppressed by $1/N_c$ expansion and OZI rule.}
Due to these reasons, the effects of quark-gluon coupling will
be omitted at the next to leading order of $1/N_c$. The main content of
the present paper is to study effects of one-loop of pseudoscalar, vector
and axial-vector mesons systematically.

Obviously, the calculation on meson one-loop effects is beyond one of ChPT
in which only pseudoscalar meson one-loop graphs are involved. So that it
will help us to understand the dynamics with energy scale lying between
perturbative QCD and ChPT more accurately. Especially, many processes
suppressed by $1/N_c$ will become calculable explicitly. However,
although in principle there are no problems on calculation of meson
one-loop graphs, practical calculation is very tedious due to complicated
dynamics structure of effective lagrangian and due to various high-order
divergences coming from propagators of spin-1 mesons(see sect. 6).
Hence we need to search a method to simplify our calculation.
Noting that we work in the framework of a non-renormalizable truncated
field theory, we have to introduce some cut-offs to regularize divergences
from meson loops. These cut-offs can be interpreted as meson-interaction
scale in the ``effective vertices'' generated by meson loops. Therefore,
the effects of meson loops describe ``very long-distance'' correction
comparing with one of quark loops. Contribution of the former should be
much smaller than the latter. In the other word, the cut-offs from meson
loops should be much smaller than one from quark loops, e.g.,
$\Lambda_{\rm CSSB}$. It indicates that momentum transfer in meson loops
is small. If these cut-offs is smaller than vector
meson masses, the dominant contributions will come from logarithmic
divergences and all high order divergences can be omitted. Sequentially,
the complicated calculation will be simplified. In sect. 6, this point
will be examined explicitly.

Although we still do not know how to obtain a low energy effective theory
from QCD directly, relationship between ChQM and ChPT is straightforward.
At very low energy, when the freedoms of spin-1 mesons are
``freezed''(i.e., integrating out these freedoms by path-integral), ChQM
should return to ChPT. It is well known that ChPT is a rigorous
consequence of the symmetry pattern in QCD and its spontaneous breaking.
So that it is necessary to examine low energy limit of ChQM and check
whether it agrees with ChPT or not. In this article, we will calculate ten
coupling constants of ChPT at $O(p^4)$ from ChQM and show that they agree
with ChPT well.

The paper is organized as follows. In sect. 2, we give a brief review on
ChQM and extend ChQM to include spin-1 meson resonances. In sect. 3, we
derive effective lagrangian of mesons generated by quark loops. In sect.
4, the correction of quark-gluon coupling is studied. We further
compare ChQM and QCD sum rules and determine the value of gluon
condensate obtained through integrating over quark and gluon fields.
General formulas for meson one-loop contribution are in sect. 5
and systematical study on meson one-loop graphs is in sect. 6. In sect 7,
we discuss low energy limit of ChQM and some phenomenological results.
Sect. 8 is devoted to summary and discussion.

\section{$SU(3)_L\times SU(3)_R$ Chiral Quark Model of Mesons}
\setcounter{equation}{0}

The QCD lagrangian with three flavor quark $\bar{\psi}=(\bar{u},
\bar{d},\bar{s})$ is,
\begin{eqnarray}
\label{2.1}
  &&{\cal L}_{QCD}(x)={\cal L}_{QCD}^{0}+\bar{\psi}(\gamma \cdot v+\gamma
    \cdot a\gamma_5)\psi-\bar{\psi}(s-i\gamma_5p)\psi, \nonumber \\
  &&{\cal L}_{QCD}^0=\bar{\psi}(i\gamma \cdot \partial+\gamma \cdot G)\psi
    -\frac{1}{4}G_{\mu\nu}^aG^{a\mu\nu}, \hspace{0.5in}(a=1,2,...,8),
\end{eqnarray}
where the $3\times 3$ gluon field matrix is given by
\begin{equation}
\label{2.2}
   G_\mu=g_s\frac{\lambda^a}{2}G_{\mu}^a(x),
\end{equation}
and
\begin{equation}
\label{2.3}
   G_{\mu\nu}^a=\partial_\mu G_\nu^a-\partial_\nu G_\mu^a
              -g_sf_{abc}G_\mu^bG_\nu^c
\end{equation}
is the gluon field strength tensor and $g_s$ is the color coupling
constant($\alpha_s=g_s^2/4\pi$). The fields $v_\mu, a_\mu$ and $p$
are $3\times3$ matrices in flavor space and denote respectively
vector, axial-vector and pseudoscalar external fields. $s={\cal
M}+s_{_{\rm ext}}$, where $s_{_{\rm ext}}$ is scalar
external fields and ${\cal M}$=diag($m_u,m_d,m_s$) is quark mass
matrix with three flavor.

The introduction of external fields $v_\mu$ and $a_\mu$ allows for
the global symmetry of the lagrangian to be invariant under local
$SU(3)_L \times SU(3)_R$, i.e., with $g_L,g_R\in SU(3)_L\times SU(3)_R$,
the explicit transformations of the different fields are
\begin{eqnarray}
\label{2.4}
      &&\psi(x) \rightarrow g_R(x)\frac{1}{2}(1+\gamma_5)\psi(x)
           +g_L(x)\frac{1}{2}(1-\gamma_5)\psi(x), \nonumber \\
      &&l_\mu\equiv v_\mu-a_\mu \rightarrow g_L(x)l_\mu g_L^{\dagger}(x)
           +ig_L(x)\partial_\mu g_L^{\dagger}(x), \nonumber \\
      &&r_\mu\equiv v_\mu+a_\mu \rightarrow g_R(x)r_\mu g_R^{\dagger}(x)
           +ig_R(x)\partial_\mu g_R^{\dagger}(x), \nonumber \\
      &&s+ip \rightarrow g_R(x)(s+ip)g_L^{\dag}(x).
\end{eqnarray}

What can be physical observable is the
generating functional of Green's function of vector, axial-vector, scalar
and pseudoscalar external fields, $v,a,s,p$. The generating functional can
be calculated in the path-integral formalism as,
\begin{equation}
\label{2.5}
    e^{iW_{QCD}[v,a,s,p]}=\frac{1}{N}\int {\cal D}G_\mu{\cal D}\bar{\psi}
    {\cal D}\psi \exp\{i\int d^4x{\cal
     L}_{QCD}(G,\psi,\bar{\psi};v,a,s,p)\}.
\end{equation}
At the low energies the coupling constant becomes strong and perturbative
QCD can no longer be done so that we need some low energy effective
models(quark model, pole model, Skyrme model,...) to approach low
energy behaviors of QCD.

The simplest pattern of chiral quark model is described by the
following chiral quark Lagrangian(M-G model) with three flavor
massless quarks\cite{GM84}
\begin{equation}\label{2.6}
{\cal L}_{M-G}=\bar{\psi}(x)(i\gamma\cdot\na -mu(x))\psi(x),
\end{equation}
where
\begin{eqnarray}\label{2.7}
 \na_\mu&=&\pa_\mu-iv_\mu-ia_\mu\gamma_5 \nonumber \\
 u(x)&=&{1\over 2}(1-\gamma_5)U(\Phi)+{1\over 2}(1+\gamma_5)
   U^{\dag}(\Phi),
    \nonumber \\
 U(\Phi)&=&\exp{\{2i\lambda^a \Phi^a(x)\}},
\end{eqnarray}
and $\lambda_a$ are Gell-Mann matrices of $SU(3)_{\rm flavor}$,
$\Phi_a$ are fields of pseudoscalar mesons octet of SU(3). Under
local $SU(3)_L
\times SU(3)_R$, the explicit transformations of pseudoscalar fields is
\begin{equation}\label{2.8}
    U(\Phi)\longrightarrow g_R(x)U(\Phi)g_L(x)
\end{equation}

We like to make several remarks about M-G model: 1) ${\cal L}_{M-G}$ is
invariant under global chiral transformations. 2) Making a
chiral rotation of quark fields, $\chi_L=\xi(\Phi)\psi_L$,
$\chi_R=\xi^{\dag}(\Phi)\psi_R$, the second term of ${\cal L}_{M-G}$
reduces to a mass term for the dressed quarks $\chi$(i.e., constituent
quarks), so that the parameter $m$ can then be interpreted as a
constituent quark mass. 3) In ${\cal L}_{M-G}$,
the operator $\bar{\psi}\psi$ acquires a vacuum expectation value.
Therefore this is an effective way to generate the order parameter due to
CSSB\cite{Esp90,Pich98}. 4) Due to the smallness of effective gluon
couplings, the contributions of gluons are perturbative correction in
${\cal L}_{M-G}$. 5) Note that there are no kinetic terms for $\Phi_a$ in
original ${\cal L}_{M-G}$. The kinetic terms of pseudoscalar fields are
reduced by quantum fluctuation effects due to
quark-loops\cite{Chan}-\cite{Li95a}. Therefore there is no so called
double counting problem. $\Phi_a$ are actually the composite fields of
quarks. 6) A very low energy strong interaction theory involving
pseudoscalar mesons only can be derived via integrating over the freedom
of quarks.

By means of M-G model, the quark mass-independent low energy
coupling constants have been derived in Refs.\cite{WY98,Esp90}. It
is remarkable that the predictions of this simple model are in agreement
with the phenomenological values of $L_i$ in ChPT. This means
the low energy limit M-G model is compatible with ChPT in chiral limit.
In the baryon physics, the skyrmion calculations show also that the
predictions from M-G model are reasonable\cite{Chan,Ait85,LY92}.
In Ref.\cite{Esp90} the leading effective gluonic corrections to $L_i$ 
are extensively discussed. Furthermore, in Ref.\cite{BBR93}
the authors proved that the interaction in ${\cal L}_{M-G}$ is equivalent
to the mean-field approximation of the Nambu-Jona-Lasinio model\cite{NJL}.
The facts indicate that M-G model is sound as a base for the
investigations of low energy meson physics.

In this paper, we like to extend M-G model to include spin-1 meson
resonances. From the viewpoint of chiral symmetry only, vector and
axial-vector mesons do not have any special status compared to
pseudoscalar mesons. As pointed out in Ref.\cite{Ecker89}, all spin-1
meson resonances
carry non-linear realizations of global chiral group $G=SU(3)_L\times
SU(3)_R$ depending on their transformation properties under the subgroup
$H=SU(3)_V$. A non-linear realization of spontaneously broken chiral
symmetry is defined in \cite{Cole69} by specifying the action of
G on elements $\xi(\Phi)$ of the coset space G/H:
\begin{equation}\label{2.9}
\xi(\Phi)\rightarrow
g_R\xi(\Phi)h^{\dag}(\Phi)=h(\Phi)\xi(\Phi)g_L^{\dag}.
\end{equation}
Explicit form of $\xi(\Phi)$ is usually taken
$$\xi(\Phi)=\exp{\{i\lambda^a \Phi^a(x)\}},\hspace{0.8in}
  U(\Phi)=\xi^2(\Phi).$$
The compensating $SU(3)_V$ transformation $h(\Phi)$ defined by
Eq.(~\ref{2.9}) is the wanted ingredient for a non-linear realization of
G. In practice, we shall only be interested in spin-1 meson resonances
transforming as octets and singlets under $SU(3)_V$. Denoting the
multiplets generically be $O_\mu$(octet) and $O_{1\mu}$(singlet), the
non-linear realization of G is given by
\begin{equation}\label{2.10}
  O_\mu\rightarrow h(\Phi)O_{\mu}h^{\dag}(\Phi), \hspace{1in}
  O_{1\mu}\rightarrow O_{1\mu}.
\end{equation}
More convenience, due to OZI rule, the vector and axial-vector octets and
singlets are combined into a single ``nonet'' matrix
$$ N_\mu=O_\mu+\frac{I}{\sqrt{3}}O_{1\mu},\hspace{0.8in}
N_\mu=V_\mu,A_\mu, $$
where
\begin{equation}
\label{vector1}
   V_\mu(x)={\bf \lambda \cdot V}_\mu =\sqrt{2}
\left(\begin{array}{ccc}
       \frac{\rho^0_\mu}{\sqrt{2}}+\frac{\omega_\mu}{\sqrt{2}}
            &\rho^+_\mu &K^{*+}_\mu   \\
    \rho^-_\mu&-\frac{\rho^0_\mu}{\sqrt{2}}+\frac{\omega_\mu}{\sqrt{2}}
            &K^{*0}_\mu   \\
       K^{*-}_\mu&\bar{K}^{*0}_\mu&\phi_\mu
       \end{array} \right),
\end{equation}
\begin{equation}
\label{avector1}
   A_\mu(x)={\bf \lambda \cdot A}_\mu =\sqrt{2}
\left(\begin{array}{ccc}
     \frac{a^0_{1\mu}}{\sqrt{2}}+\frac{f_\mu}{\sqrt{2}}
      &a^+_{1\mu}&K^{+}_{1\mu}   \\
       a^-_{1\mu}&-\frac{a^0_{1\mu}}{\sqrt{2}}+\frac{f_\mu}{\sqrt{2}}
            &K^{0}_{1\mu}   \\
       K^{-}_{1\mu}&\bar{K}^{0}_{1\mu}&f_{1\mu}
       \end{array} \right).
\end{equation}

From viewpoint of phenomenology, we can find in Eq.(~\ref{2.6})
that photon field and $W,\;Z$-fields enter dynamics of hadrons
through the coupling to quarks via covariant derivative $\na
_{\mu}$. In addition, it is well known that in the electromagnetic
and weak interactions of mesons the vector ($V$) and axial-vector
mesons ($A$) play essential role through VMD (Vector Meson
Dominate) and AVMD (Axial-Vector Meson Dominate)\cite{VMD}, and
enjoy considerable phenomenological support. This indicates that it is
available to substitute new affine connection $L_\mu+l_\mu$ and
$R_\mu+r_\mu$ for the old $l_\mu$ and $r_\mu$ in Eq.(~\ref{2.6}),
where the auxiliary fields $L_\mu=\xi^{\dag}(V_\mu-A_\mu)\xi$ and
$R_\mu=\xi(V_\mu+A_\mu)\xi^{\dag}$. Therefore, we can naturally
extend M-G model to be one including the lowest meson resonances
with spin-1 via minimum coupling principle,
\begin{equation}
\label{2.11}
{\cal L}_{\chi}=\bar{\psi}(x)(i\gamma \cdot D-mu(x))\psi(x)
 +{1\over 2}m_1^2(V_\mu^a V^{\mu a})+{1\over 2}m_2^2(A_\mu^a A^{\mu a}),
\end{equation}
where
{\footnote {\small The vector meson field $V_\mu$ and axial-vector meson
field $A_\mu$ in $D_\mu$ can combine with coupling constants $g_{_V}$ and
$g_{_A}$ respectively. However, these constants will be absorbed in mass
term of spin-1 mesons via the field rescaling
$V_\mu\rightarrow\frac{1}{g_{_V}}V_\mu$ and
$A_\mu\rightarrow\frac{1}{g_{_A}}A_\mu$.}}

\begin{equation}\label{2.12}
  \na_\mu \longrightarrow D_\mu\equiv\partial_\mu
       -i\frac{1-\gamma_5}{2}(L_\mu+l_\mu)-
        i\frac{1+\gamma_5}{2}(R_\mu+r_\mu)
\end{equation}

The transformation(~\ref{2.10}) leads to that the mass type terms of
vector and axial-vector composite fields are allowed to be added in
the lagrangian ${\cal L}_\chi$. The $L^{\prime}_\mu=L_\mu+l_\mu$
and $R^{\prime}_\mu=R_\mu+r_\mu$ transform nonlinearly as gauge
bosons under the chiral group.
\footnote{\small The $U(1)_A$ anomaly effects are absent in this present
paper, which will be studied in elsewhere.}
Making the chiral rotation from current quark fields to constituent quark
fields, $\chi_L=\xi(\Phi)\psi_L$, $\chi_R=\xi^+(\Phi)\psi_R$, the spin-1
meson fields will couple to constituent quark fields directly. Therefore,
spin-1 meson resonances in this framework are the composite fields of
constituent quark fields instead of the one of current quark fields in
Ref.\cite{Bejinen96a,Li95a}.

Finally, light quark mass-dependent term can be included in ChQM in terms
of standard form
\begin{equation}\label{2.13}
   \bar{\psi}(s-i\gamma_5p)\psi.
\end{equation}
Then general ChQM lagrangian is written
\begin{eqnarray}
\label{model}
   {\cal L}_\chi&=&{\cal L}_{QCD}^0+\bar{\psi}_L\gamma^\mu
   (L_\mu+l_\mu)\psi_L+\bar{\psi}_R\gamma^\mu (R_\mu+r_\mu)\psi_R
       \nonumber \\
   &&-m\bar{\psi}u(x)\psi-\bar{\psi}(s-i\gamma_5p)\psi
    +\frac{1}{2}m_1^2V_\mu^aV^{a\mu}+\frac{1}{2}m_2^2A_\mu^aA^{a\mu},
\end{eqnarray}
where
\begin{equation}
 \psi_L=\frac{1-\gamma_5}{2}\psi,\hspace{1in}
 \psi_R=\frac{1+\gamma_5}{2}\psi.
\end{equation}
Since the composite pseudoscalar, vector and axial-vector meson fields are
treated as background fields, there are no kinetic terms for them in
lagrangian (~\ref{model}). They will be generated by quark-loop
effects.

Our purpose is to use chiral quark model to approach low energy
behavior of QCD. In other words, we replace the generating functional of
Green's function of external fields in QCD(~\ref{2.5}) by one of chiral
quark model,
\begin{equation}
\label{Green2}
   e^{i\bar{W}[V,A,U;v,a,s,p]}=\frac{1}{N}\int {\cal D}G{\cal D}\bar{\psi}
             {\cal D}\psi \exp\{i\int d^4x{\cal L}_\chi\}.
\end{equation}
The relationship between generating functional $W_{QCD}$ and
$\bar{W}$ is as follows: As energy scale $\mu$ is higher than chiral
symmetry spontaneous breaking scale $\Lambda_{CSSB}$, the composite
mesonic fields disappear. Then we have
\begin{equation}
\label{Green3}
         W_{QCD}[v,a,s,p]=\bar{W}[0,0,1;v,a,s,p]|_{m=0}.
\end{equation}
As $\mu< \Lambda_{CSSB}$ mesonic fields will appear as dynamical freedom
in the theory. Then we obtain
\begin{equation}
\label{Green4}
   e^{iW_{QCD}[v,a,s,p]}=\frac{1}{N}\int{\cal D}U{\cal D}V_\mu{\cal D}
     A_\mu e^{i\bar{W}[V,A,U;v,a,s,p]}
\end{equation}

\section{Integrating Out Quark Fields}
\setcounter{equation}{0}

\subsection{Effective lagrangian from quark loops}

Our goal is to derive the effective lagrangian of mesons up to the next to
leading order of $1/N_c$ expansion. It is expressed as follows
\begin{equation}
\label{3.1.1}
{\cal L}_{eff}={\cal L}_{eff}^{(0)}+{\cal L}_{eff}^{(g)}+
     {\cal L}_{eff}^{(l)}
\end{equation}
where ${\cal L}_{eff}^{(0)}$ comes from quark loops,
${\cal L}_{eff}^{(g)}$ and ${\cal L}_{eff}^{(l)}$ are contributed by
quark-gluon coupling and one-loop of mesons respectively. According to
the large $N_c$ expansion argument\cite{tH74}, ${\cal L}_{eff}^{(0)}
\sim {\cal O}(N_c)$ is the leading order of ${\cal L}_{eff}$, and ${\cal
L}_{eff}^{(l)}\sim {\cal O}(1)$ is next leading order of ${\cal L}_{eff}$.
In addition, according to discussion in sect. 1, we treat ${\cal
L}_{eff}^{(g)}$ as perturbation although ${\cal L}_{eff}\sim {\cal
O}(N_c)$ is also the leading order. In this section, we derive
${\cal L}_{eff}^{(0)}$ in terms of integrating out the quark fields in
lagrangian(~\ref{model}).

A review for chiral gauge theory is in \cite{Ball89}. The effective
lagrangian of mesons in ChQM can be obtained in Euclidian space by means
of integrating over degrees of freedom of fermions
\begin{equation}
\label{3.1.2}
      \exp\{-\int d^4x{\cal L}_{eff}^{(0)}\}
      =\int {\cal D}\bar{\psi}{\cal D}\psi \exp\{-\int d^4x{\cal
       L}_\chi\}.
\end{equation}
Then we have
\begin{equation}
\label{3.1.3}
     {\cal L}_{eff}^{(0)}=-\ln{\rm det}{\cal D},
\end{equation}
with
\begin{equation}
\label{3.1.4}
{\cal D}=\gamma^\mu(\pa_\mu-i\frac{1-\gamma_5}{2}(L_\mu+l_\mu)
   -i\frac{1+\gamma_5}{2}(R_\mu+r_\mu))+mu+(s-ip\gamma_5).
\end{equation}
The gluon fields are absent in the operator ${\cal D}$ since
the quark-gluon coupling is treated as perturbation. The effective
lagrangian is separated into two parts
\begin{eqnarray}
\label{3.1.5}
&&{\cal L}_{eff}^{(0)}={\cal L}_{eff}^{Re}+{\cal L}_{eff}^{Im}
       \nonumber   \\
&&{\cal L}_{eff}^{Re}=-\frac{1}{2}\ln{\rm det}({\cal D}{\cal D}^{\dag}),
 \hspace{0.8in}
 {\cal L}_{eff}^{Im}=-\frac{1}{2}\ln{\rm det}[({\cal D}^{\dag})^{-1}
     {\cal D}]
\end{eqnarray}
where
\begin{equation}
\label{3.1.6}
 {\cal D}^{\dag}=\gamma_5\hat{{\cal D}}\gamma_5,
\end{equation}
and $\hat{B}=\frac{1}{2}(1+\gamma_5)B_L+\frac{1}{2}(1-\gamma_5)B_R$
for arbitrarily operator
$B=\frac{1}{2}(1-\gamma_5)B_L+\frac{1}{2}(1+\gamma_5)B_R$. The
effective lagrangian ${\cal L}_{eff}^{Re}$ describes the physical
processes with normal parity and ${\cal L}_{eff}^{Im}$ describes the
processes with anomal parity. In the present paper we focus our attention
on $ {\cal L}_{eff}^{Re}$. The discussion of ${\cal L}_{eff}^{Im}$ can be
found in Refs.\cite{Ball89}. In terms of Schwenger's proper time
method \cite{Schw54}, ${\cal L}_{eff}^{Re}$ is written as
\begin{equation}
\label{3.1.7}
  {\cal L}_{eff}^{Re}=-\frac{1}{2\delta(0)}\int d^4x\frac{d^4p}{(2\pi)^4}
   Tr\int_0^{\infty}\frac{d\tau}{\tau}(e^{-\tau{\cal D}^{\prime\dagger}
   {\cal D}^\prime}-e^{-\tau\Delta_0})\delta^4(x-y)|_{y\rightarrow x}
\end{equation}
with
\begin{eqnarray}
\label{3.1.8}
   &&{\cal D}^\prime={\cal D}-i\gamma \cdot p,  \hspace{0.8in}
   {\cal D}^{\prime\dag}={\cal D}^{\dag}+i\gamma\cdot p, \nonumber \\
   &&\Delta_0=p^2+M^2.
\end{eqnarray}
where M is an arbitrary parameter with dimension of mass. The
Seeley-DeWitt coefficients or heat kernel method have been used to evaluate
the expansion series of Eq.(~\ref{3.1.8}). In this paper we
will use dimensional regularization.
After completing the integration over $\tau$,
the lagrangian ${\cal L}_{eff}^{Re}$ reads
\begin{equation}
\label{3.1.9}
    {\cal L}_{eff}^{Re}=-\frac{\mu^\ep}{2\delta(0)}\int d^Dx
        \frac{d^Dp}{(2\pi)^D}\sum_{i=1}^{\infty}\frac{1}{n\Delta_0^n}
         Tr({\cal D}^{\prime\dagger}{\cal D}^\prime-\Delta_0)^n
         \delta^D(x-y)|_{y\rightarrow x},
\end{equation}
where trace is taken over the color, flavor and Lorentz space.
This effective lagrangian can be expanded in powers of derivatives,
\footnote{\small Here we need to distinguish expansion in powers of
derivatives from low energy expansion in ChPT. It will be discussed in
sect. 3.2.}
\begin{equation}\label{3.1.10}
{\cal L}_{Re}={\cal L}_2^{(0)}+{\cal L}_4^{(0)}+....
\end{equation}

In Minkowski space, effective lagrangian with two derivatives reads
\begin{eqnarray}
\label{3.1.11}
{\cal L}_{2}^{(0)}&=&\lambda(\mu) m^2<D_{\mu}UD^{\mu}U^{\dag}>
   +\quddiv{}\frac{m^3}{B_0}<\chi U^{\dag}+\chi^{\dag}U>
      \nonumber \\
 &&+\frac{1}{4}m_1^2<V_{\mu}V^{\mu}>+\frac{1}{4}m_2^2<A_{\mu}A^{\mu}>
\end{eqnarray}
where $``<...>"$ denotes trace in flavor space and
\begin{eqnarray}\label{3.1.111}
 &&\lambda(\mu)=\logdiv{} \nonumber \\
 &&\chi=2B_0(s+ip) \nonumber \\
 &&D_{\mu}U=\na_{\mu}U-2i\xi A_\mu\xi, \nonumber \\
 &&D_{\mu}U^{\dag}=\na_{\mu}U^{\dag}+2i\xi^{\dag}A_\mu\xi^{\dag}, \\
 &&\na_\mu U=\partial_\mu U+iU l_\mu-ir_\mu U.   \nonumber \\
 &&\na_\mu U^{\dag}=\partial_\mu U^{\dag}+iU^{\dag}l_\mu
          -ir_\mu U^{\dag}.  \nonumber
\end{eqnarray}

In lagrangian (~\ref{3.1.11}) the axial-vector fields $A_\mu$ mixes with
pseudoscalar fields, $\pa_\mu\Phi$. This mixing should be diagnolized via
field shift
\begin{equation}\label{3.1.12}
A_\mu\longrightarrow A_\mu-ic\Delta_\mu, \hspace{1in}
c=\frac{\lambda(\mu)m^2}{\lambda(\mu)m^2+m_2^2},
\end{equation}
where
\begin{equation}\label{3.1.13}
  \Delta_\mu=\frac{1}{2}\{\xi^{\dag}(\pa_\mu-ir_\mu)\xi
          -\xi(\pa_\mu-il_\mu)\xi^{\dag}\}
   =\frac{1}{2}\xi^{\dag}\na_\mu U\xi^{\dag}
   =-\frac{1}{2}\xi\na_\mu U^{\dag}\xi.
\end{equation}
Then effective lagrangian with two derivatives reads
\begin{equation}\label{3.1.14}
{\cal L}_2^{(0)}=\frac{f_0^2}{16}<\na_{\mu}U\na^{\mu}U^{\dag}+\xi U^{\dag}
   +\xi^{\dag}U>+\frac{1}{4}m_1^2<V_\mu V^{\mu}>
   +\frac{1}{4}\bar{m}_2^2<A_\mu A^{\mu}>.
\end{equation}
In above lagrangian we have defined the constants $f_0$, $B_0$ and
$\bar{m}_2$ to absorb divergences from quark loop integral
\begin{eqnarray}\label{3.1.15}
 \frac{f_0^2}{16}&=&\lambda(\mu)m^2(1-c), \nonumber \\
 \frac{f_0^2}{16}B_0&=&\quddiv{}m^3, \\
 \bar{m}_2^2&=&16\lambda(\mu)m^2+m_2^2. \nonumber
\end{eqnarray}
It should be noted that the field shift(~\ref{3.1.12}) is different from
that one in Refs.\cite{Li95a,Bando}. This field shift is convenient to
keep gauge symmetry explicitly. Moreover, it make
that there are no spin-1 mesons coupling to pseudoscalar fields in
${\cal L}_2^{(0)}$. We will show in sect. 6 that this result is useful
in calculation on meson loops.

The lagrangian (~\ref{3.1.11}), (~\ref{3.1.14}) and Eq.(~\ref{3.1.15})
yield two equivalent equations of motion of pseudoscalar mesons as follows
\begin{eqnarray}
\label{3.3.16}
 &&D_{\mu}(UD^{\mu}U^{\dag})+\frac{1-c}{2}
  (\chi U^{\dag}-U\chi^{\dag})=0, \nonumber \\
 &&\na_{\mu}(U\na^{\mu}U^{\dag})+\frac{1}{2}
  (\chi U^{\dag}-U\chi^{\dag})=0.
\end{eqnarray}
The first equation is in presence of axial-vector mesons and the
second one is in absence of axial-vector mesons. All pseudoscalar meson
fields satisfy these equations.

Since the non-linear realization of G on the spin-1 meson fields $O$ in
expression (~\ref{2.10}) is local we are led to define a covariant
derivative
\begin{equation}
\label{3.1.17}
   d_{\mu}{\cal O}=\pa_{\mu}{\cal O}+[\Gamma_\mu,{\cal O}],
\end{equation}
with
\begin{equation}
\label{3.1.171}
\Gamma_\mu=\frac{1}{2}\{\xi^{\dag}(\pa_\mu-ir_\mu)\xi
          +\xi(\pa_\mu-il_\mu)\xi^{\dag}\},
\end{equation}
ensuring the proper transformation
$$d_\mu {\cal O}\lra h(\Phi)d_\mu {\cal O}h^{\dag}(\Phi).$$
Without external fields, $\Gamma_\mu$ is the usual natural connection on
coset space.

Then from Eq.(~\ref{3.1.9}) and due to equation of motion(~\ref{3.3.16}),
effective lagrangian with four derivatives can be obtained in Minkowski
space as follows
\begin{eqnarray}\label{3.1.18}
{\cal L}_{4}^{(0)}&=&-\frac{\lambda_r(\mu)}{16}
   <L_{\mu\nu}L^{\mu\nu}+R_{\mu\nu}R^{\mu\nu}>
   -\frac{\gamma}{6}<L_{\mu\nu}U^{\dag}R^{\mu\nu}U> \nonumber \\
 &&-\frac{i\gamma}{3}<D_{\mu}UD_{\nu}U^{\dag}
   R^{\mu\nu}+D_{\mu}U^{\dag}D_{\nu}UL^{\mu\nu}>
  +\frac{\gamma}{12}<D_{\mu}UD_{\nu}U^{\dag}D^{\mu}UD^{\nu}U^{\dag}>
     \nonumber \\
&&+\theta_1<D_\mu UD^\mu U^{\dag}(\chi U^{\dag}+\chi^{\dag}U)>
  +\theta_2<\chi U^{\dag}\chi U^{\dag}+\chi^{\dag}U\chi^{\dag}U>
\end{eqnarray}
where
\begin{eqnarray}
\label{3.1.19}
     &&\gamma=\frac{N_c}{(4\pi)^2}, \hspace{1.0in}
     \lambda_r(\mu)=\frac{8}{3}\lambda(\mu)-\frac{4}{3}\gamma,
     \nonumber \\
     &&\theta_1=(\lambda(\mu)-\gamma)\frac{m}{2B_0},\\
     &&\theta_2=\frac{\lambda(\mu)m}{4B_0}(1-c-\frac{m}{B_0})
          -\frac{\gamma}{24}(1-c)^2. \nonumber
\end{eqnarray}
Due to field shift (~\ref{3.1.12}), in ${\cal L}_4^{(0)}$ we have defined
\begin{eqnarray}\label{3.1.20}
L_{\mu\nu}&=&(1-\frac{c}{2})F^L_{\mu\nu}+\frac{c}{2}F^R_{\mu\nu}
  +\xi^{\dag}(V_{\mu\nu}-A_{\mu\nu})\xi
  -2ic(1-\frac{c}{2})\xi^{\dag}[\Delta_\mu,\Delta_\nu]\xi \nonumber \\
 &&-(1-c)\xi^{\dag}([\Delta_\mu,V_\nu-A_\nu]
   -[\Delta_\nu,V_\mu-A_\mu])\xi. \nonumber \\
R_{\mu\nu}&=&(1-\frac{c}{2})F^R_{\mu\nu}+\frac{c}{2}F^L_{\mu\nu}
  +\xi(V_{\mu\nu}+A_{\mu\nu})\xi^{\dag}
  -2ic(1-\frac{c}{2})\xi[\Delta_\mu,\Delta_\nu]\xi^{\dag} \nonumber \\
 &&+(1-c)\xi([\Delta_\mu,V_\nu+A_\nu]
   -[\Delta_\nu,V_\mu+A_\mu])\xi^{\dag}, \nonumber \\
D_{\mu}U&=&(1-c)\na_{\mu}U-2i\xi A_\mu\xi, \nonumber \\
D_{\mu}U^{\dag}&=&(1-c)\na_{\mu}U^{\dag}+2\xi^{\dag}A_\mu\xi^{\dag},
\end{eqnarray}
with
\begin{eqnarray}\label{5.1.13}
F_{\mu\nu}^{R.L}&=&\partial_\mu(v_\nu\pm a_\nu)
      -\partial_\mu(v_\nu\pm a_\nu)-i[v_\mu\pm a_\mu,v_\nu\pm a_\nu].
         \nonumber \\
V_{\mu\nu}&=&d_{\mu}V_\nu-d_{\nu}V_\mu
   -i[V_\mu,V_\nu]-i[A_\mu,A_\nu]\nonumber \\
A_{\mu\nu}&=&d_\mu A_\nu-d_\nu A_\mu
    -i[A_\mu,V_\nu]-i[V_\mu,A_\nu].
\end{eqnarray}

\subsection{Physical effective lagrangian and beyond low energy
expansion}

In general, the effective lagrangian(it need not to be the leading order
of $1/N_c$) can be rewritten as follows
\begin{equation}\label{3.2.1}
{\cal L}_{\rm eff}={\cal L}_2^{\phi}+{\cal L}_4^{\phi}+
{\cal L}_{\rm kin}^{V.A}+{\cal L}_{\rm I}^{V.A}+...,
\end{equation}
where ${\cal L}^{\phi}$ denotes effective lagrangian describing
interaction of pseudoscalar mesons in very low energy but without spin-1
meson resonances(where $\phi$ denotes pseudoscalar fields), ${\cal L}_{\rm
kin}^{V.A}$ is kinetic terms of spin-1 mesons and ${\cal L}_{\rm
I}^{V.A}$ denotes effective lagrangian describing spin-1 mesons coupling
to pseudoscalar meson. Since there is no interaction of spin-1 mesons in
${\cal L}_2$, all terms in ${\cal L}_{\rm I}^{V.A}$ are with four
derivatives. In very low energy, equation of motion $\delta {\cal
L}/\delta {\cal O}=0,({\cal O}=V,A)$ yields classic solution of spin-1
mesons as follows
$$O_\mu=\frac{1}{m_{_{\cal O}}^2}\times O(p^3)\;{\rm terms},$$
where $p$ is momentum of pseudoscalar mesons in very low energy. In low
energy limit, degrees of freedom of spin-1 meson resonances disappear and
their dynamics is replaced by pseudoscalar fields, hence in this energy
region ${\cal L}^{V.A}$ is $O(p^6)$. It means that, up to $O(p^4)$,
${\cal L}_2^{\phi}+{\cal L}_4^{\phi}$ is just low energy limit of ChQM.
If there are processes in which all external lines are pseudoscalar
mesons, spin-1 meson fields do not appear in internal lines of tree
diagrams of these processes. It seems to be different from the results in
some Refs.\cite{Ecker89, WY98} in which chiral coupling in very low energy
receive large contribution from spin-1 meson exchange. However, we like to
point out that, this difference is caused by definition of physical meson
fields and does not change physical results. In this paper, the physical
meson fields are defined by expression (~\ref{2.7}), (~\ref{2.10}) and
field shift (~\ref{3.1.12}). 

There are divergences from quark loops in ${\cal L}^{(0)\phi}$ and
${\cal L}^{(0)V.A}$(recalling that superscript ``(0)'' denotes
effective lagrangian generated by quark loops). In sect. 3.1,
divergences in ${\cal L}_2^{(0)}$ have been absorbed by $f_0$, $B_0$
and axial-vector mass-dependent parameter $\bar{m}_2$. Here we need
to define two constants $g_\phi^2=\frac{8}{3}\lambda(\mu_\phi)$ and
$g^2=\frac{8}{3}\lambda(\mu_{_V})$ to absorb logarithmic
divergences in ${\cal L}_4^{(0)\phi}$ and ${\cal L}^{(0)V.A}$
respectively since scale factor $\mu$ of dimensional regularization
is arbitrary. Equivalently, it means that two cut-offs are needed here.
One corresponds to very low energy and regularize logarithmic
divergences in ${\cal L}_4^{(0)\phi}$, Another corresponds to
energy scale of vector meson masses and regularize logarithmic
divergences in ${\cal L}^{(0)V.A}$. It has been mentioned that
these divergences are caused by "point particles" approximation of
composite mesons in lagrangian (~\ref{model}). Therefore, the
these cut-offs can be treated as description of interaction scale of quark
pairs in mesons. Naturally, these interaction scales are different for
spin zero and spin-1 mesons. It was also shown by some model. For
instance, effective potential model\cite{Dynamics} showd that there
are obvious difference for mesons with different spin content. Thus
$g_\phi\neq g$ is a nature result.

The explicit form of ${\cal L}_2^{(0)\phi}$ has been shown in
Eq.(~\ref{3.1.14}). In general, chiral symmetry requires
${\cal L}_4^{\phi}$ taking the form as follows\cite{GL85a}
\begin{eqnarray}\label{3.2.2}
{\cal L}_4^{\phi}&=&L_1<\na_\mu U\na^\mu U^{\dag}>^2
  +L_2<\na_\mu U\na_\nu U^{\dag}\na_\mu U\na_\nu U^{\dag}>
  +L_3<\na_\mu U\na^\mu U^{\dag}\na_\nu U\na^\nu U^{\dag}>
                \nonumber \\
 &&+L_4<\na_\mu U\na^\mu U^{\dag}><\chi U^{\dag}+\chi^{\dag}U>
  +L_5<\na_\mu U\na^\mu U^{\dag}(\chi U^{\dag}+\chi^{\dag}U)>
                \nonumber \\
 &&+L_6<\chi U^{\dag}+\chi^{\dag}U>^2
   +L_7<\chi U^{\dag}-\chi^{\dag}U>^2
   +L_8<\chi U^{\dag}\chi U^{\dag}+\chi^{\dag}U\chi^{\dag}U>
              \nonumber \\
 &&-iL_9<\na_\mu U\na_\nu U^{\dag}F^{R\mu\nu}
                 +\na_\mu U^{\dag}\na_\nu UF^{L\mu\nu}>
  +L_{10}<F_{\mu\nu}^LU^{\dag}F^{R\mu\nu}U>,
\end{eqnarray}
where $L_i(i=1,2,...,10)$ are ten real constants which determine dynamics
of pseudoscalar meson interaction at very low energy, together with $f_0$
and $B_0$. From Eq.(~\ref{3.1.18}, these low energy coupling
constants read
\begin{eqnarray}\label{3.2.3}
g_\phi^2&=&\frac{8}{3}\frac{N_c}{(4\pi)^{D/2}}
    (\frac{\mu_\phi^2}{m^2})^{\epsilon/2}\Gafunc{2}, \nonumber \\
L_1^{(0)}&=&\frac{1}{2}L_2^{(0)}=\frac{g_\phi^2}{32}c^2(1-\frac{c}{2})^2+
  \frac{\gamma}{6}c(1-\frac{c}{2})(1-c)^2+\frac{\gamma}{24}(1-c)^4,
      \nonumber \\
L_3^{(0)}&=&-\frac{3}{16}g_\phi^2c^2(1-\frac{c}{2})^2-
   \gamma c(1-\frac{c}{2})(1-c)^2-\frac{\gamma}{6}(1-c)^4, \nonumber \\
L_5^{(0)}&=&(\frac{3}{8}g_\phi^2-\gamma)\frac{m}{2B_0}(1-c)^2, \\
L_8^{(0)}&=&\frac{3g_\phi^2m}{32B_0}(1-c-\frac{m}{B_0})
         -\frac{\gamma}{24}(1-c)^2, \nonumber \\
L_9^{(0)}&=&\frac{g_\phi^2}{8}c(1-\frac{c}{2})+\frac{\gamma}{3}(1-c)^2,
   \nonumber \\
L_{10}^{(0)}&=&-\frac{g_\phi^2}{8}c(1-\frac{c}{2})-\frac{\gamma}{6}(1-c)^2.
   \nonumber \\
L_4^{(0)}&=&L_6^{(0)}=L_7^{(0)}=0.  \nonumber
\end{eqnarray}
Input experimental data $L_9=(6.9\pm 0.7)\time10^{-3}$ we can obtain
$g_\phi=0.32\pm 0.02$
\footnote{\small Here $g_\phi$ is fitted only in tree level. In sect. 7,
we fit it again in one-loop level and show that the difference is small.}
(the value of $c=0.44$ will be fitted in the
following).

Since all meson fields in ${\cal L}^{\phi}$ are pseudoscalars, the low
energy expansion of ${\cal L}^{\phi}$ is well-defined. However, it is
known that the momentum expansion in ${\cal L}^{V.A}$ is very unclear.
Then how do we know that our calculation on physics in this energy region
is reliable? The problem has been discussed in
Refs.\cite{Bando,Mana95,Bejinen97}. In ChQM, the most vector- or
axial-vector-dependent processes, like vector and axial-vector meson
decays, can be calculated because of the following two reason. 1) Spin-1
meson resonances are introduced in the framework by VMD which is
phenomenological result and beyond low energy expansion. 2) ChQM is only a
phenomenological model, in which the universal coupling constant
$g^2=\frac{8}{3}\lambda(\mu_{_V})$ affects rather many processes. Although
the value of $g$ is fitted phenomenologically at leading order of
spin-1 meson resonances coupling to pseudoscalar fields in this paper, in
fact, some high order effects of momentum expansion has been included.
It will be double counting if we try to discuss high order effects of
momentum expansion in terms of $g$ determined phenomenologically by
lagrangian with four derivatives. Hence all high order derivative terms
of ${\cal L}^{V.A}$ will be omitted in this paper.

The physical vector and axial-vector fields can be obtained via the
following field rescaling in ${\cal L}^{V.A}$ which make kinetic term of
spin-1 meson fields into the standard form
\begin{eqnarray}\label{3.2.4}
V_\mu \longrightarrow\frac{1}{g}V_\mu, \hspace{0.8in}
A_\mu\longrightarrow\frac{1}{g_{_A}}A_\mu.\hspace{0.8in}
g_{_A}=g\sqrt{1-\frac{1}{2\pi^2g^2}}=g\kappa.
\end{eqnarray}
Then due to Eqs.(~\ref{3.1.12}), (~\ref{3.1.14}) and (~\ref{3.1.15}) we
have
\begin{eqnarray}\label{3.2.5}
m_1^2&=&m_{_V}^2g^2, \hspace{1in} \bar{m}_2^2=m_{_A}^2g_{_A}^2, \nonumber \\
c&=&\frac{1-\sqrt{1-\frac{4f_0^2}{m_{_A}^2g_{_A}^2}}}{2}.
\end{eqnarray}
The above equations require $m_{_A}^2g_{_A}^2\geq 4f_0^2$.

For fitting the parameters $g$ and $c$, we will calculate on $\rho$-mass
shell decays $\rho^0\rightarrow e^+e^-$ and $\rho\rightarrow\pi\pi$ in the
following. In our calculation, we set $m_u=m_d=0$ so that $f_0=f_\pi$.

The $\rho^0-\gamma$ vertex reduced by VMD reads from lagrangian
(~\ref{3.1.18}) as follows
\begin{equation}\label{3.2.6}
{\cal L}_{\rho\gamma}=g_{\rho\gamma}(q^2)\rho_\mu^0{\cal
 A}^\mu,\hspace{1in} g_{\rho\gamma}(q^2)=\frac{1}{2}egq^2,
\end{equation}
where ${\cal A}_\mu$ is photon field. This vertex vanish in
$q^2\rightarrow 0$ so
that it does not violate $U(1)_{\rm EM}$ gauge symmetry. Then
$\rho^0\rightarrow e^+e^-$ decay width can be obtained directly by
a photon field exchange
\begin{equation}\label{3.2.7}
\Gamma(\rho^0\rightarrow e^+e^-)=\frac{\pi}{3}g^2\alpha^2m_\rho.
\end{equation}
To input experimental data $6.7\pm 0.3$KeV\cite{PDG98}, we can fit
$g=0.395\pm 0.008$.

The $\rho\rightarrow\pi\pi$ vertex is obtained in the following by means
of substituting Eqs.(~\ref{3.1.12}) and (~\ref{3.2.4}) into lagrangian
(~\ref{3.1.18})
\begin{eqnarray}\label{3.2.8}
{\cal L}_{\rho\pi\pi}&=&f_{\rho\pi\pi}\ep^{ijk}\rho^\mu_i\pi_j
  \pa_\mu\pi_k, \nonumber \\
f_{\rho\pi\pi}&=&\frac{m_\rho^2}{gf_\pi^2}[2g^2c(1-\frac{c}{2})+
  \frac{(1-c)^2}{\pi^2}]
\end{eqnarray}
To take $c=0.44$ we obtain
\begin{equation}\label{3.2.9}
\Gamma(\rho\rightarrow\pi\pi)=\frac{f_{\rho\pi\pi}^2}{48\pi}m_\rho
 (1-\frac{4m_\pi^2}{m_\rho^2})^{\frac{3}{2}}=(150\pm 4)MeV.
\end{equation}
The experimental value is 150 MeV.

Furthermore, the above values of $g_\phi$, $g$ and $c$ yield constitute
quark
mass
\begin{equation}\label{3.2.10}
m=\frac{f_\pi}{g_\phi\sqrt{6(1-c)}}=320{\rm MeV},
\end{equation}
and axial-vector meson mass in chiral limit
\begin{equation}\label{3.2.11}
m_{_A}=\frac{f_\pi}{g_{_A}\sqrt{c(1-c)}}=(1154\pm 6){\rm MeV}.
\end{equation}
The prediction by the second Weinberg sum rule\cite{Wein67} is
$m_{_A}=\sqrt{2}m_\rho=1090$MeV and experimental data is $1230\pm 40$MeV.

There are six free parameters to parameterize the effective lagrangian
generated from quark loops. They are $f_\pi$(or
$m$), $B_0$, $g_\phi$, $g$, $m_{_V}$ and $m_{_A}$(or $c$).
These constants determine low energy dynamics of mesons and
it is welcome that the number of free parameters is less than ChPT.
Approximately, if we redefine logarithmic divergence from quark
loop integral in scheme of cutoff regularization, we have
\begin{equation}
\label{3.2.12}
     (\frac{\mu^2}{m^2})^{\ep/2}\Gamma(2-\frac{D}{2})\simeq
  \ln(1+\frac{\Lambda^2}{m^2})-\frac{\Lambda^2}{\Lambda^2+m^2}.
\end{equation}
Then for $N_c=3$, above values of $g_\phi$ and $g$ will yield
corresponding cut-off
$\Lambda_\phi\sim 1.3GeV$ and $\Lambda_V\sim 2GeV$ respectively, where
$\Lambda_\phi$ and $\Lambda_V$ are cut-offs corresponding ${\cal L}^\phi$
and ${\cal L}^{V.A}$. It is a very interesting result that
$\Lambda_\phi\sim\Lambda_{CSSB}$ and indicates ChQM is consistent with
original discussion on CSSB\cite{Wein67b}. In addition, as mentioned in
Introduction, it is consistent for $\Lambda_V>\Lambda_\phi$ and all spin-1
meson masses being below $\Lambda_V$.

\section{Correction of Gluon Coupling, ChQM Versus QCD Sum Rules}
\setcounter{equation}{0}

\subsection{Correction of gluon coupling}

In principle, correction of gluon interaction can be obtained via
integrating over gluon field in ChQM. Unfortunately, we do not know
how to perform analytically the remaining integration over the
gluonic fields. A available way is to make a weak gluon-field expansion
around physical vacuum expectation vaule and parameterize
phenomenologically the contribution.

\begin{figure}\label{gluon}[tp]   
   \centering
   \includegraphics[width=6.5in]{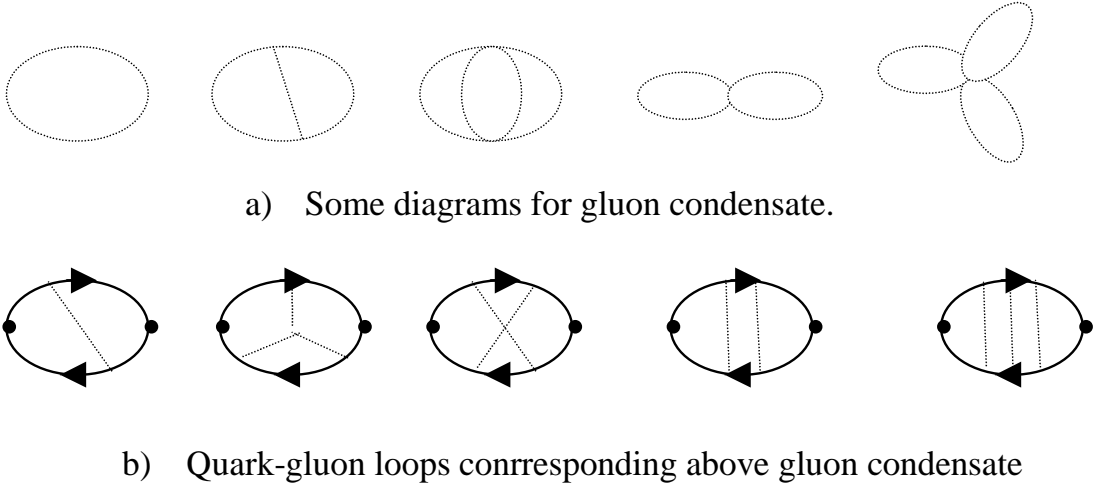}
\begin{minipage}{5in}
   \caption{Some diagrams for gluon condensate in leading order of
   $1/N_c$(fig. a) and its corresponding quark-gluon loops in ChQM(fig.
   b). Here solid line denotes quark propagator, dot line denotes gluon
   propagator and "$\bullet$" denotes other external fields(include meson
   fields, photon field, etc.).}
\end{minipage}
\end{figure}

Replacing the operator of Eq.(~\ref{3.1.4}) by one including color
coupling
\begin{eqnarray}
\label{4.1}
{\cal D}&=&\gamma^\mu(\pa_\mu-ig_s\frac{\lambda_c^a}{2}G_\mu^a
   -i\frac{1-\gamma_5}{2}(L+l)_\mu
   -i\frac{1+\gamma_5}{2}(R+r)_\mu) \nonumber \\
   &&+mu+(s-ip\gamma_5),
\end{eqnarray}
where $\lambda_c^a(a=1,2,...,8)$ denote color SU(3) Gell-Mann matrices.
In terms of Eqs.(~\ref{3.1.7})-(~\ref{3.1.9}), a effective
action including gluon interaction can be obtained via
integrating over quark fields. In general, the corresponding effective
action with normal parity can be written as
\begin{equation}
     I_{Re}=I_{M}+I_{G}+I_{MG}.
\end{equation}
Here, $I_M=\int d^4x{\cal L}_{eff}^{(0)}$, where
${\cal L}_{eff}^{(0)}$ has been obtained in section 3 in which only
meson interactions are involved. $I_{G}$ is nothing other than the
effective action of gluonic fields introduced by the fermion loop
and $I_{MG}$ includes interaction that gluonic fields couple to
mesonic fields. Then we can write the path-integral form for
gluonic fields as
\begin{eqnarray}\label{4.2}
   e^{iI_{eff}}&=&\int {\cal D}G_\mu^a\cdot
     \exp{\{i(-\frac{1}{4}\int d^4xG_{\mu\nu}^aG^{\mu\nu a}
        +I_{M}+I_{G}+I_{MG})\}} \nonumber \\
   &=&\frac{1}{N}e^{iI_M}\int {\cal D}G_\mu^a\cdot e^{iI_{MG}}
      \exp{\{i(-\frac{1}{4}\int d^4xG_{\mu\nu}^aG^{\mu\nu a}+I_{G})\}}
            \nonumber \\
   &=&\frac{1}{N}\exp{\{i(I_{M}+<0|I_{MG}|0>+\frac{1}{2}(<0|I^2_{MG}|0>
         -<0|I_{MG}|0>^2)+...)\}}
\end{eqnarray}
where brackets denote the expectation value of operator. After we
trace over color space, $I_{M}$ is proportional to $N_c$ but
$I_{MG}$ is not because $G_\mu=g_s\frac{\lambda^a}{2}G_{\mu}^a(x)$
is color-$\lambda^a$-dependent. Therefore $I_{M}$ is $O(N_c)$ but
$I_{MG}$ is $O(1/N_c)$ at least due to color coupling constant $g_s\sim
O(1/\sqrt{N_c})$. However, since expectation value of opeator sum
over all possible connect loop diagrams of operator, the power
counting of $1/N_c$ become very complicated for $<0|I_{MG}|0>$,
$<0|I^2_{MG}|0>$ and $<0|I_{MG}|0>^2$ etc.. In general, the numbers
of $O(N_c)$ terms in Eq.(~\ref{4.2}) are infinite(e.g., see
fig.~\ref{gluon}). Thus it is impossible to sum over contribution
of all possible $O(N_c)$ terms in Eq.(~\ref{4.2}). A available way is to
capture dominant contribution and ignore other minor ingredients.

The success of the QCD sum rules\cite{Shifman79} implies that the
contribution from quadratic gluon condensate
$<0|(\alpha_s/\pi)G_{\mu\nu}^aG^{a\mu\nu}|0>$ is dominant at low
energy(detail discussion for QCD sum rules is in next subsection).
Moreover, it was shown in ref.\cite{Esp90} that contribution of the
term with triple gluon condensate is only around $5\%$ of quadratic
one(if we believe the dilute instantons gas estimate of the triple
condensate). So that the contribution of gluon coupling will be
calculated to $O(\alpha_s)$ in this paper. Up to this order, the
effective action reduced by integrating over gluon fields in
$I_{MG}$ is proportional a constant
\begin{equation}
\label{4.3}
     k=\frac{1}{24m^4}<0|\frac{\alpha_s}{\pi}G_{\mu\nu}^aG^{a\mu\nu}|0>.
\end{equation}
The explicit form of this effective lagrangian is
\begin{eqnarray}
\label{4.4}
{\cal L}^{(g)}&=&{\cal L}_{\rm kin}^{(g)}+{\cal L}_{\rm I}^{(g)}
        \nonumber \\
{\cal L}_{\rm kin}^{(g)}&=&\frac{k}{4}m^2<D_\mu UD^\mu U^{\dag}>
   +km^3<\chi U^{\dag}+\chi^{\dag}U>
 +\frac{k}{20}<L_{\mu\nu}L^{\mu\nu}+R_{\mu\nu}R^{\mu\nu}>\nonumber \\
{\cal L}_{\rm I}^{(g)}&=&-\frac{k}{40}<L_{\mu\nu}U^{\dag}R^{\mu\nu}U>
          -\frac{k}{40}<D_\mu UD^\mu U^{\dag}D_\nu UD^{\nu}U^{\dag}>
                    \nonumber \\
&&+[\frac{km}{8B_0}(1-c-\frac{2m}{B_0})-\frac{k}{160}(1-c)^2]
  <\chi U^{\dag}\chi U^{\dag}+\chi^{\dag}U\chi^{\dag}U>
                   \nonumber \\
  &&-\frac{km}{8B_0}<D_\mu UD^\mu U^{\dag}
          (\chi U^{\dag}+\chi^{\dag}U)>,
\end{eqnarray}
where ${\cal L}_{\rm kin}^{(g)}$ is nothing but to modify free parameters
$f_0$, $B_0$ and $g_\phi$ or $g$. The effective lagrangian ${\cal
L}_I^{(g)}$ contribute to low energy coupling constants of ChPT, $L_3$,
$L_5$, $L_8$ and $L_{10}$. The results are as follows
\begin{eqnarray}\label{4.41}
 L_3^{(g)}&=&-\frac{k}{40}(1-c)^4, \nonumber \\
 L_5^{(g)}&=&-\frac{km}{8B_0}(1-c)^2, \nonumber \\
 L_8^{(g)}&=&\frac{km}{8B_0}(1-c-\frac{2m}{B_0})-\frac{k}{160}(1-c)^2, \\
 L_{10}^{(g)}&=&-\frac{k}{40}(1-c)^2. \nonumber 
\end{eqnarray}
The above results coincide with one given in Ref.\cite{Esp90} if spin-1
meson resonances disappear. However, in presence of spin-1 meson 
resonances, these coupling constants are suppressed by factor $(1-c)$
which come from diagnolization of $A_{\mu}-\pa^{\mu}\Phi$. Recalling
$c=0.44$, we find that here the contribution from gluon coupling is much
smaller than one in ref.\cite{Esp90}. In addition, from Eq.(~\ref{3.2.3})
we can find that $L_3^{(0)}$ and $L_{10}^{(0)}$ are also compensated by
exchange effects of spin-1 meson resonances. It make those low energy
coupling constants obtained from ChQM agree with ChPT very well(see sect.
7). 

The value of gluon condensate has been estimated in QCD sum rules. For
determined the value of $k$ in ChQM, we like to comparing ChQM and QCD sum
rules in the following subsection.

\subsection{ChQM versus QCD sum rules}

In this subsection, we will use method of
Shifman-Vainshtein-Zakharov(SVZ) sum rules\cite{Shifman79} to study
relation between ChQM and QCD sum rules in terms of $\rho$
meson spectral distribution. Review paper on QCD sum rules of
$\rho$ meson is in \cite{SVZrho}. It must be claimed that the
comparison is available only for energy scale $\mu<\Lambda_{\rm
CSSB}$, since ChQM is legitimate only at low energy.

We start with the time order current-current correlator
\begin{equation}\label{4.6}
 \Pi_{\mu\nu}=j\int d^4xe^{iq\cdot x}
  <0|T\{j_\mu(x)^{\rm em}j_\nu(0)^{\rm em}\}|0>,
\end{equation}
where $q$ is the total momentum of the quark-antiquark pair injected in
the vacuum, where $j_\mu^{\rm em}$ is electromagnetic current of the
$\rho$ meson,
\begin{equation}\label{4.7}
   j_{\mu}^{\rm em}=\frac{1}{2}(\bar{u}\gamma_{\mu}u
      -\bar{d}\gamma_{\mu}d).
\end{equation}
Due to the current conservation $\Pi_{\mu\nu}$ is transversal and, hence,
\begin{equation}\label{4.8}
  \Pi_{\mu\nu}=(q_{\mu}q_{\nu}-q^2g_{\mu\nu})\Pi(q^2).
\end{equation}
The spectral density of $\rho$-meson is defined by
\begin{equation}\label{4.9}
  \rho(s)=b{\rm Im}\Pi(s),\hspace{1in} s=q^2,
\end{equation}
where normalization constant $b$ is determined by requirement of
Eq.(~\ref{4.9}) coinciding with the cross-section of $e^+e^-$
annihilation into hadrons(measured in the units $\sigma(e^+e^-\rightarrow
\mu^+\mu^-)$). Here we focus our attention on the following integral
of spectral density with weight $e^{-s/M^2}$,
\begin{equation}\label{4.10}
  I(M^2)=\frac{1}{M^2}\int dse^{-s/M^2}\rho(s).
\end{equation}
The prediction of QCD sum rules for this integral is
\begin{equation}\label{4.11}
  I(M^2)=I_{\rm npc}(M^2)+I_{\rm pc}(M^2),
\end{equation}
where $I_{\rm npc}(M^2)$ is non-perturbative contribution and $I_{\rm
pc}(M^2)$ is correction of perturbative QCD. $I_{\rm npc}(M^2)$ was
obtained by orginal work of Shifman ${\sl et\;al.}$, with
$m_u=m_d=0$\cite{Shifman79},
\begin{eqnarray}\label{4.12}
  I_{\rm npc}(M^2)&=&1+\frac{\pi^2}{3M^4}<0|\frac{\alpha_s}{\pi}
    G_{\mu\nu}^aG^{a\mu\nu}|0>-\frac{8\pi^2}{M^6}<0|\alpha_s
    (\bar{u}\gamma_{\alpha}\gamma_5\lambda^au-
     \bar{d}\gamma_{\alpha}\gamma_5\lambda^ad)^2|0> \nonumber \\
    &&-\frac{16\pi^2}{9M^6}<0|\alpha_s(\bar{u}\gamma_{\alpha}
      \lambda^au
     +\bar{d}\gamma_{\alpha}\lambda^ad)\sum_{q=u,d,s}
     (\bar{q}\gamma_{\alpha}\lambda^aq)|0>,
\end{eqnarray}
where
\begin{eqnarray}\label{4.13}
&&<0|{\cal O}_G|0>\equiv<0|(\alpha_s/\pi)G_{\mu\nu}^aG^{a\mu\nu}|0>
   =x_G1.2\times 10^{-2}GeV^4,
      \nonumber \\
&&<0|\alpha_s(\bar{u}\gamma_{\alpha}\gamma_5\lambda^au-
   \bar{d}\gamma_{\alpha}\gamma_5\lambda^ad)^2|0>\simeq x_{4q} 
   6.5\times 10^{-4}GeV^4, \\
&&<0|\alpha_s(\bar{u}\gamma_{\alpha}\lambda^au
     +\bar{d}\gamma_{\alpha}\lambda^ad)\sum_{q=u,d,s}
     (\bar{q}\gamma_{\alpha}\lambda^aq)|0>\simeq -x_{4q}6.5\times
    10^{-4}GeV^4. \nonumber
\end{eqnarray}
In Eq.(~\ref{4.13}), the parameters $x_G$ and $x_{4q}$ are allowed to
float in vicinity of unity. This parameterization will be used for
matching $\rho$-meson sum rule with data.

The perturbative correction up to third order in $\alpha_s(s)$ is given in
Refs.\cite{Surg91, Gorish91}
\begin{equation}\label{QSR1}
 I_{\rm pc}(M^2)=1+\frac{4}{9}a(\frac{M^2}{e^\gamma})\{1
   +0.729[a(\frac{M^2}{e^\gamma})]
   -0.386[a(\frac{M^2}{e^\gamma})]^2\},
\end{equation}
with
\begin{eqnarray}\label{QSR2}
a(\mu^2)&\equiv&(\frac{11}{12}N_c-\frac{2}{12}N_f)
   \frac{\alpha_s(\mu^2)}{\pi}
  =\frac{1}{\ln{(\mu^2/\Lambda^2)}}
   -0.79\frac{\ln{\ln{(\mu^2/\Lambda^2)}}}{\ln^2{(\mu^2/\Lambda^2)}}
      \nonumber \\
   &+&\frac{0.79^2}{\ln^3{(\mu^2/\Lambda^2)}}
   [(\ln{\ln{(\mu^2/\Lambda^2)}})^2-\ln{\ln{(\mu^2/\Lambda^2)}}+0.415]
   +O(\frac{1}{\ln^4{(\mu^2/\Lambda^2)}}),
\end{eqnarray}
where $\Lambda$ is the scale parameter of QCD introduced in a standard
way\cite{Hinch96}, in this paper we take $\Lambda=0.2$GeV.

\begin{figure}[tp]
   \centering
   \label{SVZ1}
   \includegraphics[width=4in]{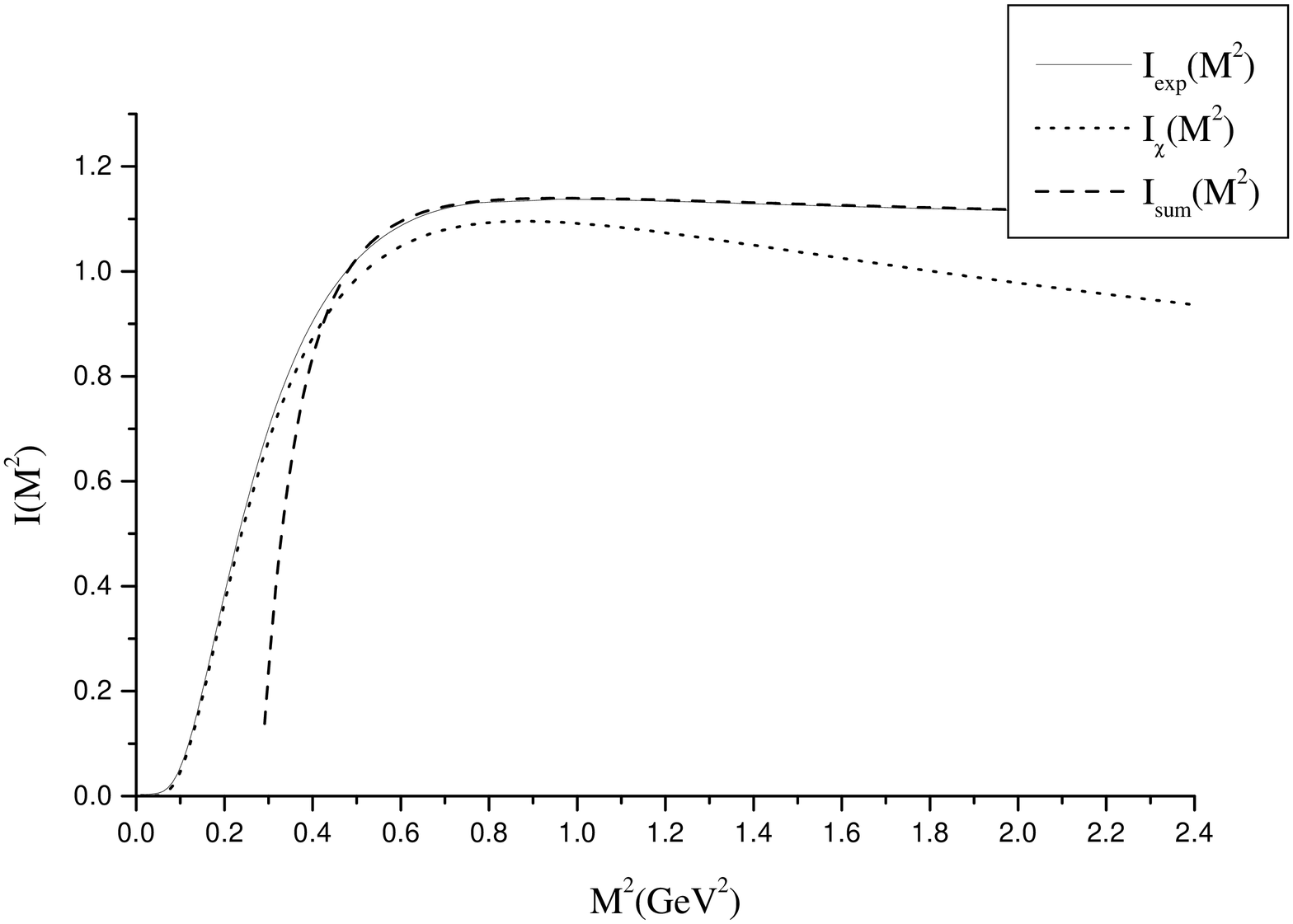}
\begin{minipage}{5in}
   \caption{$I(M^2)$ versus $M^2$ in the $\rho$-meson channel: the solid
   line denotes experiment data, the dot line denotes prediction by ChQM
   (without perturbative correction) and the dash line denotes predicetion
   by SVZ sum rules with $x_G=x_{4q}=1$(without perturbative correction).}
\end{minipage}
   \vspace{0.4in}
   \label{SVZ2}
   \includegraphics[width=4in]{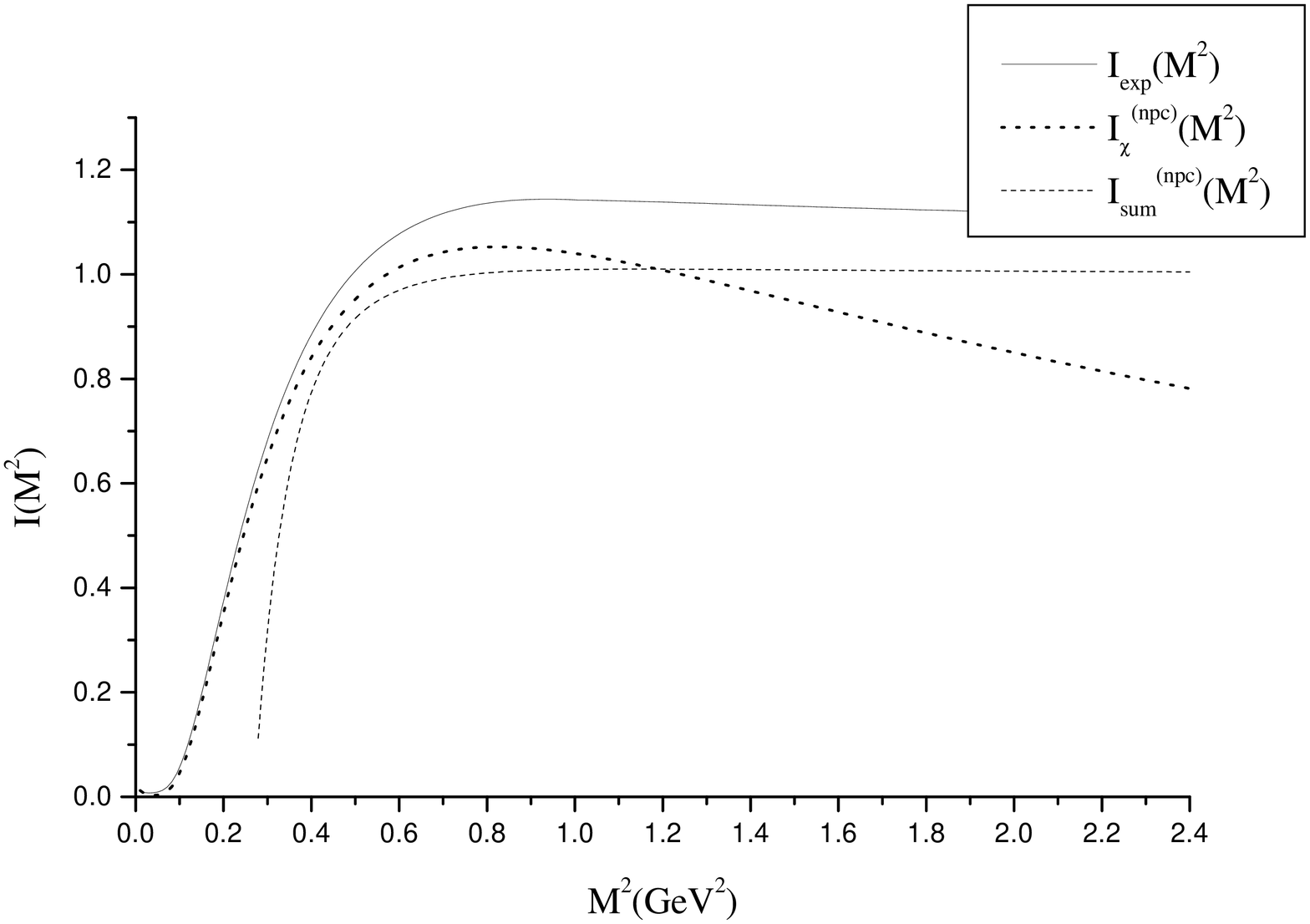}
\begin{minipage}{5in}
   \caption{$I(M^2)$ versus $M^2$ in the $\rho$-meson channel: the solid
   line denotes experiment data, the dot line denotes prediction by ChQM
   (with perturbative correction) and the dash line denotes predicetion
   by SVZ sum rules with $x_G=0.8$, $x_{4q}=1.3$(with perturbative
   correction).}
\end{minipage}
\end{figure}

Now let us return to framework of ChQM. Due to VMD, the electromagnetic
current of $\rho$ meson is obtained easily
\begin{equation}\label{4.14}
j_\mu^{\rm em}=\frac{g}{2}(\pa^2\rho_\mu^0-\pa_\mu\pa_\nu\rho^{0\nu}).
\end{equation}
Then we obtain polarization operator
\begin{equation}\label{4.15}
\Pi(q^2)=\frac{g^2}{4}\frac{q^2}{q^2-m_\rho^2+i\sqrt{q^2}\Gamma(q^2)},
\end{equation}
where $\Gamma(q^2)$ is just width of $\rho\rightarrow\pi\pi$ decay when
$q^2=m_\rho^2$, so that
\begin{eqnarray}\label{4.16}
\Gamma(q^2)&=&\frac{f_{\rho\pi\pi}^2(q^2)}{48\pi}\sqrt{q^2}
 (1-\frac{4m_\pi^2}{q^2})^{\frac{3}{2}}\theta(q^2-4m_\pi^2),
   \nonumber \\
f_{\rho\pi\pi}(q^2)&=&\frac{q^2}{gf_\pi^2}[2g^2c(1-\frac{c}{2})
   +\frac{(1-c)^2}{\pi^2}].
\end{eqnarray}
It should be pointed out that the momentum-dependence of $f_{\rho\pi\pi}$
is yielded in this model, since $\rho-\pi\pi$ vertex come from
${\cal L}_4$ instead of ${\cal L}_2$. Then the spectral density of
$\rho$-meson can be written as follows
\begin{equation}\label{4.17}
\rho(s)=2.3\pi g^2\frac{s\sqrt{s}\Gamma(s)}{(s-m_\rho^2)^2+s\Gamma^2(s)},
\end{equation}
where the spectral density has been normalized for coincides with
$$R^{I=1}=\frac{\sigma(e^+e^-\rightarrow{\rm hadrons}, I=1)}
  {\sigma(e^+e^-\rightarrow \mu^+\mu^-)}.$$
In fig.~\ref{SVZ1} we show the $I(M^2)$ curve which are obtained from QCD
sum rules(without perturbative QCD correction), ChQM and experiment
respectively. It is not surprising that the $I_{\chi}(M^2)$
obtained from ChQM agree with experiment excellently in $M^2<0.6{\rm
GeV}^2$ but do not match data when $M^2>1{\rm GeV}^2$. The reason
is that ChQM is a low energy model so that other heavier
vector-isovector meson resonances, such as $\rho(1450)$ and
$\rho(1700)$, are not included in this model. For improving this shortage,
we like to add preturbative QCD correction into spectral density through
the following rough way,
\begin{equation}\label{4.18}
 \rho(s)=2.3\pi g^2\frac{s\sqrt{s}\Gamma(s)}{(s-m_\rho^2)^2+s\Gamma^2(s)}
    +0.4(1+\frac{\alpha_s(s)}{\pi})\theta(s-s_0),
\end{equation}
where the part of perturbative correction has been normalized the same as
non-perturbative part, $\alpha_s(s)$ can be obtained from
Eq.(~\ref{QSR2}) and $s_0=2.5$GeV$^2$ is suggested by the experimental
data\cite{Barate97}. The result is shown in fig.4.3. We
can find now $I_{\chi}(M^2)$ agree with data well in $M^2<1.2$GeV$^2$, and
indeed perturbative correction in Eq.(~\ref{4.18}) improve high energy
behavior of $I_{\chi}(M^2)$ significantly. The $I_{\rm SVZ}(M^2)$
obtained from QCD sum rules and including perturbative correction is also
shown in fig.4.3, where $x_G=0.8$ and $x_{4q}=1.3$. It is shown
that ChQM coincides with QCD sum rules at $\rho$-meson energy scale.
Therefore, the gluon condensate-dependent parameter $k$ can be determined
as follows(with $x_G=0.8$ and $m\simeq 0.32$GeV),
\begin{equation}\label{4.19}
k=\frac{<0|{\cal O}_G|0>}{24m^4}=\frac{x_G0.012GeV^4}{24m^4}\simeq 0.038.
\end{equation}

Two more remarks are needed here. 1) From fig.4.3 we can find QCD
sum rules agree with data excellent when $M^2>0.6$GeV$^2$ even without
triple or more higher power gluon condensate. It means that quadratic
condensate is dominant in gluon vacuum condensate when $M^2>0.6$GeV$^2$.
In addition, it can be obtained from Eq.(~\ref{4.41}) that the quadratic
gluon condensate always accompanies the factor $1/(192m^4)$. Furthermore,
calculation will shown that the triple gluon condensate accompanies the
factor $1/(2880m^6)$ at least. Therefore, in ChQM, the operaor product
expansion in gluon sector is appoximately in powers of $1/(14m^2)$. Due to
$14m^2\simeq 1.4$GeV$^2\gg 0.6$GeV$^2$, the result of QCD sum rules
implies that here it is enough to consider quadratic gluon condensate only
and to ignore other vacuum condensate with triple or more higher powers of
gluon field strength. 2) Although phenomenological comparison between
$I_{\chi}$ and $I_{\rm SVZ}$ is successful, unfortunately, it is
open to analytically determine relation between $I_{\chi}$ and
$I_{\rm SVZ}$. From viewpoint of operator product expansion, it is because
there are two extra pamameters with dimension 1 in ChQM, constituent quark
mass $m$ and $\rho$-meson mass $m_{\rho}$. It make the operator product
expansion in framework of ChQM become impracticable. For example, we can
construct many operators with dimension 4, such as $<0|GG|0>$,
$<0|(\bar{\psi}\psi)^2|0>/m^2$, .... In general, both of these paramters
are functions of vacuum condenstate of QCD sum rules. However, the
relation between them and vacuum condenstate is unknown at all. In fact,
it is the most important but the hardest problem in studies on low energy
QCD, that we know nothing about how quark and gluon fields become hardons
at low energy. 

\section{General Formula for One Loop of Mesons}
\setcounter{equation}{0}

For convenience to study one loop contribution of mesons systematically.
In this section we like to derive general formula for one
loop of mesons with any spin by means of background field method.

Considering an action of mesons
$I[L_\mu,R_\mu,U]=\int d^4x{\cal L}(x)$, we expand meson fields in this
action around their classic solutions
\begin{eqnarray}
\label{5.1}
  &&V_{\mu}(x)=\bar{V}_{\mu}(x)+v_{\mu}(x), \nonumber \\
  &&A_{\mu}(x)=\bar{A}_{\mu}(x)+a_{\mu}(x), \\
  &&U(x)=\xi e^{i\vphi}\xi,  \hspace{0.8in}\bar{U}(x)=\xi^2(\Phi),
          \nonumber
\end{eqnarray}
where background fields $\bar{V}_\mu,\;\bar{A}_\mu$ and $\bar{U}(x)$ are
solutions of classics motion equation of mesons, i.e.,
$\delta I/\delta V_{\mu}(x)=0,\;\delta I/\delta A_{\mu}(x)=0$ and
$\delta I/\delta U(x)=0$, respectively.
$v_{\mu}(x),\;a_{\mu}(x)$ and $\vphi(x)$ are quantum fluctuation fields
around those classic solutions. In following two sections external vector
and axial-vector fields are included in background fields $\bar{V}_\mu$
and $\bar{A}_\mu$. Corresponding expansion of fields(~\ref{5.1}), the
action is written as
$$I[V_\mu,A_\mu,U]=\bar{I}[\bar{V}_\mu,\;\bar{A}_\mu,\bar{U}]
       +\delta I[\bar{V}_\mu,\bar{A}_\mu,\bar{U};v_\mu,a_\mu,\vphi].$$
Then the quantum correction of previous action,
$\Gamma[\bar{V}_\mu,\bar{A}_\mu,\bar{U}]$, can be evaluated by means of
integrating over the quantum fields
\begin{equation}
\label{5.2}
    e^{i\Gamma[\bar{V}_\mu,\bar{A}_\mu,\bar{U}]}=
      \int {\cal D}v_{\mu}(x){\cal D}a_{\mu}(x){\cal D}\vphi(x)
       e^{i\delta I[\bar{V}_\mu,\bar{A}_\mu,\bar{U};v_\mu,a_\mu,\vphi]}
\end{equation}
In particular, the one loop contribution is obtained via integrating
over the quantum fields in quadratic terms of
$\delta I[\bar{V}_\mu,\bar{A}_\mu,\bar{U};v_\mu,a_\mu,\vphi]$
\begin{eqnarray}
\label{5.3}
    e^{i\Gamma_{one-loop}[\bar{V}_\mu,\bar{A}_\mu,\bar{U}]}&=&
      \int {\cal D}v_{\mu}^a(x){\cal D}a_{\mu}^a(x){\cal D}\vphi^a(x)
      exp\{i\int d^4x
    (\vphi^{a}{\cal H}_{0\vphi\vphi}^{ab}(x)\vphi^{b} \nonumber \\
     &&+v_\mu^{a}{\cal H}_{0vv}^{\mu\nu,ab}(x)v_\nu^{b}
       +a_\mu^{a}{\cal H}_{0aa}^{\mu\nu,ab}(x)a_\nu^{b}
       +v_\mu^{a}{\cal H}_{0va}^{\mu\nu,ab}(x)a_\nu^{b} \nonumber \\
       &&+v_\mu^{a}{\cal H}_{0v\vphi}^{\mu,ab}(x)\vphi^b
       +a_\mu^{a}{\cal H}_{0a\vphi}^{\mu,ab}(x)\vphi^b)\},
\end{eqnarray}
where
\begin{eqnarray}
\label{5.4}
     {\cal H}_{0st}^{ab}(x)\delta^4(x-y)=\frac{1}{2!}
       \frac{\delta^2I}{\delta S^{a}(x)\delta T^{b}(y)}
            |_{V_\mu=\bar{V}_\mu,A_\mu=\bar{A}_\mu,U=\bar{U}},
\end{eqnarray}
where $s,t=\vphi,\;v_\mu,\;a_\mu$ and $S,T=\Phi,\;V_\mu,\;A_\mu$. In
expansion of the action, the terms to be proportional to quantum field
disappear due to classic equation of motion of mesons.

In principle, the integration in Eq.(~\ref{5.3}) can be
performed explicitly in terms of Gauss integral formula, e.g., for two
quantum field case
\begin{equation}\label{5.5}
 \int{\cal D}s{\cal D}te^{sAs+tBt+sCt}\;\propto\;
    \frac{1}{{\rm Det}(A+CB^{-1}C){\rm Det}B}.
\end{equation}
However, for three quantum field case, the result of functional
integral (~\ref{5.3}) becomes so complicated that calculations of
those determinants are impractical. Instead, like usual treatment
in functional integral, we introduce external sources for quantum
fields into the action technically for calculating path-integral of
Eq.(~\ref{5.3}).

At first, we consider a lagrangian of bosons, which has general form as
follows
\begin{eqnarray}
\label{5.6}
{\cal L}(x)&=&-\frac{1}{2}\phi_{A}(x)\na_{x}^{(\phi)AB}\phi_{B}(x)
     -\frac{1}{2}\Psi_{A}(x)\na_{x}^{(\Psi)AB}\Psi_{B}(x)
     -\frac{1}{2}\varphi_{A}(x)\na_{x}^{(\varphi)AB}\varphi_{B}(x)
                    \nonumber \\
   &&+\phi_{A}(x){\cal H}_{\phi}^{AB}(x)\phi_{B}(x)
     +\Psi_{A}(x){\cal H}_{\Psi}^{AB}(x)\Psi_{B}(x)
     +\varphi_{A}(x){\cal H}_{\varphi}^{AB}(x)\varphi_{B}(x)
            \nonumber \\
   &&+\phi_{A}(x)\Omega_{\phi\Psi}^{AB}(x)\Psi_{B}(x)
     +\phi_{A}(x)\Omega_{\phi\varphi}^{AB}(x)\varphi_{B}(x)
     +\varphi_{A}(x)\Omega_{\varphi\Psi}^{AB}(x)\Psi_{B}(x)
            \nonumber \\
   &=&-\frac{1}{2}\phi_{A}(x)\na_{x}^{(\phi)AB}\phi_{B}(x)
      -\frac{1}{2}\Psi_{A}(x)\na_{x}^{(\Psi)AB}\Psi_{B}(x)
      -\frac{1}{2}\varphi_{A}(x)\na_{x}^{(\varphi)AB}\varphi_{B}(x)
            \nonumber \\
      &&+V[\phi(x),\Psi(x),\varphi(x)],
\end{eqnarray}
where $\phi(x),\;\varphi(x)$ and $\Psi(x)$ are bosons with
arbitrary spin and parity, the index $A,\;B$ may be Lorentz index,
gauge group index,..., $\na_{(j)x}(j=\phi,\;\varphi,\;\Psi)$ are
free field operators of $j(x)$ fields, e.g.,
$\delta^{ab}(\pa^2_{x}+m^2)$ for fields with zero spin and
$\delta^{ab}[\delta^{\mu\nu}(\pa^2_{x}+m^2)-\pa_{x}^{\mu}\pa_{x}^{\nu}]$
for fields with spin-1. ${\cal H}_{j}(x)$ and
$\Omega_{ij}(x)$($i,j=\phi,\;\varphi,\;\Psi;i\neq j$) are function of
external fields, and in which differential operators may be included(for
our purpose in this paper, it is enough to consider one differential
operator included only). $V[\phi(x),\Psi(x),\varphi(x)]$ denotes
interaction between quantum fields and external fields. It is needed to
define the adjoint operators of ${\cal H}_j$ and $\Omega_{ij}$ as follows
for arbitrary operators $F(x)$ and $G(x)$ in which there are no
differential operators
\begin{eqnarray}
\label{5.7}
&&\int d^4x F(x)[{\cal H}_{j}(x)G(x)]=\int d^4x[{\cal H}_{j}^*(x)F(x)]G(x)
     \nonumber \\
&&\int d^4x F(x)[\Omega(x)_{ij}G(x)]=\int d^4x[\Omega_{ij}^*(x)F(x)]G(x)
        \equiv\int d^4x[\Omega_{ji}(x)F(x)]G(x).
\end{eqnarray}
where ${\cal H}^*_j$ and $\Omega^*_{ij}=\Omega_{ji}$ are adjoint operators
of ${\cal H}_j$ and $\Omega_{ij}$ respectively.

The one loop correction of
lagrangian(~\ref{5.6}) can be derived by means of path-integral with
corresponding external source fields, $\zeta,\;\eta$ and $\gamma$
\begin{eqnarray}
\label{5.8}
 &&e^{i\Gamma_{one-loop}}=\int{\cal D}\phi(x){\cal D}\Psi(x){\cal D}
       \varphi(x)\exp{\{i\int d^4x{\cal L}(x)\}} \nonumber \\
   &=&\exp{\{i\int d^4xV[\frac{1}{i}\frac{\delta}{\delta\zeta(x)},
    \frac{1}{i}\frac{\delta}{\delta\eta(x)},
    \frac{1}{i}\frac{\delta}{\delta\gamma(x)}\}}
   \int{\cal D}\phi(x){\cal D}\Psi(x){\cal D}\varphi(x)\times\nonumber\\
   &&{\rm exp}\{i\int d^4x[-\frac{1}{2}\sum_{j=\phi,\Psi,\varphi}
       j_{A}(x)\na_{x}^{(j)AB}j_{B}(x)+\zeta_{A}(x)\phi^{A}(x)
             \nonumber \\
    &&+\eta_{A}(x)\Psi^{A}(x)+\gamma_{A}(x)\varphi^{A}(x)]\}
          |_{\zeta=\eta=\gamma=0}            \nonumber \\
    &=&\frac{1}{N}\exp{\{i\int d^4xV[\frac{1}{i}\frac{\delta}
     {\delta\zeta(x)},\frac{1}{i}\frac{\delta}{\delta\eta(x)},
     \frac{1}{i}\frac{\delta}{\delta\gamma(x)}]\}}
     e^{iW_{\phi}[\zeta]}e^{iW_{\Psi}[\eta]}e^{iW_{\varphi}[\gamma]}
     |_{\zeta=\eta=\gamma=0},
\end{eqnarray}
where the constant $N$ do not incorporate dynamics so that we omit
it in the following calculation. The generating functional of
connect Green function, $W_{\phi}[\zeta],\;W_{\Psi}[\eta]$ and
$W_{\varphi}[\gamma]$, are written as
\begin{eqnarray}
\label{5.9}
     W_{\phi}[\zeta]&=&\frac{1}{2}\int d^4x\zeta_{A}(x)
           \Delta_{\phi}^{AB}(x-y)\zeta_{B}(y) \nonumber \\
     W_{\Psi}[\eta]&=&\frac{1}{2}\int d^4x\eta_{A}(x)
           \Delta_{\Psi}^{AB}(x-y)\eta_{B}(y)  \\
     W_{\varphi}[\gamma]&=&\frac{1}{2}\int d^4x\gamma_{A}(x)
           \Delta_{\varphi}^{AB}(x-y)\gamma_{B}(y) \nonumber
\end{eqnarray}
where $\Delta_{j}^{AB}(x-y)(j=\phi,\;\varphi,\;\Psi)$ are
propagators of $j$ fields(inverse of free field operator
$\na_{x}^{(j)AB}$). Substituting Eq.(~\ref{5.9}) into
Eq.(~\ref{5.8}) and performing functional differential
explicitly, we obtain effective action generated by one loop of
fields $\phi,\;\Psi$ and $\varphi$ as follows
\begin{equation}\label{5.10}
\Gamma_{one-loop}=\sum_{i\neq j}(\Gamma_{one-loop}^{(jj)}
   +\Gamma_{one-loop}^{(ij)})
   +\Gamma_{one-loop}^{(\phi\Psi\varphi)} \hspace{0.5in}
    (j=\vphi,\;\phi,\;\Psi)
\end{equation}
where
\begin{eqnarray}
\label{5.11}
&&i\Gamma_{one-loop}^{(jj)}=\int d^4x {\cal H}_j^{AB}(x)
          \Delta_{AB}(x-x) \nonumber \\
&&+\frac{1}{2}\int d^4x_{1}d^4x_{2}
    [{\cal H}_j^{AB}(x_1)\Delta^{(j)}_{BC}(x_1-x_2)]
 [({\cal H}_j^{CD}(x_2)+{\cal H}_j^{*DC}(x_2))\Delta_{DA}^{(j)}(x_2-x_1)]
            \nonumber \\
&&+\frac{1}{3}\int d^4x_{1}d^4x_{2}d^4x_{3}
        [{\cal H}_j^{AB}(x_1)\Delta_{BC}^{(j)}(x_1-x_2)]
 [({\cal H}_j^{CD}(x_2)+{\cal H}_j^{*DC}(x_2))\Delta_{DE}^{(j)}(x_2-x_3)]
            \nonumber \\  &&  \hspace{1.0in} \times
 [({\cal H}_j^{EF}(x_3)+{\cal H}_j^{*FE}(x_3))\Delta_{FA}^{(j)}(x_3-x_1)]
           \nonumber \\
&&+\frac{1}{4}\int d^4x_{1}d^4x_{2}d^4x_{3}d^4x_{4}
        [{\cal H}_j^{AB}(x_1)\Delta_{BC}^{(j)}(x_1-x_2)]
 [({\cal H}_j^{CD}(x_2)+{\cal H}_j^{*DC}(x_2))\Delta_{DE}^{(j)}(x_2-x_3)]
           \nonumber \\ && \hspace{0.2in} \times
 [({\cal H}_j^{EF}(x_3)+{\cal H}_j^{*FE}(x_3))\Delta_{FG}^{(j)}(x_3-x_4)]
 [({\cal H}_j^{GH}(x_4)+{\cal H}_j^{*HG}(x_4))\Delta_{HA}^{(j)}(x_4-x_1)]
           \nonumber \\
&&+...,  \hspace{2.0in} (j=\phi,\;\varphi,\;\Psi)
\end{eqnarray}
\begin{eqnarray}
\label{5.12}
&&i\Gamma_{one-loop}^{(ij)}=\frac{1}{2}\int d^4x_{1}d^4x_{2}
     [\Omega_{ij}^{AB}(x_1)\Delta_{BC}^{(j)}(x_1-x_2)]
     [\Omega_{ji}^{DC}(x_2)\Delta_{DA}^{(i)}(x_2-x_1)] \nonumber \\
     &&+\frac{1}{2}\int d^4x_{1}d^4x_{2}d^4x_{3}\sum_{k=i,j}
       [\Omega_{ij}^{AB}(x_1)\Delta_{BC}^{(j)}(x_1-x_2)]
       [\Omega_{ji}^{DC}(x_2)\Delta_{DE}^{(i)}(x_2-x_3)]
            \nonumber \\ && \hspace{1.0in} \times
[({\cal H}_k^{EF}(x_3)+{\cal H}_k^{*FE}(x_3))\Delta_{FA}^{(k)}(x_3-x_1)]
           \nonumber \\
&&+\frac{1}{4}\int d^4x_{1}d^4x_{2}d^4x_{3}d^4x_{4}
     [\Omega_{ij}^{AB}(x_1)\Delta_{BC}^{(j)}(x_1-x_2)]
     [\Omega_{ji}^{DC}(x_2)\Delta_{DE}^{(i)}(x_2-x_3)]
            \nonumber \\ && \hspace{1.0in} \times
     [\Omega_{ij}^{EF}(x_3)\Delta_{FG}^{(j)}(x_3-x_4)]
     [\Omega_{ji}^{HG}(x_4)\Delta_{HA}^{(i)}(x_4-x_1)]
            \nonumber \\
&&+\frac{1}{2}\int d^4x_{1}d^4x_{2}d^4x_{3}d^4x_{4}
     [\Omega_{ij}^{AB}(x_1)\Delta_{BC}^{(j)}(x_1-x_2)]
[({\cal H}_j^{CD}(x_2)+{\cal H}_j^{*DC}(x_2))\Delta_{DE}^{(j)}(x_2-x_3)]
            \nonumber \\  &&  \hspace{0.6in} \times
     [\Omega_{ji}^{FE}(x_3)\Delta_{FG}^{(i)}(x_3-x_4)]
[({\cal H}_i^{GH}(x_4)+{\cal H}_i^{*HG}(x_4))\Delta_{HA}^{(i)}(x_4-x_1)]
            \nonumber \\
 &&+\frac{1}{2}\int d^4x_{1}d^4x_{2}d^4x_{3}d^4x_{4}\sum_{k=i,j}
     [\Omega_{ij}^{AB}(x_1)\Delta_{BC}^{(j)}(x_1-x_2)]
     [\Omega_{ji}^{DC}(x_2)\Delta_{DE}^{(i)}(x_2-x_3)]
          \nonumber \\ &&\hspace{0.3in} \times
 [({\cal H}_k^{EF}(x_3)+{\cal H}_k^{*FE}(x_3))\Delta_{FG}^{(k)}(x_3-x_4)]
 [({\cal H}_k^{GH}(x_4)+{\cal H}_k^{*HG}(x_4))\Delta_{HA}^{(k)}(x_4-x_1)]
           \nonumber \\
      &&+....  \hspace{2.0in} (i\neq j)
\end{eqnarray}
\begin{eqnarray}
\label{5.13}
i\Gamma_{one-loop}^{(\phi\Psi\varphi)}&=&\int d^4x_{1}d^4x_{2}d^4x_{3}
    [\Omega_{\phi\varphi}^{AB}(x_1)\Delta_{(\varphi)BC}(x_1-x_2)]
    [\Omega_{\varphi\Psi}^{CD}(x_2)\Delta_{(\Psi)DE}(x_2-x_3)]
        \nonumber \\&&\times
    [\Omega_{\Psi\phi}^{EF}(x_3)\Delta_{(\phi)FA}(x_3-x_1)]
         +....
\end{eqnarray}
Here $\Gamma_{one-loop}^{(jj)}$ denotes the effective action
generated by one loop in which all internal lines are propagators
of $j$ field(see fig.~\ref{Ploop}). Hereafter we call this kind of loops
as ``pure loop'' of $j$ field. $\Gamma_{one-loop}^{(ij)}$ denotes
the effective action generated by one loop with internal lines of
two different fields $i,\;j$(see fig.~\ref{Ploop}). To call it as ``mixing
loop'' of $i,\;j$ fields hereafter. Similarly,
$\Gamma_{one-loop}^{(\phi\Psi\varphi)}$ denotes effective action
generated by mixing loops with internal lines of three different
fields.

An useful formula is as follows. If $F_1(x),\;F_2(x),\;...,\;F_n(x)$
and $G_1(x),\;G_2(x),\;...,\;G_n(x)$ are operators in D dimension
spacetime and are independent on differential operator. Their Fourier
transformation are
\begin{equation}\label{5.14}
  F_i(x)=\int\frac{d^Dp}{(2\pi)^D}f_i(p)e^{-ip\cdot x} \hspace{0.8in}
  G_i(x)=\int\frac{d^Dp}{(2\pi)^D}g_i(p)e^{-ip\cdot x}. \nonumber
\end{equation}
In addition, $\Delta(x-y)$ is propagator of a field and $\Delta(p)$ is
its Fourier transformation,
\begin{equation}\label{5.15}
  \Delta(x-y)=\int\frac{d^Dp}{(2\pi)^D}\Delta(p)e^{-ip\cdot (x-y)}
       \nonumber
\end{equation}
Then we have the following formula which is available in low energy
region.
\begin{eqnarray}
\label{5.16}
  &&\int d^Dx_1 d^Dx_2...d^Dx_n
    [(F_1^{A_1}(x_1)\frac{\pa}{\pa x_1^{A_1}}+G_1(x_1))
     \Delta(x_1-x_2)]\times  \nonumber \\
  &&[(F_2^{A_2}(x_2)\frac{\pa}{\pa x_2^{A_2}}+G_1(x_2))\Delta(x_2-x_3)]
    ...
    [(F_n^{A_n}(x_n)\frac{\pa}{\pa x_n^{A_n}}+G_1(x_n))\Delta(x_n-x_1)]
         \nonumber \\
 &=&\int d^Dx\int\frac{d^Dq}{(2\pi)^D}
    [F_1^{A_1}(x)(\frac{\pa}{\pa x^{A_1}}+iq_{A_1})+G_1(x)]
       \Delta(q-i\pa) \nonumber \\ && \times
    [F_2^{A_2}(x)(\frac{\pa}{\pa x^{A_2}}+iq_{A_2})+G_2(x)]
       \Delta(q-i\pa)...
    [F_{n-1}^{A_{n-1}}(x)(\frac{\pa}{\pa x^{A_{n-1}}}+iq_{A_{n-1}})
        +G_{n-1}(x)]  \nonumber \\
  &&\Delta(q-i\pa)
    [F_{n}^{A_{n}}(x)+iq_{A_{n}}+G_{n}(x)]\Delta(q-i\pa),
\end{eqnarray}
where the formal operator symbol $\Delta(q-i\pa)$ has been introduced,
which can be expanded in power of external momentum in low energy region.
For instance, for zero spin fields it is
\begin{equation}
\label{5.17}
  \Delta (q-i\pa)\propto\frac{1}{q^2-m^2+i\ep}(1
      +\frac{\pa^2+2iq\cdot\pa}{q^2-m^2+i\ep}
      -\frac{4(q\cdot\pa)^2}{(q^2-m^2)^2+i\ep}+...) \nonumber
\end{equation}
and for spin-1 fields it is
\begin{eqnarray}
\label{5.18}
 &&\Delta_{\mu\nu}(q-i\pa)=\Delta_{\mu\nu}(q)+\Delta_{\mu\rho}(q)
     {\cal P}^{\rho\sigma}\Delta_{\sigma\nu}(q)+
    \Delta_{\mu\rho}(q){\cal P}^{\rho\sigma}\Delta_{\sigma\lambda}(q)
     {\cal P}^{\lambda\tau}\Delta_{\tau\nu}(q)+...,
        \nonumber \\
 &&\Delta_{\mu\nu}(q)\propto\frac{1}{q^2-m^2+i\ep}(\delta_{\mu\nu}
        +\frac{q_{\mu}q_{\nu}}{q^2-m^2})  \\
 &&{\cal P}_{\mu\nu}=(\pa^2+2iq\cdot\pa)\delta_{\mu\nu}
        +\pa_\mu\pa_\nu  \nonumber
\end{eqnarray}
The formula can be checked directly by inserting Eqs.(~\ref{5.14}) and
(~\ref{5.15}) into left side of Eq.(~\ref{5.16}).
Employing the formula(~\ref{5.16}) in Eq.(~\ref{5.11})-(~\ref{5.13}),
the effective action $\Gamma_{one-loop}$ can be expressed by explicit
integral of loop-momentum(to see appendix). It should be pointed out
that low energy theorem allows to expand integrand in Feynman-integral of
meson-loops in powers of external momentum. For example, setting $k=q+p$
(where $p$ is external momentum, $k$ and $q$ are momentums carried by
internal lines of loop), then we have
\begin{equation}\label{5.19}
\frac{1}{(q^2+m^2)((k^2+m^2)}=\frac{1}{(q^2+m^2)^2}
 +\frac{2q\cdot p+p^2}{(q^2+m^2)^3}+\frac{4(q\cdot p)^2}{(q^2+m^2)^4}
 +....
\end{equation}

\section{One Loop of Mesons in ChQM}
\setcounter{equation}{0}
\setcounter{figure}{0}

It has been well known that, at very low energy, the effective lagrangian
${\cal L}_2^{\phi}$ is not only treated at tree level but its one-loop
graphs contribute to ${\cal L}_4^{\phi}$. Simultaneously, ${\cal
L}_4^{\phi}$ is treated at tree level only\cite{Wein79,GL85a,GL85b}
because its one-loop contribution belongs to higher order terms only.
These conclusion and their proof are still suited for here since low
energy expansion on ${\cal L}^{\phi}$ is well-defined in ChQM. It will be
checked again in section 6.1 by another method. Obviously, besides
pseudoscalar, in this framework internal line of one-loop graphs can be
spin-1 meson fields too. It has been pointed out that, effective
lagrangian ${\cal L}^{V.A}$ with six or higher
derivative terms is not well-defined because the parameter $g$ is
determined phenomenologically by leading order of spin-1 meson coupling to
pseudoscalar fields. Therefore in this paper we only consider spin-1
meson-dependent one-loop contribution generated by ${\cal L}^{V.A}$ with
four derivatives, i.e., leading coupling of spin-1 mesons and pseudoscalar
mesons.

There are several remarks relating to our following calculations:
1) Since the contribution from gluon coupling is much smaller than one
from quark loops, we omit the one-loop effects
generated by lagrangian(~\ref{4.4}). 2) Pseudoscalar mesons are
treated as massless particles for keeping chiral symmetry of
results. 3) In following discussion we appoint that $p$ is external
momentum and $q$ is momentum carried by internal lines(hereafter we
call it as loop-momentum). Due to Eq.(~\ref{5.19}), it is
convenience in following analysis that we use a momentum to label
all momentums carried by every internal lines, because the
difference is only high order of external momentum, and which can
be taken into account in calculation on high order diagram. 4) In
our calculation the dimensional regularization is used for keeping
chiral symmetry. However, sometimes for convenience we will
approximately use cut-off regularization to discuss problems.

\subsection{Pure one-loops of pseudoscalar mesons}

\begin{figure}\label{Ploop}[tp]
   \centering
   \includegraphics[width=5in]{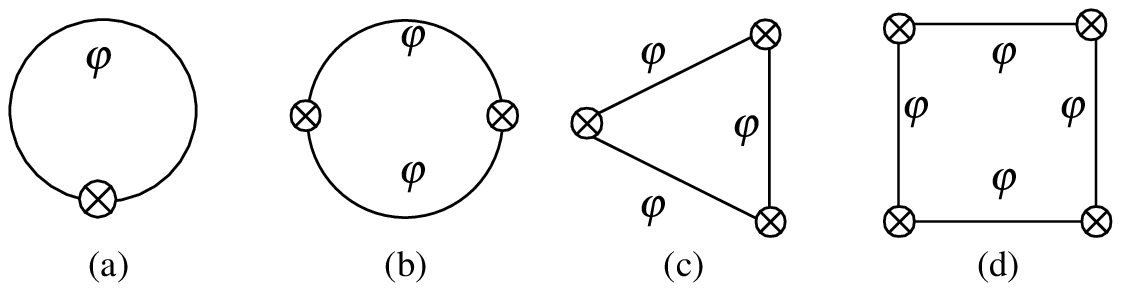}
\begin{minipage}{5in}
   \caption{Pure one-loop graphs of $0^-$ mesons. Here ``$\otimes$"
  denotes classic external fields and $\vphi$ denotes internal lines of
  $0^-$ mesons.}
\end{minipage}
\end{figure}

To $O(p^4)$ there are four kinds of potential one-loop graphs which
involve pseudoscalar meson internal lines only(fig.~\ref{Ploop})

\begin{enumerate}
\item Recalling that we treat pseudoscalar as zero mass particles for
keeping chiral symmetry, the potential "tadpole"
contribution(fig.~\ref{Ploop}-a)
can be omitted because of in dimensional regularization
\begin{equation}
\label{6.1.1}
  \int\frac{d^D q}{(2\pi)^D}\frac{1}{q^2+i\ep}\equiv 0.
\end{equation}
\item If there are some vertices generated by ${\cal L}_4$ in figure
\ref{Ploop}-b, \ref{Ploop}-c and \ref{Ploop}-d, the internal lines in
these figures must carry momentum since external momentum cannot be
higher than $p^4$. It make integral form of one loop be same as
Eq.(~\ref{6.1.1}) or yield more high
order divergences(higher than quadratic). These contribution can be
omitted in dimensional regularization too. Therefore merely one-loop
graphs generated by pseudoscalar mesons in ${\cal L}_2$ are needed to
calculate in this paper. This conclusion is same as one obtained
by Weinberg power counting rules\cite{Wein79,GL85a}.
\end{enumerate}
Inserting Eq.(~\ref{5.1}) in ${\cal L}_2$(~\ref{3.1.14}) and retaining
terms to quadratic form of quantum fields we obtain
\begin{equation}
\label{6.1.2}
 {\cal L}_2=\bar{{\cal L}}_2+\frac{f_0^2}{16}<d_\mu\varphi d^\mu\varphi
    -[\Delta_\mu,\varphi][\Delta^\mu,\varphi]
    -\frac{1}{4}\{\varphi,\varphi\}(\xi\chi^{\dag}\xi
       +\xi^{\dag}\chi\xi^{\dag})>
\end{equation}
where the antihermitian matrix $\Delta_\mu$ and $\Gamma_\mu$ and
covariant derivative have been defined in sect. 3.1.

Employing the completeness relation of generators
$\lambda^a(a=1,2,...,N^2-1)$ of $SU(N)$\cite{GL85a}
\begin{eqnarray}
\label{6.1.3}
    &&<\lambda^aA\lambda^aB>=-\frac{2}{N}<AB>+2<A><B> \nonumber \\
    &&<\lambda^aA><\lambda^aB>=2<AB>-\frac{2}{N}<A><B>
\end{eqnarray}
we can write the vertex included quadratic form of $\varphi$ in
more explicit form in terms of the components $\varphi^a$
\begin{eqnarray}
\label{6.1.4}
{\cal L}_{\varphi\varphi}(x)&=&\vphi^{a}{\cal
         H}_{0\vphi\vphi}^{ab}(x)\vphi^{b}
    =-\frac{1}{2}\varphi^a(x)\na^{(\varphi)ab}_x\varphi^b(x)+
    \varphi^a(x){\cal H}_{\varphi}^{ab}(x)\varphi^b(x) \nonumber \\
  \na^{ab}&=&\frac{f_0^2}{4}\delta^{ab}\pa^2      \nonumber \\ {\cal
   H}_{\varphi}^{ab}&=&-\frac{f_0^2}{8}(\{\pa^{\mu},\Gamma_{\mu}^{ab}\}
    +\Gamma_{\mu}^{ac}\Gamma^{\mu,cb})
 -\frac{f_0^2}{16}<[\Delta_\mu,\lambda^a][\Delta^\mu,\lambda^b]>
       \nonumber \\
  &&-\frac{f_0^2}{64}<\{\lambda^a,\lambda^b\}
     (\xi\chi^{\dag}\xi+\xi^{\dag}\chi\xi^{\dag})>
\end{eqnarray}
where
\begin{equation}
\label{6.1.5}
    \Gamma_{\mu}^{ab}=-\frac{1}{2}<[\lambda^a,\lambda^b]\Gamma_\mu>.
\end{equation}
From Eq.(~\ref{6.1.4}) we can find that the adjoint operator of
${\cal H}_{\varphi}^{ab}$ defined in sect. 5 is
\begin{equation}\label{6.1.6}
 {\cal H}_{\varphi}^{*ab}={\cal H}_{\varphi}^{ba}
\end{equation}
The free field operator $\na^{(\varphi)ab}_x$ in Eq.(~\ref{6.1.4})
yield propagator of $\varphi$ field
\begin{equation}
\label{6.1.7}
 \Delta^{(\varphi)ab}(x-y)=\int \frac{d^4q}{(2\pi)^4}\delta^{ab}
     \Delta^{(\varphi)}(q)e^{-iq\cdot (x-y)},\hspace{0.5in}
     \Delta^{(\varphi)}(q)=-\frac{4}{f_0^2}\frac{1}{q^2+i\ep}
\end{equation}
In fact, due to massless pseudoscalar fields, there are infrared
divergences when we substitute the propagator of $\vphi$(~\ref{6.1.7})
into Eq.(~\ref{8.2}) and perform the integral of loop-momentum explicitly.
Because the effective lagrangian generated by one-loops of
pseudoscalar mesons is $O(p^4)$, we can introduce a
external momentum scale factor $\mu_p$to regularize this infrared
divergence, i.e.,
\begin{equation}\label{6.1.8}
  \frac{1}{q^2+i\ep}\longrightarrow\frac{1}{q^2+\mu_p^2+i\ep}
       =\frac{1}{q^2+i\ep}(1-\frac{\mu_p^2}{q^2+i\ep}+...)
       \hspace{0.4in}\mu_p^2\simeq m_\phi^2
\end{equation}
From above equation we can easily find that the contribution of
$\mu_p^2$ is $O(p^6)$, so that above method to regularize infrared
divergences is available.

Now substituting the vertex(~\ref{6.1.4}), and propagator (~\ref{6.1.8})
into general formula(~\ref{8.2}) and performing the integral of loop
momentum, we obtain
\begin{equation}\label{6.1.9}
 {\cal L}_{one-loop}^{(\vphi\vphi)}=\frac{1}{4(4\pi)^{D/2}}
     (\frac{\mu^2}{\mu_p^2})^{\ep/2}\Gamma(2-\frac{D}{2})
     (\frac{1}{6}\Gamma_{\mu\nu}^{ab}\Gamma_{\mu\nu}^{ba}
      +\sigma^{ab}\sigma^{ba}),
\end{equation}
where
\begin{eqnarray}\label{6.1.10}
\Gamma_{\mu\nu}^{ab}&=&-\frac{1}{2}<\Gamma_{\mu\nu}[\lambda^a,\lambda^b]>
            \nonumber \\
\Gamma_{\mu\nu}&=&\pa_\mu\Gamma_\nu-\pa_\nu\Gamma_\mu
       +[\Gamma_\mu,\Gamma_\nu]=[d_\mu,d_\nu] \nonumber \\
  &=&-[\Delta_\mu,\Delta_\nu]-\frac{i}{2}(\xi F^L_{\mu\nu}\xi^{\dag}
       +\xi^{\dag}F^R_{\mu\nu}\xi)  \\
\sigma^{ab}&=&\frac{1}{2}<[\Delta^{\mu},\lambda^a]
       [\Delta_{\mu},\lambda^b]>+\frac{1}{8}<\{\lambda^a,\lambda^b\}
       (\xi\chi^{\dag}\xi+\xi^{\dag}\chi\xi^{\dag})> \nonumber
\end{eqnarray}
Then we can insert Eqs.(~\ref{6.1.10}) in Eq.(~\ref{6.1.9}) and simplify
the result by employing completeness relation of generators of
$SU(3)$(Eq.(~\ref{6.1.3}), $N=3$) and the following
identity\cite{GL85a}
\begin{equation}\label{6.1.11}
  <ABAB>=-2<A^2B^2>+\frac{1}{2}<A^2><B^2>+<AB>^2.
\end{equation}
In explicit form, the effective lagrangian generated by one-loop
graphs in fig.~\ref{Ploop} reads
\begin{eqnarray}\label{6.1.12}
{\cal L}_{one-loop}^{(\vphi\vphi)}&=&\frac{1}{4(4\pi)^{D/2}}
     (\frac{\mu^2}{\mu_p^2})^{\ep/2}\Gamma(2-\frac{D}{2})\{
     -\frac{1}{4}<F^L_{\mu\nu}F^{L\mu\nu}
     +F^R_{\mu\nu}F^{R\mu\nu}+2F^L_{\mu\nu}\bar{U}^{\dag}
      F^{R\mu\nu}\bar{U}> \nonumber \\
&&-\frac{i}{2}<F^{L\mu\nu}\na_{\mu}\bar{U}^{\dag}\na_{\nu}\bar{U}
   +F^{R\mu\nu}\na_{\mu}\bar{U}\na_{\nu}\bar{U}^{\dag}>
   +\frac{3}{16}<\na_{\mu}\bar{U}^{\dag}\na^{\mu}\bar{U}>^2
     \nonumber \\
&&+\frac{3}{8}<\na_{\mu}\bar{U}^{\dag}\na_{\nu}\bar{U}>
     <\na^{\mu}\bar{U}^{\dag}\na^{\nu}\bar{U}>  \nonumber \\
&&+\frac{1}{4}<\na_{\mu}\bar{U}^{\dag}\na^{\mu}\bar{U}>
      <\chi\bar{U}^{\dag}+\chi^{\dag}\bar{U}>
  +\frac{3}{4}<\na_{\mu}\bar{U}^{\dag}\na^{\mu}\bar{U}
      (\chi\bar{U}^{\dag}+\chi^{\dag}\bar{U})> \nonumber \\
&&+\frac{11}{72}<\chi\bar{U}^{\dag}+\chi^{\dag}\bar{U}>^2
  +\frac{5}{24}<\chi\bar{U}^{\dag}\chi\bar{U}^{\dag}
         +\chi^{\dag}\bar{U}\chi^{\dag}\bar{U}>\}
\end{eqnarray}

The effective lagrangian (~\ref{6.1.12}) is just result of
ChPT\cite{GL85a, Dynamics}. Of course, in framework of truncated field
theory, it is different from ChPT that the divergences in
Eq.(~\ref{6.1.12}) have to be parameterized, i.e., we have to define
\begin{equation}\label{6.1.13}
 g_1=\frac{1}{(4\pi)^{D/2}}
     (\frac{\mu^2}{\mu_p^2})^{\ep/2}\Gamma(2-\frac{D}{2})
\end{equation}
to absorb the divergence in Eq.(~\ref{6.1.12}). It also means that a
cut-off for pseudoscalar meson loop integral is introduced here.
Phenomenologically, contribution from meson loops should be small than one
from quark loops. We can find all coefficients in lagrangian
(~\ref{6.1.12}) are not small sufficient to provide suppression.
Thus the parameter $g_1$ must be smaller than $g_\phi^2$ which absorb the
divergences from quark loops. Usually, it is argued simply that $g_1$
is suppressed by $1/N_c$ due to $g_1/g_\phi^2\sim O(1/N_c)$. 
In practice, the parameter $g_1$ should be very small.
For instance, comparing with experimental data of $L_1$ we have $g1\ll
64L_1/3\simeq 0.015$. However, for $N_c=3$ in real world, $1/N_c$
expanosion can not yield a suppression for $g_1$ as large as our
expectation. In fact, the scale also play important role in calculation of
meson loops. It means that the truncated point in meson loops should be
smaller than one in quark loops. Approximately, if we suppose
$\mu_p\rightarrow m_\pi$ in very low energy limit and define $g_1$ in
scheme of cut-off regularization
\begin{equation}\label{6.1.14}
g_1=\frac{1}{16\pi^2}[\ln{(1+\frac{\Lambda^2}{\mu_p^2})}-
   \frac{\Lambda^2}{\Lambda^2+\mu_p^2}].
\end{equation}
Then we have $\Lambda\ll 700$MeV. This cut-off is much smaller than one
from quark loops. It indicates square of momentum transfer is very small
in pure pseudoscalar meson one-loop. The similar result will be obtained
in other meson one-loop calculation.

Finally, we point out that flavor number do not play crucial role
in effective lagrangian reduced by meson loops. We can perform
calculation in $N$ flavors and take $N=3$ finally.

\subsection{Pure one-loops of spin-1 mesons}

Inserting Eq.(~\ref{5.1}) into ${\cal L}^{V.A}$ and retaining
terms to quadratic form of quantum fields $v_{\mu}$ or $a_{\mu}$ we obtain
\begin{eqnarray}
\label{6.2.1}
{\cal L}_{(vv)}&=&-\frac{1}{4}<(\bar{d}_\mu v_\nu-\bar{d}_\nu v_\mu)^2>
+\frac{1}{8}i<[v_{\mu},v_{\nu}](\xi\bar{L}^{\mu\nu}\xi^{\dag}
    +\xi^{\dag}\bar{R}^{\mu\nu}\xi)> \nonumber \\
&&-\frac{\kappa^2}{4}<([\bar{\Delta}_{\mu},v_{\nu}]
  -[\bar{\Delta}_{\nu},v_{\mu}])
   [\bar{\Delta}^{\mu},v^{\nu}]>
  +\frac{4\gamma}{3}<[v_{\mu},v_{\nu}]
   [\bar{\Delta}^\mu,\bar{\Delta}^\nu]>
   \nonumber \\
&&+\frac{1}{4}m_{_V}^2<v_\mu v^\mu>
\end{eqnarray}
and
\begin{eqnarray}
\label{6.2.2}
{\cal L}_{(aa)}&=&-\frac{1}{4}<(\bar{d}_\mu a_\nu-\bar{d}_\nu a_\mu)^2>
  +(\frac{1}{8\kappa^2}-\frac{2\gamma}{3g_{_A}^2})i
 <[a_\mu,a_\nu](\xi\bar{L}^{\mu\nu}\xi^{\dag}
    +\xi^{\dag}\bar{R}^{\mu\nu}\xi)> \nonumber \\
&&+(\frac{8\gamma}{3g_{_A}^2}-\frac{1}{4\kappa^2})
   <([\bar{\Delta}_{\mu},a_{\nu}]
  -[\bar{\Delta}_{\nu},a_{\mu}])[\bar{\Delta}^{\mu},a^{\nu}]>
   \nonumber \\
&&+\frac{2\theta}{g_{_A}^2}<\{a_\mu,a_\mu\}
      (\xi\chi^{\dag}\xi+\xi^{\dag}\chi\xi^{\dag})> \nonumber \\
&&-\frac{4\gamma}{3g_{_A}^2}<\{a_\mu,a_\mu\}
     \{\bar{\Delta}_\mu,\bar{\Delta}_\nu\}
    +2\bar{\Delta}_\mu a_\nu\bar{\Delta}^\mu a^\nu>
    +\frac{1}{4}m_{_A}^2<a_\mu a^\mu>,
\end{eqnarray}
where $g_{_A}$ and $\kappa$ have been define in Eq.(~\ref{3.2.4}) and
\begin{eqnarray}\label{6.2.3}
\theta&=&\theta_1=\frac{3g^2m}{16B_0}-\frac{\gamma m}{2B_0}
  \nonumber \\
\bar{d}_\mu t_\nu&=&d_\mu t_\nu-i[\bar{V},t_\mu], \hspace{0.5in}
    (t=v,a) \nonumber \\
\bar{\Delta}_\mu&=&(1-c)\Delta_\mu-i\bar{A}_\mu
   =\frac{1}{2}\xi^{\dag}D_\mu U\xi^{\dag}=-\frac{1}{2}\xi D_\mu U\xi,
    \nonumber \\
\bar{L}_{\mu\nu}&=&(1-\frac{c}{2})F^L_{\mu\nu}+\frac{c}{2}F^R_{\mu\nu}
  +\xi^{\dag}(\bar{V}_{\mu\nu}-\bar{A}_{\mu\nu})\xi
  -2ic(1-\frac{c}{2})\xi^{\dag}
  [\bar{\Delta}_\mu,\bar{\Delta}_\nu]\xi \nonumber \\
 &&-(1-c)\xi^{\dag}
   ([\bar{\Delta}_\mu,\bar{V}_\nu-\bar{A}_\nu]
   -[\bar{\Delta}_\nu,\bar{V}_\mu-\bar{A}_\mu])\xi \nonumber \\
\bar{R}_{\mu\nu}&=&(1-\frac{c}{2})F^R_{\mu\nu}+\frac{c}{2}F^L_{\mu\nu}
  +\xi(\bar{V}_{\mu\nu}+\bar{A}_{\mu\nu})\xi^{\dag}
  -2ic(1-\frac{c}{2})\xi
  [\bar{\Delta}_\mu,\bar{\Delta}_\nu]\xi^{\dag} \nonumber \\
 &&+(1-c)\xi([\bar{\Delta}_\mu,\bar{V}_\nu+\bar{A}_\nu]
   -[\bar{\Delta}_\nu,\bar{V}_\mu+\bar{A}_\mu])\xi^{\dag} \nonumber \\
\end{eqnarray}
with
\begin{eqnarray}\label{6.2.4}
\bar{V}_{\mu\nu}&=&d_{\mu}\bar{V}_\nu-d_{\nu}\bar{V}_\mu
   -i[\bar{V}_\mu,\bar{V}_\nu]-i[\bar{A}_\mu,\bar{A}_\nu]\nonumber \\
\bar{A}_{\mu\nu}&=&d_\mu\bar{A}_\nu-d_\nu\bar{A}_\mu
    -i[\bar{A}_\mu,\bar{V}_\nu]-i[\bar{V}_\mu,\bar{A}_\nu].
\end{eqnarray}
The kinetic terms of quantum fields $v_\mu$ and $a_\mu$ in
Eqs.(~\ref{6.2.1}) and (~\ref{6.2.2}) yield their propagators in momentum
space as follows
\begin{equation}\label{6.2.5}
\Delta_{\mu\nu}^{(t)}(q)=\frac{1}{q^2-m_{_T}^2+i\ep}
   (\delta_{\mu\nu}-\frac{q_\mu q_\nu}{m_{_T}^2}),
\end{equation}
where $t=v,\;a$ and $T=V,\;A$. It seem that the propagator of spin-1
mesons yield bad ultraviolet behavior in loop-momentum integral. However,
it is only ostensible conclusion in
the present framework. For illustrating it, we like to study kinetic term
of vector mesons generated by meson one-loop. The completeness
relation of generators $\lambda^a(a=1,2,...,N^2-1)$ of $SU(N)$
(~\ref{6.1.3}), vertices (~\ref{6.2.1}), (~\ref{6.2.2}) and chiral
symmetry require kinetic term of vector mesons generated by vector
meson one-loop has form as follows
\begin{equation}\label{6.2.6}
 f(\frac{\Lambda^2}{m_{_V}^2})[c_1N_f<\bar{V}_{\mu\nu}\bar{V}^{\mu\nu}>
   -c_1<\bar{V}_{\mu\nu}><\bar{V}^{\mu\nu}>]
\end{equation}
where for convenience to discuss, we introduce ultraviolet cut-off to
regularize divergence, and all possible high order powers of $\Lambda$ are
also included in $f(\frac{\Lambda^2}{m_{_V}^2})$. The kinetic term of
vector mesons generated by axial-vector meson one-loop has similar form.
OZI rule indicates the coefficient of
$<\bar{V}_{\mu\nu}><\bar{V}^{\mu\nu}>$ should be very small
comparing with $g^2\simeq 0.16$. However, since the coefficient of
$\xi\bar{L}^{\mu\nu}\xi^{\dag}+\xi^{\dag}\bar{R}^{\mu\nu}\xi$ in vertices
(~\ref{6.2.1}) are not small, the constants $c_1$ are not small
sufficient too. Therefore, factor $f(\frac{\Lambda^2}{m_{_V}^2})$
should be very small here(similarly, $f(\frac{\Lambda^2}{m_{_A}^2})$ is
very small simultaneously). In this present model, it happen only for very
low energy cut-off(i.e., $f(\frac{\Lambda^2}{m_{_T}^2})$ vanish for
$\Lambda\rightarrow 0$, where $T=V,\;A$). It means that square of the
momentum transfer of meson loop is small even for higher energy
scale(e.g., $\mu\sim m_\rho$). Sequentially,
$q_\mu q_\nu/m_{_T}^2$ in vector and axial-vector propagators is
suppressed due to small momentum transfer. Thus this terms 
can be omitted here and high order divergences will disappear.

Instead of above intuitive analysis, we can discuss this problem through
another way. To consider operator equation of quantum field
$t_\mu$($t=v,a$)
\begin{equation}\label{6.2.7}
  \frac{\delta {\cal L}}{\delta t_\mu}=0.
\end{equation}
We writing this equation in explicit form
\begin{equation}\label{6.2.8}
  d_\nu t^{\mu\nu}+m_{_T}^2t^\mu=j_{_T}^{\mu}(v,a,\varphi).
\end{equation}
Acting covariant differentiators $d_\mu$ to two side of above
equation, we obtain
\begin{equation}\label{6.2.9}
 d_\mu t^\mu=-\frac{1}{m_{_T}^2}d_\mu \bar{j}_{_T}^\mu.
\end{equation}
On the other hand, acting $d_\nu$ to two side of Eq.(~\ref{6.2.8}), we
obtain
\begin{equation}\label{6.2.10}
 \pa^2d_\mu\bar{t}_\nu+m_{_T}^2d_\mu t_\nu=d_\mu \tilde{j}^T_\nu.
\end{equation}
Here number of derivatives in $j_{_T}^\mu$, $\bar{j}_{_T}^\mu$ and
$\tilde{j}_T^\mu$ are the same. Solution of equation(~\ref{6.2.10}) is
\begin{equation}\label{6.2.11}
  d_\mu\bar{t}_\nu(x)=\int d^4y\Delta^{(T)}(x-y)
  \tilde{j}^{T}_{\mu\nu}(y)
\end{equation}
where
\begin{eqnarray}\label{6.2.12}
&&\tilde{j}^{T}_{\mu\nu}(y)=d_\mu \tilde{j}^T_\nu(y) \nonumber \\
&&\Delta^{(T)}(x-y)=\int\frac{d^4p}{(2\pi)^4}\Delta^{(T)}(p)
 e^{ip\cdot(x-y)},\hspace{0.8in}\Delta^{(T)}(p)=\frac{-1}{p^2-m_{_T}^2}.
\end{eqnarray}
Comparing Eq.(~\ref{6.2.9}) with Eq.(~\ref{6.2.11}) we can find
that the Lorentz covariant term $d_\mu t^\mu$ is high order of momentum
expansion comparing with $d_\mu t_\nu$. Therefore the terms
$<(d_\mu t^\mu)^2>$ can be omitted in our calculation.

Applying above argument in vertices(~\ref{6.2.1}) and (~\ref{6.2.2}) we
have
\begin{eqnarray}\label{6.2.13}
{\cal L}_{(tt)}&=&\frac{1}{2}t_{\mu}^ad_{\nu}^{ac}d^{\nu,cb}t_{\mu}^b
   +t_{\mu}^a{\cal S}_{t}^{\mu\nu,ab}t_{\nu}^b
   +t_{\mu}^a{\cal A}_{t}^{\mu\nu,ab}t_{\nu}^b
   +\frac{m_t^2}{2}t_{\mu}^at^{\mu a} \nonumber \\
&=&-\frac{1}{2}t_{\mu}^a\na^{(t)\mu\nu}t_{\nu}^a
   +t_{\mu}^a{\cal H}_{t}^{\mu\nu,ab}t_{\nu}^b,
   \hspace{0.8in}(t=v,a),
\end{eqnarray}
where ${\cal S}_{t}^{\mu\nu,ab}$ is
symmetrical and ${\cal A}_{t}^{\mu\nu,ab}$ is antisymmetrical under
interchange of Lorentz index $\mu$ and $\nu$, or gauge group index
$a$ and $b$ respectively. Explicitly, they read
\begin{eqnarray}\label{6.2.14}
d_{\mu}^{ab}&=&\pa_\mu\delta^{ab}+\Gamma_{\mu}^{ab}         \\
{\cal S}_{\rm
   v}^{\mu\nu,ab}&=&\frac{\kappa^2}{8}
   <[\bar{\Delta}^{\mu},\lambda^a]
    [\bar{\Delta}^{\nu},\lambda^b]+[\bar{\Delta}^{\nu},\lambda^a]
    [\bar{\Delta}^{\mu},\lambda^b]>-\frac{\kappa^2}{4}\delta^{\mu\nu}
  <[\bar{\Delta}_{\rho},\lambda^a][\bar{\Delta}^{\rho},\lambda^a]> \\
{\cal A}_{\rm v}^{\mu\nu,ab}&=&
   \frac{3}{16}i<[\lambda^a,\lambda^b](\xi\bar{L}^{\mu\nu}\xi^{\dag}
    +\xi^{\dag}\bar{R}^{\mu\nu}\xi)>+(\frac{1}{8}+\frac{4\gamma}{3g^2})
   <[\lambda^a,\lambda^b][\bar{\Delta}^{\mu},\bar{\Delta}^{\nu}]>
       \nonumber \\
&&-\frac{\kappa^2}{8}
   <[\bar{\Delta}^{\mu},\lambda^a][\bar{\Delta}^{\nu},\lambda^b]
   -[\bar{\Delta}^{\mu},\lambda^a][\bar{\Delta}^{\nu},\lambda^b]>.
        \nonumber \\
{\cal S}_{a}^{\mu\nu,ab}&=&(\frac{1}{8\kappa^2}-\frac{4\gamma}{3g_{_A}^2})
   <[\bar{\Delta}^{\mu},\lambda^a][\bar{\Delta}^{\nu},\lambda^b]
   +[\bar{\Delta}^{\nu},\lambda^a][\bar{\Delta}^{\mu},\lambda^b]>
    \nonumber \\
&&-(\frac{1}{4\kappa^2}-\frac{8\gamma}{3g_{_A}^2})\delta^{\mu\nu}
    <[\bar{\Delta}_{\rho},\lambda^a][\bar{\Delta}^{\rho},\lambda^b]>
    -\frac{8\gamma}{3g_{_A}^2}\delta^{\mu\nu}
   <\bar{\Delta}_{\rho}\lambda^a\bar{\Delta}^{\rho}\lambda^b>
        \nonumber \\
&&-\frac{4\gamma}{3g_{_A}^2}<\{\lambda^a,\lambda^b\}
   \{\bar{\Delta}^{\mu},\bar{\Delta}^{\nu}\}>
   -\frac{2\theta}{g_{_A}^2}\delta^{\mu\nu}<\{\lambda^a,\lambda^b\}
    (\xi\chi^{\dag}\xi+\xi^{\dag}\chi\xi^{\dag})>, \\
{\cal A}_{a}^{\mu\nu,ab}&=&(\frac{1}{16}+\frac{1}{8\kappa^2}-
    \frac{2\gamma}{3g_{_A}^2})i<[\lambda^a,\lambda^b]
   (\xi\bar{L}^{\mu\nu}\xi^{\dag}+\xi^{\dag}\bar{R}^{\mu\nu}\xi)>
+\frac{1}{8}<[\lambda^a,\lambda^b][\bar{\Delta}^{\mu},\bar{\Delta}^{\nu}]>
        \nonumber \\
&&-(\frac{1}{8\kappa^2}-\frac{4\gamma}{3g_{_A}^2})
   <[\bar{\Delta}^{\mu},\lambda^a][\bar{\Delta}^{\nu},\lambda^b]
   -[\bar{\Delta}^{\mu},\lambda^a][\bar{\Delta}^{\nu},\lambda^b]>.
\end{eqnarray}
and
\begin{eqnarray}\label{6.2.15}
\na_{\mu\nu}^{(t)}&=&-\delta_{\mu\nu}(\pa^2+m_t^2) \\
{\cal H}_{t}^{\mu\nu,ab}&=&\frac{1}{2}\delta^{\mu\nu}
   (\{\pa^{\rho},\Gamma_{\rho}^{ab}\}
    +\Gamma_{\rho}^{ac}\Gamma^{\rho,cb})+{\cal S}_{t}^{\mu\nu,ab}
    +{\cal A}_{t}^{\mu\nu,ab}
\end{eqnarray}
The modified kinetic terms of vector mesons and axial-vector mesons
in Eqs.(~\ref{6.2.13}) and (~\ref{6.2.15}) yield their propagators in
momentum space as follows
\begin{equation}\label{6.2.16}
\Delta_{\mu\nu}^{(t)}(q)=\frac{1}{q^2-m_{_T}^2+i\ep}
\end{equation}
This propagator now have a good ultraviolet behavior and chiral symmetry
can keep well in follow calculation
\footnote{\small It is obvious that Eq.(~\ref{6.2.16}) is just leading
order contribution of Eq.(~\ref{6.2.5}) because of small momentum
transfer. However, we can not obtain Eq.(~\ref{6.2.16}) simply from
Eq.(~\ref{6.2.5}), since it will break chiral symmetry of effective
lagrangian generated by meson loops.}.
The adjoint operators of ${\cal H}_{j}^{\mu\nu,ab}$ can be obtained
from Eq.(~\ref{6.2.15}) as follow
\begin{equation}\label{6.2.17}
 {\cal H}_{t}^{*\mu\nu,ab}={\cal H}_{t}^{\nu\mu,ba}
\end{equation}

\begin{figure}\label{Vloop}[tp]
   \centering
   \includegraphics[width=5in]{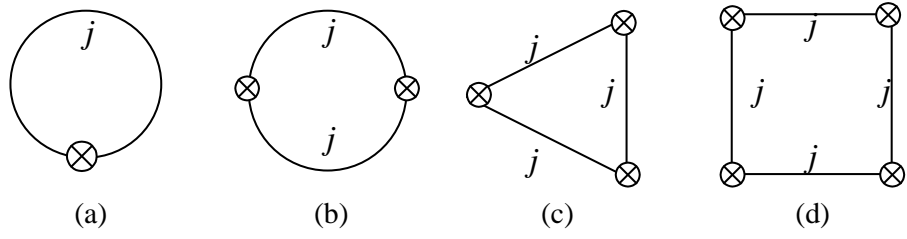}
\begin{minipage}{5in}
   \caption{Pure one-loop graphs of $1^-$ or $1^+$ mesons. Here
  ``$\otimes$" denotes classic external fields and $j$ denotes
   internal lines of $1^-$ or $1^+$ mesons.}
\end{minipage}
\end{figure}

Up to $O(p^4)$ there are four kinds of potential one-loop
graphs(fig.~\ref{Vloop}) which involve vector meson or axial-vector meson
internal lines only. From propagator of spin-1 mesons in
Eq.(~\ref{6.2.16}), it can be easily found that quadratic
divergence only come from ``tadpole'' graph(fig.~\ref{Vloop}-a)). In fact,
the effective lagrangian generated by ``tadpole'' graph is
$O(p^2)$, Therefore, it is nothing but to modify the free
parameters ``$f_0$'', ``$B_0$'' and axial-vector meson
mass-dependent parameter ``$\bar{m}_2$''. In addition, it is
determined by Eq.(~\ref{6.2.16}) that much higher order divergences
do not appear here. According to previous discussion,
the neglected higher order divergences is smaller than logarithmic
one due to small square of momentum transfer. Sequentially, we only
need to focus our attention on logarithmic divergence and finite
terms in our calculation.

Now substituting Eq.(~\ref{6.2.15})-(~\ref{6.2.17}) into general
formula(~\ref{8.2}) and performing loop-momentum integral explicitly, we
obtain
\begin{equation}\label{6.2.18}
{\cal L}_{one-loop}^{(tt)}=\frac{1}{(4\pi)^{D/2}}(\frac{\mu^2}{m_j^2})^2
\Gamma(2-\frac{D}{2})\{\frac{1}{6}\Gamma_{\mu\nu}^{ab}\Gamma^{\mu\nu,ba}
      -{\cal A}_{t\mu\nu}^{ab}{\cal A}_t^{\mu\nu,ba}
      +{\cal S}_{t\mu\nu}^{ab}{\cal S}_t^{\mu\nu,ba}\}
\end{equation}
Inserting Eq.(~\ref{6.1.10}) and (~\ref{6.2.14}) into (~\ref{6.2.18}) and
choosing $N_f=3$ we can obtain the effective lagrangian generated
by vector meson one-loop as follows
\begin{eqnarray}\label{6.2.19}
{\cal L}_{one-loop}^{(vv)}&=&
\frac{1}{(4\pi)^{D/2}}(\frac{\mu^2}{m_{_V}^2})^2\Gamma(2-\frac{D}{2})
     \{\frac{19}{32}
<\bar{L}_{\mu\nu}\bar{L}^{\mu\nu}+\bar{R}_{\mu\nu}\bar{R}^{\mu\nu}
+2\bar{L}_{\mu\nu}\bar{U}^{\dag}\bar{R}^{\mu\nu}\bar{U}>
  \nonumber \\
&&-\frac{19}{24}<\bar{V}_{\mu\nu}><\bar{V}^{\mu\nu}>
+(\frac{15\gamma}{2g^2}-\frac{1}{2})i
  <\bar{L}^{\mu\nu}D_{\mu}\bar{U}^{\dag}D_{\nu}\bar{U}
+\bar{R}^{\mu\nu}D_{\mu}\bar{U}D_{\nu}\bar{U}^{\dag}>
   \nonumber \\
&&+(\frac{1}{24}+\frac{\kappa^4}{24}-\frac{25\gamma^2}{9g^4})
(<D_{\mu}\bar{U}^{\dag}D^{\mu}\bar{U}>^2+2<D_{\mu}\bar{U}^{\dag}
  D_{\nu}\bar{U}><D^{\mu}\bar{U}^{\dag}D^{\nu}\bar{U}>)
      \nonumber \\
&&+(\frac{9}{64}\kappa^4-\frac{3}{8}+\frac{25\gamma^2}{g^4})
 <D_{\mu}\bar{U}^{\dag}D^{\mu}\bar{U}D_{\nu}\bar{U}^{\dag}
   D^{\nu}\bar{U}>\}
\end{eqnarray}
Similarly, the effective lagrangian generated by axial-vector meson
one-loop reads
\begin{eqnarray}\label{6.2.20}
{\cal L}_{one-loop}^{(aa)}&=&
\frac{1}{(4\pi)^{D/2}}(\frac{\mu^2}{m_{_A}^2})^2\Gamma(2-\frac{D}{2})
\{A_1<\bar{L}_{\mu\nu}\bar{L}^{\mu\nu}+\bar{R}_{\mu\nu}\bar{R}^{\mu\nu}
+2\bar{L}_{\mu\nu}\bar{U}^{\dag}\bar{R}^{\mu\nu}\bar{U}>
    \nonumber \\
&&-\frac{4}{3}A_1<\bar{V}_{\mu\nu}><\bar{V}^{\mu\nu}>
 +A_2i<\bar{L}^{\mu\nu}D_{\mu}\bar{U}^{\dag}D_{\nu}\bar{U}
   +\bar{R}^{\mu\nu}D_{\mu}\bar{U}D_{\nu}\bar{U}^{\dag}>
    \nonumber \\
&&+A_3<D_{\mu}\bar{U}^{\dag}D^{\mu}\bar{U}>^2
+A_4<D_{\mu}\bar{U}^{\dag}D_{\nu}\bar{U}>
   <D^{\mu}\bar{U}^{\dag}D^{\nu}\bar{U}> \nonumber \\
&&+A_5<D_{\mu}\bar{U}^{\dag}D^{\mu}\bar{U}D_{\nu}\bar{U}^{\dag}
   D^{\nu}\bar{U}>+A_6<D_{\mu}\bar{U}^{\dag}D^{\mu}\bar{U}>
  <\chi\bar{U}^{\dag}+\chi^{\dag}\bar{U}> \nonumber \\
&&+A_7<D_{\mu}\bar{U}^{\dag}D^{\mu}\bar{U}
   (\bar{U}^{\dag}\chi+\chi^{\dag}\bar{U})>
  +\frac{1408\theta^2}{9g_{_A}^4}
    <\chi\bar{U}^{\dag}+\chi^{\dag}\bar{U}>^2 \nonumber \\
&&+\frac{640\theta^2}{3g_{_A}^4}<\chi\bar{U}^{\dag}\chi\bar{U}^{\dag}
    +\chi^{\dag}\bar{U}\chi^{\dag}\bar{U}>\}
\end{eqnarray}
where
\begin{eqnarray}\label{6.2.21}
A_1&=&\frac{27}{32\kappa^4}(1-\frac{40\gamma}{9g^2})^2-\frac{1}{4}
   \nonumber \\
A_2&=&\frac{9\gamma}{2\kappa^4g^2}(1-\frac{40\gamma}{9g^2})-\frac{1}{2}
   \nonumber \\
A_3&=&\frac{1}{16}+\frac{1}{\kappa^4}(\frac{15}{128}-
   \frac{11\gamma}{6g^2}+\frac{1261\gamma^2}{162 g^4}) \nonumber \\
A_4&=&\frac{1}{8}+\frac{1}{\kappa^4}(\frac{15}{64}-\frac{4\gamma}{3g^2}
    +\frac{173 r^2}{81 g^4})  \\
A_5&=&-\frac{3}{8}+\frac{1}{\kappa^4}(\frac{3}{64}-\frac{2\gamma}{g^2}
    +\frac{563\gamma^2}{27g^4}) \nonumber \\
A_6&=&\frac{2\theta}{\kappa^4g^2}(\frac{448\gamma}{9g^2}-3) \nonumber \\
A_7&=&\frac{6\theta}{\kappa^4g^2}(\frac{256\gamma}{9g^2}-3) \nonumber
\end{eqnarray}

\subsection{Mixing loops of vector and axial-vector mesons}

\begin{figure}\label{Mloop}[tp]
   \centering
   \includegraphics[width=4.5in]{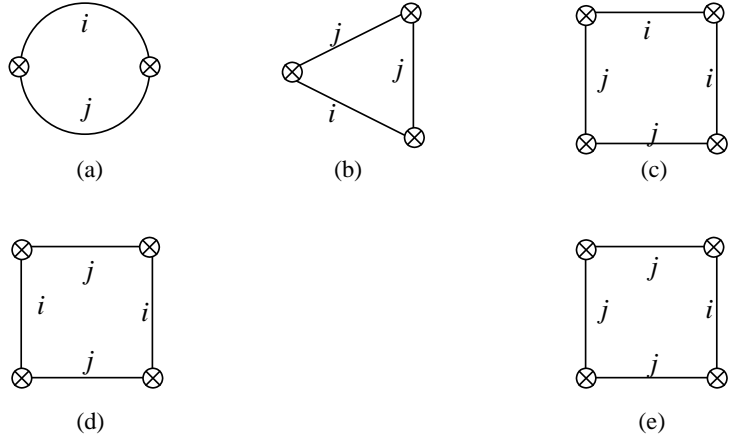}
\begin{minipage}{5in}
   \caption{Mixing one-loop graphs of $1^-$ and $1^+$ mesons. Here
  ``$\otimes$" denotes classic external fields. $i,\;j$ denote
  internal lines of $1^-$ and $1^+$ mesons and $i\neq j$.}
\end{minipage}
\end{figure}

Due to discussion in above subsection, the vertices include a quantum
field of vector mesons and a quantum field of axial-vector mesons read
from lagrangian(~\ref{3.1.18}) as follow
\begin{eqnarray}\label{6.3.1}
  {\cal L}_{(va)}&=&v_{\mu}^a\Omega^{\mu\nu,ab}a_{\nu}^a \nonumber \\
\Omega^{\mu\nu,ab}&=&B_1\delta^{\mu\nu}
  \bar{\Delta}_{\rho}^{ac}d^{\rho,cb}
  +\Sigma^{\mu\nu,ab} \nonumber \\
\Sigma^{\mu\nu,ab}&=&-\frac{4\gamma}{3\kappa g^2}
    (d^{\nu}\bar{\Delta}^{\mu})^{ab}-
   \frac{\kappa}{2}i<[\lambda^a,\lambda^b]
  (\xi\bar{L}^{\mu\nu}\xi^{\dag}-\xi^{\dag}\bar{R}^{\mu\nu}\xi)>
    \nonumber \\
  &&+B_2<[\lambda^a,\lambda^b]
    (\xi\chi^{\dag}\xi-\xi^{\dag}\chi\xi^{\dag})>.
\end{eqnarray}
where
\begin{eqnarray}\label{6.3.2}
B_1&=&\frac{1}{\kappa}(1-\frac{4\gamma}{g^2}) \hspace{0.8in}
B_2=\frac{1-c}{8\kappa}(1-\frac{16\gamma}{3g^2})+
  \frac{2\theta_3}{\kappa g^2} \nonumber \\
\bar{\Delta}_{\mu}^{ab}&=&<\bar{\Delta}_{\mu}[\lambda^a,\lambda^b]>,
   \hspace{0.8in}
(d_{\mu}\bar{\Delta}_{\nu})^{ab}=
  <d_\mu\bar{\Delta}_{\nu}[\lambda^a,\lambda^b]>.
\end{eqnarray}
The adjoint operator of $\Omega^{\mu\nu,ab}$ is
\begin{equation}\label{6.3.201}
\Omega^{*\mu\nu,ab}=B_1\delta^{\mu\nu}d^{\rho,bc}\bar{\Delta}_{\rho}^{ca}
  +\Sigma^{\mu\nu,ab}
\end{equation}
To $O(p^4)$, the vertices (~\ref{6.2.14})-(6.31) and (~\ref{6.3.2})
generate the following kinds of mixing one-loop graphs(fig.~\ref{Mloop})

Inserting vertices (~\ref{6.2.14})-(6.31), (~\ref{6.3.2}) and
propagator(~\ref{6.2.16}) into general formula (~\ref{8.3}), we can
perform the integral of loop-momentum explicitly. The effective
lagrangian generated by one-loop in fig.~\ref{Mloop}. is
\begin{eqnarray}\label{6.3.3}
{\cal L}_{one-loop}^{(va)}&=&\frac{1}{(4\pi)^{D/2}}\Gamma(2-\frac{D}{2})
 \int_{0}^{1}dt(\frac{\mu^2}{f(t)})^{\ep/2}
 \{\frac{1}{2}\Sigma_{\mu\nu}^{ab}\Sigma^{\mu\nu,ab}
  +tB_1\delta^{\mu\nu}\Sigma_{\mu\nu}^{ab}
  (d_\rho\bar{\Delta}^\rho)^{ba} \nonumber \\
&&-B_1^2\delta_{\mu\nu}\bar{\Delta}_{\rho}^{ab}\bar{\Delta}_{\rho}^{ab}
   [t{\cal S}_V^{\mu\nu,ca}+(1-t){\cal S}_A^{\mu\nu,ca}]
  +t(1-t)D\frac{4\gamma^2}{9}(d_\mu\bar{\Delta}_\nu)^{ab}
    (d_\mu\bar{\Delta}_\nu)^{ba} \nonumber \\
&&+(\frac{1}{2}-t+t^2)[(d_\mu\bar{\Delta}^\mu)^{ab}
  (d_\nu\bar{\Delta}^\nu)^{ba}-(d_\mu\bar{\Delta}_\nu)^{ab}
  (d^\nu\bar{\Delta}^\mu)^{ba}] \nonumber \\
&&+\frac{1}{4}t(1-t)B_1^4
  (\delta_{\mu\nu}\delta_{\rho\sigma}
   +\delta_{\mu\rho}\delta_{\nu\sigma}
   +\delta_{\mu\sigma}\delta_{\nu\rho})\bar{\Delta}^{\mu,ab}
    \bar{\Delta}^{\nu,bc}\bar{\Delta}^{\rho,cd}\bar{\Delta}^{\sigma,da},
\end{eqnarray}
where
\begin{equation}\label{6.3.4}
f(t)=tm_{_V}^2+(1-t)m_{_A}^2.
\end{equation}
It should be noted that, in Eq.(~\ref{6.3.3}), the local
$SU(3)_L\times SU(3)_R$ chiral symmetry is still kept even though we have
performed complicated calculation. The explicit form of
Eq.(~\ref{6.3.3}) with three flavors is
\begin{eqnarray}\label{6.3.5}
{\cal L}_{one-loop}^{(va)}&=&\frac{1}{(4\pi)^{D/2}}
 (\frac{\mu^2}{m^2})^{\ep/2}\Gamma(2-\frac{D}{2})\int_{0}^{1}
   dt(\frac{m^2}{f(t)})^{\ep/2} \nonumber\times \\
&&\{C_1(t)<\bar{L}_{\mu\nu}\bar{L}^{\mu\nu}
  +\bar{R}_{\mu\nu}\bar{R}^{\mu\nu}-2\bar{L}_{\mu\nu}\bar{U}^{\dag}
   \bar{R}^{\mu\nu}\bar{U}>-
  \frac{4}{3}C_1(t)<\bar{A}_{\mu\nu}><\bar{A}^{\mu\nu}> \nonumber \\
&&+C_2(t)i<\bar{L}^{\mu\nu}D_{\mu}\bar{U}^{\dag}D_{\nu}\bar{U}
   +\bar{R}^{\mu\nu}D_{\mu}\bar{U}D_{\nu}\bar{U}^{\dag}>
  +C_3(t)<D_{\mu}\bar{U}^{\dag}D^{\mu}\bar{U}>^2 \nonumber \\
&&+C_4(t)<D_{\mu}\bar{U}^{\dag}D_{\nu}\bar{U}>
   <D^{\mu}\bar{U}^{\dag}D^{\nu}\bar{U}>
 +C_5(t)<D_{\mu}\bar{U}^{\dag}D^{\mu}\bar{U}D_{\nu}\bar{U}^{\dag}
   D^{\nu}\bar{U}> \nonumber \\
&&+C_6(t)<D_{\mu}\bar{U}^{\dag}D^{\mu}\bar{U}>
  <\chi\bar{U}^{\dag}+\chi^{\dag}\bar{U}>
  +3C_6(t)<D_{\mu}\bar{U}^{\dag}D^{\mu}\bar{U}
    (\bar{U}^{\dag}\chi+\chi^{\dag}\bar{U})> \nonumber \\
&&+16B_2^2<\chi\bar{U}^{\dag}-\chi^{\dag}\bar{U}>^2
  +C_7(t)<\chi\bar{U}^{\dag}\chi\bar{U}^{\dag}
    +\chi^{\dag}\bar{U}\chi^{\dag}\bar{U}> \}
\end{eqnarray}

where
\begin{eqnarray}\label{6.3.6}
C_1(t)&=&3(1-\frac{3\gamma}{g^2}+\frac{8\gamma^2}{9\kappa^2g^4})
   -3B_1^2(t-t^2) \nonumber \\
C_2(t)&=&\frac{8\gamma^2}{3\kappa^2g^4}+\frac{3}{2}B_1^2(1-2t)^2
    \nonumber \\
C_3(t)&=&-\frac{2\gamma^2}{3\kappa^2g^4}-B_1^2(\frac{3}{8}+
  \frac{3}{4\kappa^2}-\frac{32\gamma}{3g_{_A}^2})+
 tB_1^2(\frac{3}{4\kappa^2}+\frac{4\gamma}{g^2}-\frac{32\gamma}{3g_{_A}^2}
  +\frac{9}{4}B_1^2) \nonumber \\
&&-\frac{3}{4}t^2B_1^2(2-3B_1^2) \nonumber \\
C_4(t)&=&-\frac{4\gamma^2}{3\kappa^2g^4}-B_1^2(\frac{3}{4}+
  \frac{3}{2\kappa^2}-\frac{16\gamma}{3g_{_A}^2})+
  tB_1^2(\frac{3}{2\kappa^2}+\frac{8\gamma}{g^2}-\frac{64\gamma}{3g_{_A}^2}
  +\frac{9}{2}B_1^2) \nonumber \\
&&-\frac{3}{2}t^2B_1^2(2-3B_1^2) \nonumber \\
C_5(t)&=&\frac{4\gamma^2}{\kappa^2g^4}-3B_1^2(\frac{3}{4}+
  \frac{3}{4\kappa^2}-\frac{32\gamma}{3g_{_A}^2})-3
  tB_1^2(9+\frac{9}{4}\kappa^2-\frac{3}{4\kappa^2}
  +\frac{32\gamma}{3g_{_A}^2})+9t^2B_1^2 \nonumber \\
C_6(t)&=&(1-t)\frac{32\theta}{g_{_A}^2} \nonumber \\
C_7(t)&=&-12[2B_2+\frac{\gamma}{3\kappa g^2}(1-c)]^2
  +6tB_1(4B_2-\frac{4\gamma}{3\kappa g^2}+\frac{1-c}{4})
  -\frac{3}{2}t^2B_1^2(1-c)
\end{eqnarray}

\subsection{Mixing loops of pseudoscalar and spin-1 mesons}

The definition of spin-1 meson field (~\ref{2.10}) and the field shift
(~\ref{3.1.12}) lead to there are no coupling between spin-1 mesons and
pseudoscalar fields in ${\cal L}_2^{(0)}$. It make calculation on meson
one-loop be very easily, especially, for calculation on mixing loops of
$0^-$ and $1^\pm$ mesons.

Since all vertices combining with one spin-1 meson internal line
and one pseudoscalar meson internal line come from ${\cal
L}_4^{(0)}$, every vertices is composed of momentum factor
$4-n_{_I}^{V.A}$, where $n_{_I}^{V.A}$ is number of spin-1 meson
internal lines attaching to this vertex. Our goal in this paper is
to calculate effective lagrangian generated by meson one-loop
graphs and expand it up to four powers of external derivatives.
Then in a Feynman diagram, momentum powers carried by all internal
lines are $4N_{_V}-2N_{_I}^{V.A}-4,$ so that integral of
loop-momentum must be proportional to
\begin{equation}\label{6.4.1}
\int\frac{d^4p}{(2\pi)^{4}}\frac{q^{4N_{_V}-2N_{_I}^{V.A}-4}}
 {(q^2+i\ep)^{N_{_I}^\phi}(q^2-m_{_T}^2+i\ep)^{N_{_I}^{V.A}}}
\end{equation}
where $N_{_V}$ is number of vertices, $N_{_I}^{V.A}$ is number of all
spin-1 meson internal lines and $N_{_I}^\phi$ is number of all $0^-$
meson internal lines in the Feynman diagram. Due to relationship for case
of one-loop
$$N_{_I}^{V.A}=N_{_L}+N_{_V}-N_{_I}^\phi-1=N_{_V}-N_{_I}^\phi,$$
loop-momentum integral(~\ref{6.4.1}) is equal to
\begin{equation}\label{6.4.2}
\int\frac{d^4p}{(2\pi)^{4}}\frac{q^{2(N_{_V}-2)}}
 {(q^2-m_{_T}^2+i\ep)^{N_{_V}-N_{_I}^\phi}}.
\end{equation}

From the following two aspects we
will illustrate the contribution proportioning loop-momentum integral can
be omitted: (a) Because $N_{_I}^\phi\geq 1$ in this kind of mixing loops.
the divergences in Eq.(~\ref{6.4.1}) is quadratic or higher order.
According to discussion in section 6.2, it is smaller than logarithmic
divergence due to small square of momentum transfer. (b) Since external
momentum is $O(p^4)$, this part contribution is equivalent to "tadpole"
contribution of spin-1 mesons which is generated by ${\cal L}_6$. We have
point out that in this formalism these interaction with higher order
derivatives is not defined. Therefore we have to ignore it here.
Sequentially, all contributions from mixing loops of pseudoscalar and
spin-1 mesons are omitted in this paper.

\subsection{Factorization of Divergences}

ChQM is a non-renormalizable truncated field theory, since there are no
enough counterterms to cancel divergences from meson loops. So that the
divergences from meson loops have to be parameterized(or introduce
ultra-violet cut-off).

Besides ultra-violet divergences, there are infrared divergences in pure
pseudoscalar meson loops. Therefore, we have defined a independent
parameter in sect. 6.1 to put the ultra-violet and the infrared
divergences into a ``bag'' simultaneously
\begin{equation}\label{6.5.1}
g_1=\lim_{\mu_p\rightarrow 0}
  \frac{1}{(4\pi)^{D/2}}(\frac{\mu^2}{\mu_p^2})^{\ep/2}
  \Gamma(2-\frac{D}{2})
\end{equation}

In the case of presence of spin-1 mesons, since momentum transfer is
smaller than spin-1 meson masses, we like to introduce a
ultra-violet cut-off $\Lambda_{_M}$ to ``renormalization'' all
divergences from spin-1 meson loops approximately.
\footnote{\small In fact, similar to sect. 3, it should be different that
the cut-offs correspond to pseudoscalar interaction in very low energy and
to spin-1 mesons coupling to pseudoscalar in energy scale $\mu\sim
m_\rho$. According to sect. 6.1 and 6.2, we have evaluated that both of
cut-offs are smaller than $700$MeV. In this energy region, the difference
is not large, so that we take a cut-off here.}

Explicitly, we define
\begin{eqnarray}\label{6.5.2}
 g_2&=&\frac{1}{(4\pi)^{D/2}}(\frac{\mu^2}{m_{_V}^2})^{\ep/2}
  \Gamma(2-\frac{D}{2})\simeq\frac{1}{16\pi^2}
  [\ln{(1+\frac{\Lambda_{_M}^2}{m_{_V}^2})}-\frac{\Lambda_{_M}^2}{\Lambda^2+m_{_V}^2}]
   \\
 g_3&=&\frac{1}{(4\pi)^{D/2}}(\frac{\mu^2}{m_{_A}^2})^{\ep/2}
  \Gamma(2-\frac{D}{2})\simeq\frac{1}{16\pi^2}
  [\ln{(1+\frac{\Lambda_{_M}^2}{m_{_A}^2})}-\frac{\Lambda_{_M}^2}{\Lambda^2+m_{_A}^2}]
\end{eqnarray}
and
\begin{eqnarray}\label{6.5.3}
M&=&\frac{1}{(4\pi)^{D/2}}\Gamma(2-\frac{D}{2})\int_0^1 dt
  (\frac{\mu^2}{f(t)})^{\ep/2}\simeq
    \frac{1}{16\pi^2}\int_0^1 dt
   [\ln{(1+\frac{\Lambda_{_M}^2}{f(t)})}-
    \frac{\Lambda_{_M}^2}{\Lambda_{_M}^2+f(t)}]
    \nonumber \\
 &=&\frac{1}{16\pi^2}\{\ln{(1+\frac{\Lambda_{_M}^2}{m_{_V}^2})}+
   \frac{m_{_A}^2}{m_{_A}^2-m_{_V}^2}(\ln{\frac{m_{_V}^2}{m_{_A}^2}}+
   \ln{(\frac{\Lambda_{_M}^2+m_{_A}^2}{\Lambda_{_M}^2+m_{_V}^2})})\}
     \\
P&=&\frac{1}{(4\pi)^{D/2}}\Gamma(2-\frac{D}{2})\int_0^1 dt\C t
  (\frac{\mu^2}{f(t)})^{\ep/2}\simeq
  \frac{1}{16\pi^2}\int_0^1 dt\C t
  [\ln{(1+\frac{\Lambda_{_M}^2}{f(t)})}-
  \frac{\Lambda_{_M}^2}{\Lambda_{_M}^2+f(t)}]
      \nonumber \\
&=&\frac{1}{32\pi^2}\{\ln{(1+\frac{\Lambda_{_M}^2}{m_{_V}^2})}+
    \frac{m_{_A}^4}{(m_{_A}^2-m_{_V}^2)^2}\ln{\frac{m_{_V}^2}{m_{_A}^2}}+
    \frac{\Lambda_{_M}^2}{m_{_A}^2-m_{_V}^2} \nonumber \\
 &&+\frac{m_{_A}^4-\Lambda_{_M}^4}{(m_{_A}^2-m_{_V}^2)^2}
   \ln{(\frac{\Lambda_{_M}^2+m_{_A}^2}{\Lambda_{_M}^2+m_{_V}^2})}\} \\
Q&=&\frac{1}{(4\pi)^{D/2}}\Gamma(2-\frac{D}{2})\int_0^1 dt\C t^2
  (\frac{\mu^2}{f(t)})^{\ep/2}\simeq
  \frac{1}{16\pi^2}\int_0^1 dt\C t^2[\ln{(1+\frac{\Lambda_{_M}^2}{f(t)})}
  -\frac{\Lambda_{_M}^2}{\Lambda_{_M}^2+f(t)}]
     \nonumber \\
&=&\frac{1}{48\pi^2}\{\ln{(1+\frac{\Lambda_{_M}^2}{m_{_V}^2})}+
     \frac{m_{_A}^6}{(m_{_A}^2-m_{_V}^2)^3}\ln{\frac{m_{_V}^2}{m_{_A}^2}}+
   \frac{\Lambda_{_M}^2}{m_{_A}^2-m_{_V}^2} \nonumber \\
 &&+\frac{\Lambda_{_M}^2(2\Lambda_{_M}^2+m_{_A}^2)}{(m_{_A}^2-m_{_V}^2)^2}
   -\frac{2\Lambda_{_M}^6+3\Lambda_{_M}^4 m_{_A}^2-m_{_A}^6}{(m_{_A}^2-m_{_V}^2)^3}
     \ln{(\frac{\Lambda_{_M}^2+m_{_A}^2}{\Lambda_{_M}^2+m_{_V}^2})}\}
\end{eqnarray}

Substituting Eqs.(~\ref{6.5.1})-(~~\ref{6.5.3}) into
effective lagrangian generated by meson one-loop, all divergences are
canceled.

\section{Physics Beyond Leading Order of $1/N_c$}
\setcounter{equation}{0}

In this section we like to summarize the results in sect. 3, sect. 4 and
sect.6 and discuss some physics up to the next to leading order of
$1/N_c$ expansion.

\subsection{Effective Lagrangian}

It should be noted that the correction of meson one-loop to the term
$<\bar{L}_{\mu\nu}\bar{L}^{\mu\nu}+\bar{R}_{\mu\nu}\bar{R}^{\mu\nu}>$ is
nothing since it can be absorbed by $g_\phi$ and $g$. Then
up to the next to leading order of $1/N_c$ expansion, the effective
lagrangian with four derivatives can be written
\begin{eqnarray}\label{7.1}
{\cal L}_{4}&=&{\cal L}_{one-loop}^{(\varphi\varphi)}
  -\frac{\lambda_r(\mu)}{16}
  <L_{\mu\nu}L^{\mu\nu}+R_{\mu\nu}R^{\mu\nu}>
  -\frac{f_1}{8}<V_{\mu\nu}><V^{\mu\nu}> \nonumber \\
&&-\frac{f_2}{8}<A_{\mu\nu}><A^{\mu\nu}>
   +R_1<L_{\mu\nu}U^{\dag}R^{\mu\nu}U> \nonumber \\
&&-iR_2<D_{\mu}UD_{\nu}U^{\dag}R^{\mu\nu}+D_{\mu}U^{\dag}
   D_{\nu}UL^{\mu\nu}> +R_3<D_{\mu}UD^{\mu}U^{\dag}>^2 \nonumber \\
&&+R_4<D_{\mu}UD_{\nu}U^{\dag}><D^{\mu}UD^{\nu}U^{\dag}>
  +R_5<D_{\mu}UD^{\mu}U^{\dag}><D_{\nu}UD^{\nu}U^{\dag}> \nonumber \\
&&+R_6<D_{\mu}UD^{\mu}U^{\dag}><\chi U^{\dag}+U\chi^{\dag}>
  +R_7<D_{\mu}UD^{\mu}U^{\dag}(\chi U^{\dag}+U\chi^{\dag} U)>
    \nonumber \\
&&+R_8<\chi U^{\dag}+U\chi^{\dag}>^2+R_9<\chi U^{\dag}-U\chi^{\dag}>^2
  +R_{10}<\chi U^{\dag}\chi U^{\dag}+\chi^{\dag}U\chi^{\dag}U>,
\end{eqnarray}
where
\begin{eqnarray}\label{7.101}
f_1&=&8(\frac{19}{24}g_2+\frac{4}{3}A_1g_3), \nonumber \\
f_2&=&32(1-\frac{3\gamma}{g^2}+\frac{8\gamma^2}{9\kappa^2g^4})M
   -32B_1^2(P-Q),
\end{eqnarray}
and $\lambda_r(\mu)$, $L_{\mu\nu}$, $R_{\mu\nu}$, $D_\mu U$ and $D_\mu
U^{\dag}$ are defined in Eqs.(~\ref{3.1.111}), (~\ref{3.1.19}) and
(~\ref{3.1.20}). According to sect. 3.1, the constants $R_i(i=1,2,...,11)$
contain three different classes of contributions, i.e.,
$$R_i=R_i^{(0)}+R_i^{(g)}+R_i^{(l)}.$$
The contribution from quark loops, $R_i^{(0)}$, can be obtained
from ${\cal L}_4^{(0)}$,
\begin{eqnarray}\label{7.2}
R_1^{(0)}&=&-\frac{\gamma}{6}, \hspace{1in}
R_2^{(0)}=\frac{\gamma}{3}, \nonumber \\
R_3^{(0)}&=&\frac{1}{2}R_4^{(0)}=-\frac{1}{4}R_5^{(0)}=\frac{\gamma}{24},
   \nonumber \\
R_7^{(0)}&=&\theta_1, \hspace{1.2in}
R_{10}^{(0)}=\theta_2, \nonumber \\
R_6^{(0)}&=&R_8{(0)}=R_9^{(0)}=0.
\end{eqnarray}
The contribution from quark-gluon coupling, $R_i^{(g)}$, read from
${\cal L}_{\rm I}^{(g)}$ as follow
\begin{eqnarray}\label{7.3}
R_1^{(g)}&=&-\frac{k}{40}, \hspace{1.in}
R_5^{(g)}=-\frac{k}{40},
   \nonumber \\
R_7^{(g)}&=&-\frac{km}{8B_0}, \hspace{0.9in}
R_{10}^{(g)}=\frac{km}{8B_0}(1-c-\frac{2m}{B_0})-\frac{k}{160}(1-c)^2
         \nonumber \\
R_2^{(g)}&=&R_3^{(g)}=R_4^{(g)}=R_6^{(g)}=R_8{(0)}=R_9^{(0)}=0.
\end{eqnarray}
$R_i^{(l)}$ generated by meson one-loop effects read explicitly
\begin{eqnarray}\label{7.4}
R_1^{(l)}&=&\frac{3}{8}(f_1-f_2), \nonumber \\
R_2^{(l)}&=&-(\frac{15\gamma}{2g^2}-\frac{1}{2})g_2-A_2g_3
   -\frac{8\gamma^2}{3\kappa^2g^4}M-\frac{3}{2}B_1^2(M-4P+4Q),
      \nonumber \\
R_3^{(l)}&=&(\frac{1}{24}+\frac{\kappa^4}{24}-\frac{25\gamma^2}{9g^4})g_2
   +A_3g_3-[\frac{2\gamma^2}{3\kappa^2g^4}+B_1^2(\frac{3}{8}+
  \frac{3}{4\kappa^2}-\frac{32\gamma}{3g_{_A}^2})]M \nonumber \\
&&+PB_1^2(\frac{3}{4\kappa^2}+\frac{4\gamma}{g^2}-\frac{32\gamma}{3g_{_A}^2}
  +\frac{9}{4}B_1^2)-\frac{3}{4}QB_1^2(2-3B_1^2), \nonumber \\
R_4^{(l)}&=&(\frac{1}{12}+\frac{\kappa^4}{12}-\frac{50\gamma^2}{9g^4})g_2
  +A_4g_3-[\frac{4\gamma^2}{3\kappa^2g^4}+B_1^2(\frac{3}{4}+
  \frac{3}{2\kappa^2}-\frac{16\gamma}{3g_{_A}^2})]M \nonumber \\
&&+PB_1^2(\frac{3}{2\kappa^2}+\frac{8\gamma}{g^2}-\frac{64\gamma}{3g_{_A}^2}
  +\frac{9}{2}B_1^2)-\frac{3}{2}QB_1^2(2-3B_1^2), \nonumber \\
R_5^{(l)}&=&(\frac{9}{64}\kappa^4-\frac{3}{8}+\frac{25\gamma^2}{g^4})g_2
  +A_5g_3+[\frac{4\gamma^2}{\kappa^2g^4}-3B_1^2(\frac{3}{4}+
  \frac{3}{4\kappa^2}-\frac{32\gamma}{3g_{_A}^2})]M \nonumber \\
&&-3PB_1^2(9+\frac{9}{4}\kappa^2-\frac{3}{4\kappa^2}
  +\frac{32\gamma}{3g_{_A}^2})+9QB_1^2, \nonumber \\
R_6^{(l)}&=&A_6g_3+(M-P)\frac{32\theta}{g_{_A}^2}, \nonumber \\
R_7^{(l)}&=&A_7g_3+3(M-P)\frac{32\theta}{g_{_A}^2}, \nonumber \\
R_8^{(l)}&=&\frac{1408\theta^2}{9g_{_A}^2}g_3, \nonumber \\
R_9^{(l)}&=&16MB_2^2, \nonumber \\
R_{10}^{(l)}&=&\frac{640\theta^2}{3g_{_A}^2}g_3
  -12M[2B_2+\frac{\gamma}{3\kappa g^2}(1-c)]^2
  +6PB_1(4B_2-\frac{4\gamma}{3\kappa g^2}+\frac{1-c}{4}) \nonumber \\
 &&-\frac{3}{2}QB_1^2(1-c).
\end{eqnarray}

\subsection{Low energy limit}

The low energy limit of ChQM can be obtained via integrating out
the degrees of freedom of spin-1 meson resonances. It means that,
at very low energy, the dynamics of spin-1 meson resonances are
replaced by pseudoscalar meson. In Ref.\cite{WY98}, the authors
have employed this method to a similar model. We have shown in
sect. 3.2 that ${\cal L}^{\phi}$ is just low energy limit of this
model. Since there are no spin-1 mesons coupling to pseudoscalar
field which make the classical solutions of spin-1 mesons be
$O(p^3)$ at very low energy. It indicates that physical vector and
axial-vector meson fields defined by non-linear realization of G
and the field shift (~\ref{3.1.12}) do not contribute to low energy
coupling constants $L_i$ by virtual spin-1 field exchange and
virtual spin-1 field vertices contributions. It is different from
the discussion in Ref.\cite{WY98} and Ref.\cite{Ecker89}. We must
point out that these contributions from virtual spin-1 meson
resonances are model-dependent, which depend on definition of
physical spin-1 meson resonances. However, the physical results
should not be changed by the definition. For Example, as
Ref.\cite{Li95a} the definitions spin-1 meson fields(${\cal V}_\mu,\;
{\cal A}_\mu$) are
$$L_\mu={\cal V}_\mu-{\cal A}_\mu,\hspace{0.6in}
  R_\mu={\cal V}_\mu+{\cal A}_\mu$$
and diagonalization of ${\cal A}_\mu-\pa_\mu\Phi$ mixing is
$${\cal A}_\mu\longrightarrow {\cal A}_\mu-c\pa_\mu\Phi.$$
Recalling the definition of spin-1 meson fields in this paper
$$L_\mu=\xi^{\dag}(V_\mu-A_\mu)\xi,\hspace{0.6in}
  R_\mu=\xi(V_\mu+A_\mu)\xi^{\dag},$$
the measures of integral for spin-1 meson fields are the same for two
different definitions. Therefore, we obtain the same generating functional
of Green's function from two different definitions
\begin{eqnarray}
 e^{iW[v,a,s,p;U]}&=&\frac{1}{N}\int{\cal D}V_\mu{\cal D}A_\mu
  \exp{\{iI_{\rm eff}[v,a,s,p;V_\mu,A_\mu,U]\}} \nonumber \\
  &=&\frac{1}{N}\int{\cal D}{\cal V}_\mu{\cal D}{\cal A}_\mu
  \exp{\{iI_{\rm eff}[v,a,s,p;{\cal V}_\mu,{\cal A}_\mu,U]\}}.
   \nonumber
\end{eqnarray}
Of course, the definition in this paper has attractive symmetrical
properties and is very convenient for our calculation.

The explicit expression of the coupling constants $L_i$(besides of $L_7$)
read
\begin{eqnarray}\label{7.5}
L_1&=&\frac{g_\phi^2}{32}c^2(1-\frac{c}{2})^2+\frac{R_2}{2}c(1-\frac{c}{2})
      (1-c)^2+R_3(1-c)^4+\frac{3}{64}g_1, \nonumber \\
L_2&=&\frac{g_\phi^2}{16}c^2(1-\frac{c}{2})^2+R_2c(1-\frac{c}{2})(1-c)^2
       +R_4(1-c)^4+\frac{3}{32}g_1, \nonumber \\
L_3&=&-\frac{3}{16}g_\phi^2c^2(1-\frac{c}{2})^2-3R_2c(1-\frac{c}{2})(1-c)^2
       +R_5(1-c)^4, \nonumber \\
L_4&=&R_6(1-c)^2+\frac{g_1}{16}, \nonumber \\
L_5&=&R_7(1-c)^2+\frac{3}{16}g_1, \\
L_6&=&R_8+\frac{11}{288}g_1, \nonumber \\
L_8&=&R_{10}+\frac{5}{96}g_1, \nonumber \\
L_9&=&\frac{g_\phi^2}{8}c(1-\frac{c}{2})+R_2(1-c)^2+\frac{g_1}{8},
   \nonumber \\
L_{10}&=&-\frac{g_\phi^2}{8}c(1-\frac{c}{2})+R_1(1-c)^2-\frac{g_1}{8}.
\end{eqnarray}

The constants $L_7$ has been known to get dominant contribution from
$\eta_0$\cite{GL85a} and this contribution is suppressed by $1/N_c$ too.
$\eta_0$ participate the dynamics in ChQM via
$\Phi(x)\rightarrow \Phi(x)+\Phi_0(x)(\Phi_0(x)=\frac{1}{\sqrt{3}}\eta_0)$
and $U(1)$ axial anomaly of QCD,
${\cal L}_\chi \rightarrow {\cal L}_\chi-\frac{1}{2}\tau <\Phi_0>^2$,
where
\begin{equation}
  \tau \propto <0|(\alpha_s/\pi)G_{\mu\nu}^a\td{G}^{a\mu\nu}|0>
        \hspace{0.7in}
   (\td{G}^{a\mu\nu}\equiv \epsilon^{\mu\nu\alpha\beta}G_{\alpha\beta})
\end{equation}
relate to gluon anomaly and is a free parameter here. The equation
of motion of $\eta_0$ is
\begin{equation}
  (\pa^2+m_{\eta_0}^2(\tau))\eta_0=\frac{\sqrt{6}}{8}i
     <\chi^{\dag}U-U^{\dag}\chi>.
\end{equation}
The lowest order solution of above equation is $O(p^2)$. So that $L_7^{g}$
can be obtained simply by $\eta_0$ exchange via integrating over freedom
of $\eta_0$
\begin{equation}
  L_7^{g}=-\frac{f_\pi^2}{128m_{\eta^{\prime}}^2},
\end{equation}
where we have ignored the $\eta-\eta^{\prime}$ mixing so that we can input
$m_{\eta_0}\simeq m_{\eta^\prime}$. The total value of $L_7$ is
\begin{equation}
  L_7=L_7^{g}+R_9.
\end{equation}
The latter comes from the mixing loops of vector and axial-vector mesons.

Form effective lagrangian(~\ref{7.1}), we obtain decay width of
$\phi\rightarrow\pi^+\pi^-$ straightforward via a $\omega$ exchange
\begin{equation}\label{OZI1}
 \Gamma(\phi\rightarrow\pi^+\pi^-)=\frac{2f_1^2m_\phi^4}{g^4m_\omega^4}
 \Gamma(\omega\rightarrow\pi^+\pi^-).
\end{equation}
Input data $\Gamma(\omega\rightarrow\pi^+\pi^-)=(185\pm 10){\rm KeV}$,
$\Gamma(\phi\rightarrow\pi^+\pi^-)=(0.35\pm 0.17){\rm KeV}$ and
$g=0.39$, we obtain
$$f_1\simeq 0.0028\pm 0.0014.$$
Then we can fit $\Lambda_{_M}\simeq 550$MeV and $g_2\simeq 0.47\times
10^{-3},\;g_3\simeq 0.12\times 10^{-3}$. It is consistent with our
previous discussion for $\Lambda_{_M}<m_\rho$.

The above results show that, the contributions yielded by spin-1 meson
loops is very small and the important contributions come from pseudoscalar
one-loop. Therefore, here we can ignore the former and input $L_1$ and
$L_9$(the effective gluon coupling do not contribute to $L_1$
and $L_9$) to fit $g_1$ and $g_\phi$. The results is $g_1=0.003\gg
g_2,\;g_3$ and $g_\phi^2=0.096$.

The numerical results on these low energy coupling constants are in
table 1, where we take $f_0=f_\pi=185$MeV, $m_{_V}=m_\rho=770$MeV
and $m_{_A}\simeq 1150$MeV obtained in sect. 3. In addition, to
leading order of $m_u$ and $m_d$, and taking $m_u+m_d\simeq
9$MeV\cite{Gao97}, we have $B_0=\frac{m_\pi^2}{m_u+m_d}\simeq
2$GeV. We find that all values of $L_i$ agree with data well.

\begin{table}[tbp]
\centering
 \begin{tabular}{cccccc}   \hline
     &ChPT &{leading order}&{one-loop}&gluon correction(k=0.038)&Total
        \\ \hline
   $L_1$&$0.7\pm 0.3$ &0.77&0.14&0&0.91 \\
   $L_2$&$1.3\pm 0.7$ &1.54&0.28&0&1.82 \\
   $L_3$&$-4.4\pm 2.5$&-4.48&$0.02^{a)}$&-0.09&-4.55 \\
   $L_4$&$-0.3\pm 0.5$&0 &0.13&0&0.13  \\
   $L_5$&$1.4\pm 0.5$&0.43&0.59&-0.21&0.81 \\
   $L_6$&$-0.2\pm 0.3$&0&0.07&0&0.07  \\
   $L_7$&$-0.4\pm 0.15$&0&$5\times 10^{-7}$&$-0.4\pm 0.1^{b)}$&
            $-0.4\pm 0.1$  \\
   $L_8$&$0.9\pm 0.3$ &0.38&0.16&0.16&0.70 \\
   $L_9$&$6.9\pm 0.7 $&6.1&0.38&0&6.48 \\
   $L_{10}$&$-5.2\pm 0.3$&-5.1&-0.38&-0.30&-5.78 \\
\hline
   \end{tabular}
\begin{minipage}{5in}
\caption {\small $L_i$ in units of $10^{-3}$, ${\mu}=m_\rho$.
   a)contribution from spin-1 meson loops. b)contribution from gluon
   anomaly.}
\end{minipage}
\end{table}

In the following we like to fit the parameter $k$ phenomenologically and
check whether it coincides with the result from QCD sum rules. Since
quadratic gluon condensate is weakly dependent on scale, we choose a
scale-independent quantity, $L_9+L_{10}$, to fit it here. The experimental
data require $L_9+L_{10}\geq 0.7\times 10^{-3}$, if we ignore contribution
from spin-1 meson loops, from Eq.(~\ref{7.4}) we obtain
$$\frac{\gamma}{6}(1-c)^2-\frac{k}{40}(1-c)^2\geq 0.7\times 10^{-3}.$$
So that we have $k\leq 0.038$. This value do not conflict with the result
from QCD sum rules.

It has been pointed out in sect. 3 that, there are six free parameters to 
parameterize effective lagrangian generated from quark loops. 
Up to the next leading order of $1/N_c$, four extra free parameters are
needed. They are $k$, $g_1$, $\Lambda_M$ and $\tau$. Then up to
powers four of derivatives and the next leading order of $1/N_c$, the
total ten real free parameters determine dynamics with energy scale below
axial-vector meson masses completely.

\subsection{Reexamining chiral sum rules}
\begin{description}
\item[1)] {\bf $\rho\rightarrow\pi\pi$ decay and KSRF sum rules}

Up to the next leading order of $1/N_c$, $f_{\rho\pi\pi}$ reads
\begin{equation}\label{7.3.1}
f_{\rho\pi\pi}=\frac{m_\rho^2}{gf_\pi^2}[16R_2(1-c)^2
  +2g^2c(1-\frac{c}{2})]=5.92.
\end{equation}
Then width of $\rho\rightarrow\pi\pi$ decay is yielded as follows
\begin{equation}\label{7.3.2}
\Gamma(\rho\rightarrow\pi\pi)=\frac{f_{\rho\pi\pi}^2}{48\pi}
   m_{\rho}(1-\frac{4m_\pi^2}{m_\rho^2})^{\frac{3}{2}}=145MeV.
\end{equation}
It agree with experimental data $150$MeV well. The KSRF (I) sum
rules\cite{KSRF}
$$g_{\rho\gamma}(m_\rho^2)=\frac{1}{2}f_{\rho\pi\pi}f_\pi^2$$
is the result of current algebra and PCAC.
$g_{\rho\gamma}=\frac{1}{2}gm_\rho^2$ has been obtained in
Eq.(~\ref{3.2.6}). The KSRF (I) sum rule yield
$$g=f_{\rho\pi\pi}\frac{f_\pi^2}{m_\rho^2}=0.343.$$ The error bar is about
$10\%$. Considering that error bar between KSRF (I) sum rule and experiment
is about $10\%$ too, our result agree with KSRF (I) sum rule well.
Furthermore, $\frac{1}{2}f_\pi^2f_{\rho\pi\pi}^2=(772{\rm MeV})^2$
yields the KSRF (II) sum rule\cite{KSRF}
$m_\rho^2=\frac{1}{2}f_\pi^2f_{\rho\pi\pi}^2$.

\item[2)] {Axial vector meson mass and Weinberg sum rule}

The axial-vector meson mass relation has been given in Eqs.(~\ref{3.1.15})
and (~\ref{3.2.11}). In leading order of $1/N_c$, $m_{_A}=1150$MeV close to
prediction by the second Weinberg sum rule $m_{_A}=\sqrt{2}m_{_V}$. To
the next leading to order of $1/N_c$, $R_1=-0.036$ and $g_{_A}^2=g^2+8R_1$
yield $m_{_A}=1230$MeV which very close to experimental data
$m_{_A}=1230\pm 40$MeV.

The first Weinberg sum rule is\cite{Wein67}
\begin{equation}\label{7.3.3}
\frac{f_\rho^2}{m_\rho^2}-\frac{f_a^2}{m_{_A}^2}=\frac{1}{4}f_\pi^2
\end{equation}
where $f_\rho$ and $f_a$ are defined by matrix element of vector and
axial-vector currents
\begin{eqnarray}\label{7.3.4}
<0|\bar{\psi}\frac{\tau^i}{2}\gamma_\mu\psi|\rho^{j\lambda}>&=&
 f_\rho\ep^\lambda_\mu\delta^{ij}, \nonumber \\
<0|\bar{\psi}\frac{\tau^i}{2}\gamma_\mu\gamma_5\psi|a_1^{j\lambda}>&=&
 f_a\ep^\lambda_\mu\delta^{ij}.
\end{eqnarray}
In this model, these currents are obtained easily through functional
differential for corresponding external fields,
\begin{eqnarray}\label{7.3.5}
  \bar{\psi}\frac{\tau^i}{2}\gamma_\mu\psi&=&
  \frac{\delta {\cal L}_\chi}{\delta v_i^\mu}, \nonumber \\
\bar{\psi}\frac{\tau^i}{2}\gamma_\mu\gamma_5\psi&=&
 \frac{\delta {\cal L}_\chi}{\delta a_i^\mu},
\end{eqnarray}
where $v$ and $a$ are vector and axial-vector external fields
respectively. To replace ${\cal L}_\chi$ in Eq.(~\ref{7.3.5}) by effective
lagrangian (~\ref{3.1.14}) and normalize vector and axial-vector mesonic
fields we obtain
\begin{eqnarray}\label{7.3.6}
\bar{\psi}\frac{\tau^i}{2}\gamma_\mu\psi&=&
 -\frac{1}{2}gm_\rho^2\rho_\mu^i+..., \nonumber \\
\bar{\psi}\frac{\tau^i}{2}\gamma_\mu\gamma_5\psi&=&
 (\frac{8\lambda(\mu)}{g_{_A}}-\frac{1}{2}g_{_A}m_{_A}^2)a_\mu^i+...=
 -\frac{m_2^2}{2g_{_A}}a_\mu^i+...,
\end{eqnarray}
where the mass relation(~\ref{3.1.15}) has been used. Comparing
Eq.(~\ref{7.3.4}) and Eq.(~\ref{7.3.6}) we have
\begin{equation}\label{7.3.7}
f_\rho=-\frac{1}{2}gm_\rho^2,\hspace{0.5in}
f_a=-\frac{m_2^2}{2g_{_A}}.
\end{equation}
Moreover, due to Eqs.(~\ref{3.1.15}) and (~\ref{3.2.11}) we have
\begin{equation}\label{7.3.8}
m_2^2=m_{_A}^2g_{_A}^2-16\lambda(\mu)m^2=\frac{f_\pi^2}{c}\simeq m_\rho^2g^2
\end{equation}
Substituting Eq.(~\ref{7.3.7}) into Eq.(~\ref{7.3.3}) and using the mass
relation (~\ref{7.3.8}) and Eq.(~\ref{3.2.4}) we can prove the
first Weinberg sum rule is satisfied in ChQM. It is not surprising since
the first Weinberg sum rule is derived from chiral symmetry, PCAC and VMD,
and the present theory is just a realization of chiral symmetry, PCAC and
VMD.
\end{description}

\section{Summary}

It was well known that, in very low energy, ChPT is a rigorous
consequence of the symmetry pattern in QCD and its spontaneous
breaking. Effective lagrangian of ChPT depends on a number of
low-energy coupling constants which can not be determined from the
symmetries of the fundamental theory only. As soon as we start
going beyond the lowest order, the number of free parameters
increases very rapidly, it makes calculation beyond the lowest few
order rather impractical. In addition, at energy lying scale
between perturbative QCD and ChPT, there is no well-defined method
to yield a explicitly convergence expansion. Therefore, instead of
ChPT, some phenomenological models are needed to capture the
physics between perturbative QCD and ChPT. ChQM is just the one of
such models with fewer free parameters. Since the low energy
coupling constants, $L_i(i=1,\;2,...,10)$, yielded by fewer
parameters in ChQM agree with ChPT, it reflects dynamics
constraint of these low energy constants.

In ChQM, the dynamics of composite meson fields are generated by the
quark loops. Of course, it is better if we can start with a
lagrangian which is purely fermionic and hadronic fields are generated by
the model itself. However, it is rather impractical since we know nothing
about how quarks and anti-quarks are bounded into hadrons. Furthermore,
the most ambitious attempts are to find a chirally symmetric solution
to the Schwinger-Dyson equations\cite{SD}. These methods are typically
plagued by instabilities in the solution of the equation.
So that the composite meson fields still have to be added explicitly by
hand in this paper.

We have presented in this article a systematic treatment of all spin-1
meson resonances $V$ and $A$ in framework of ChQM. All possible chiral
couplings between spin-1 meson resonances and pseudoscalar fields were
derived up to lowest order in the chiral expansion and the next to leading
order of $1/N_c$ expansion. Therefore, it become possible to study low
energy physics of mesons in a unified model with finite number of free
parameters. It should be noted that the $A_\mu-\pa_\mu\Phi$ mixing play a
crucial role in the model. This mixing is the energy transition
caused by quark loops instead of input from symmetry. Because the
gap between $m_{_A}^2g_{_A}^2$ and $f_\pi^2$ is not very
large($m_{_A}^2g_{_A}^2\simeq 4f_\pi^2$), the constant $c\simeq 0.45$
defined in the field shift(~\ref{3.1.12}) play a significant role. It
means that it can not be consistent if we introduce only vector meson
resonances but without their chiral partners in ChQM.

Through using method of SVZ sum rules in framework of ChQM, it was shown
that ChQM is indeed a phenomenologically successful model at low energy.
It is rather difficult to analytically determine the relation between ChQM
and various vacuum condensates in SVZ sum rules. However, it can be found
that the following basic idea of ChQM is surprisingly similar to one of
SVZ sum rules. 1)The mesonic dynamics in ChQM is generated by
quark loops(and gluon loops when quark-gluon coupling is included).
Correspondingly, in SVZ sum rules, the hadronic properties
are studied through parameterizing the effects caused by the vacuum
fields. Here, the effects of quark loops in ChQM correspond to various
fermion condensate of SVZ sum rules, and the effects of gluon loops in
ChQM correspond to various gluon condensate of SVZ sum rules. 2)
Physically, for avoiding double counting, we require that the gluon fields
must be {\sl soft} so that exchange effects of gluons between quark fields
do not creat bound states. The similar conclusion is obtained in SVZ sum
rules clearly\cite{Novikov85}, that "{\sl only soft(low frequency) modes
are to be retained in the (gluon) condensates}"\cite{Shifman99}. A result
is yielded directly from the above conclusion, that contribution from the
triple gluon condensate $<0|g_s^3GGG|0>$ is smaller than one from triple
gluon terms in quadratic gluon condenstate $<0|g_s^2GG|0>$. The reason is
that momentum power counting of the former is higher than one of the
latter. This result also coincides with phenomenological analysis
previously.

We have shown how to systematically calculate meson one-loop graphs
in ChQM. As well as our expectation, the contributions from
soft gluon coupling and meson one-loop effects are both 
smaller than one from quark loops. We must point out that, instead of
$1/N_c$ suppression, the contribution from meson loops are suppressed by
``very long-distance'' property of meson interaction in
meson loops(or scale property). Theoretically, the large $N_c$ expansion
is still an attractive argument\cite{tH74}. However, since in practical we
only take $N_c=3$ instead of $N_c\rightarrow\infty$, the next to leading
order contribution is not suppressed intensively by $1/N_c$ expansion.
Especially, many contributions from meson one-loop are proportional to
flavor number $N_f=3$, so that in fact they do not receive any suppression
from a theoretically well-defined expansion. We have to be aware that it
is a difficulty how the large $N_c$ argument is applied in case of
$N_c=3$ for the real world.

There are some conclusions for meson one-loop correction. 1)
It is a important feature on meson loops that the square of momentum
transfer in meson loops is very small. This feature indicates that
the contributions from the ``low-frequency'' of quantum meson fields are
dominant. It is common conclusion of low energy theory. 2) Although
the contribution from meson loops is small, many nontrivial results are
yielded. For example, $L_4$, $L_6$ and some processes forbidden by OZI
rules do not vanish here(further studies on OZI forbidden processes will
be done elsewhere). 3) Due to large mass gap between pseudoscalar
and spin-1 mesons, contribution from one-loop graphs of $0^-$ mesons is
much larger than the one from one-loops of $1^\pm$ mesons. 4) The
calculation on meson one-loop is formal in this paper. There are still
some problems that need to further study, e.g., imaginary part of $S$
matrix element appears only via direct calculation on relevant Feynman
diagrams. It is important to examine unitary of $S$ matrix element and
will be found in other papers.

\begin{center}
{\bf ACKNOWLEDGMENTS}
\end{center}
We would like to thank Prof. D.N. Gao and Dr. J.J. Zhu for their helpful
discussion. This work is partially supported by NSF of China through C. N.
Yang and the Grant LWTZ-1298 of Chinese Academy of Science.

\section*{Appendix Momentum-integral Formulas for Meson One-loops}
\setcounter{equation}{0}

It has been pointed out that the operators in
Eqs.(~\ref{5.11})-(~\ref{5.16}),
${\cal H}_j^{AB}(x),\;\Omega_{ij}^{AB}(x)$ and their adjoint operators
include one differential operator at the most. Therefore we set
\begin{eqnarray}\label{8.1}
{\cal H}_j^{AB}(x)&=&{\cal C}_j^{\mu,AB}(x)\pa_\mu+... \nonumber \\
{\cal H}_j^{*AB}(x)&=&{\cal C}_j^{*\mu,AB}(x)\pa_\mu+... \nonumber \\
\Omega_{ij}^{AB}(x)&=&{\cal P}_{ij}^{\mu,AB}(x)\pa_\mu+... \\
\Omega_{ij}^{*AB}(x)&=&{\cal P}_{ij}^{*\mu,AB}(x)\pa_\mu+... \nonumber
\end{eqnarray}
where $j=\phi(x),\;\vphi(x)$ and $\Psi(x)$ are fields. Employing
Eq.(~\ref{5.16}), Eqs.(~\ref{5.11})-(~\ref{5.13}) are rewritten
\begin{eqnarray}\label{8.2}
&&i{\cal L}_{one-loop}^{(jj)}(x)=\int\frac{d^Dq}{(2\pi)^D}\{
   {\cal H}_j^{AB}(x,q)\Delta_{AB}(q) \nonumber \\
&+&\frac{1}{2}{\cal H}_j^{AB}(x,q)\Delta^{(j)}_{BC}(q-i\pa)]
 ({\cal H}_j^{CD}(x,q)+{\cal H}_j^{*DC}(x,q))\Delta_{DA}^{(j)}(q)
            \nonumber \\
&+&\frac{1}{3}{\cal H}_j^{AB}(x,q)\Delta_{BC}^{(j)}(q-i\pa)
  ({\cal H}_j^{CD}(x,q)+{\cal H}_j^{*DC}(x,q))\Delta_{DE}^{(j)}(q-i\pa)
       \nonumber \\ && \times
  ({\cal H}_j^{EF}(x,q)+{\cal H}_j^{*FE}(x,q))\Delta_{FA}^{(j)}(q)
           \nonumber \\
&+&\frac{1}{4}{\cal H}_j^{AB}(x,q)\Delta_{BC}^{(j)}(q-i\pa)
 ({\cal H}_j^{CD}(x,q)+{\cal H}_j^{*DC}(x,q))\Delta_{DE}^{(j)}(q-i\pa)
  \nonumber \\ && \times
 ({\cal H}_j^{EF}(x,q)+{\cal H}_j^{*FE}(x,q))
 \Delta_{FG}^{(j)}(q-i\pa)
({\cal H}_j^{GH}(x,q)+{\cal H}_j^{*HG}(x,q))\Delta_{HA}^{(j)}(q)
  +...\},
\end{eqnarray}
\vspace{0.3in}
\begin{eqnarray}
\label{8.3}
&&i{\cal L}_{one-loop}^{(ij)}(x)=\int\frac{d^Dq}{(2\pi)^D}\{
     \frac{1}{2}\Omega_{ij}^{AB}(x,q)\Delta_{BC}^{(j)}(q-i\pa)
     \Omega_{ji}^{DC}(x,q)\Delta_{DA}^{(i)}(q) \nonumber \\
&+&\frac{1}{2}\sum_{k=i,j}
       \Omega_{ij}^{AB}(x,q)\Delta_{BC}^{(j)}(q-i\pa)
       \Omega_{ji}^{DC}(x,q)\Delta_{DE}^{(i)}(q-i\pa)
({\cal H}_k^{EF}(x,q)+{\cal H}_k^{*FE}(x,q)\Delta_{FA}^{(k)}(q)
           \nonumber \\
&+&\frac{1}{4}\Omega_{ij}^{AB}(x,q)\Delta_{BC}^{(j)}(q-i\pa)
     \Omega_{ji}^{DC}(x,q)\Delta_{DE}^{(i)}(q-i\pa)
     \Omega_{ij}^{EF}(x,q)\Delta_{FG}^{(j)}(q-i\pa)
     \Omega_{ji}^{HG}(x,q)\Delta_{HA}^{(i)}(q)
            \nonumber \\
&+&\frac{1}{2}\Omega_{ij}^{AB}(x,q)\Delta_{BC}^{(j)}(q-i\pa)
({\cal H}_j^{CD}(x,q)+{\cal H}_j^{*DC}(x,q))\Delta_{DE}^{(j)}(q-i\pa)
            \nonumber \\  &&  \hspace{0.3in} \times
     \Omega_{ji}^{FE}(x,q)\Delta_{FG}^{(i)}(q-i\pa)
({\cal H}_i^{GH}(x,q)+{\cal H}_i^{*HG}(x,q))\Delta_{HA}^{(i)}(q)
            \nonumber \\
&+&\frac{1}{2}\sum_{k=i,j}
     \Omega_{ij}^{AB}(x,q)\Delta_{BC}^{(j)}(q-i\pa)
     \Omega_{ji}^{DC}(x,q)\Delta_{DE}^{(i)}(q-i\pa)
 ({\cal H}_k^{EF}(x,q)+{\cal H}_k^{*FE}(x,q))
           \nonumber \\ && \hspace{0.3in} \times
      \Delta_{FG}^{(k)}(q-i\pa)
({\cal H}_k^{GH}(x,q)+{\cal H}_k^{*HG}(x,q))\Delta_{HA}^{(k)}(q)
      +....\}  \hspace{1.0in} (i\neq j)
\end{eqnarray}
\vspace{0.3in}
\begin{eqnarray}
\label{8.4}
i{\cal L}_{one-loop}^{(\phi\Psi\varphi)}(x)
         &=&\int\frac{d^Dq}{(2\pi)^D}
    \Omega_{\phi\varphi}^{AB}(x,q)\Delta_{(\varphi)BC}(q-i\pa)
    \Omega_{\varphi\Psi}^{CD}(x,q)\Delta_{(\Psi)DE}(q-i\pa)
    \nonumber \\
    &&\times\Omega_{\Psi\phi}^{EF}(x,q)\Delta_{(\phi)FA}(q)
      +....
\end{eqnarray}
where
\begin{eqnarray}\label{8.5}
{\cal H}_j^{AB}(x,q)&=&iq_\mu{\cal C}_j^{\mu,AB}(x)+
        {\cal H}_j^{AB}(x) \nonumber \\
{\cal H}_j^{*AB}(x,q)&=&iq_\mu{\cal C}_j^{*\mu,AB}(x)+
       {\cal H}_j^{*AB}(x) \nonumber \\
\Omega_{ij}^{AB}(x,q)&=&iq_\mu{\cal P}_{ij}^{\mu,AB}(x)+
         \Omega_{ij}^{AB}(x) \\
\Omega_{ji}^{AB}(x,q)&=&\Omega_{ij}^{*AB}(x)=iq_\mu
  {\cal P}_{*ij}^{\mu,AB}(x)\pa_\mu+\Omega_{ji}^{AB}(x) \nonumber
\end{eqnarray}

\end{document}